\documentclass[aps,prc,superscriptaddress,showpacs,nofootinbib,floatfix,twocolumn]{revtex4}

\usepackage{graphicx}
\usepackage{dcolumn}

\usepackage{xspace}

\newcommand{\GeVc}      {\ensuremath{\rm{GeV}\!/c}\xspace}
\newcommand{\pT}        {\ensuremath{p_{\rm T}}\xspace}
\newcommand{\pt}        {\ensuremath{p_{\rm T}}\xspace}

\newcommand{\dphi}      {\ensuremath{\Delta\phi}\xspace}

\newcommand{\vtwo}      {\ensuremath{v_{2}}\xspace}
\newcommand{\vTwo}      {\ensuremath{v_{2}}\xspace}

\newcommand{\vtworaw}   {\ensuremath{v_{2}^{\rm meas}}\xspace}

\newcommand{\vtwocorr}  {\ensuremath{v_{2}^{\rm corr}}\xspace}

\newcommand{\RAA}       {\ensuremath{R_{\rm AA}}\xspace}

\newcommand{\RAAphi}   {\ensuremath{R_{\rm AA}(\Delta \phi)}\xspace}

\newcommand{\RAApt}    {\ensuremath{R_{\rm AA}(p_{\rm T})}\xspace}

\newcommand{\leff}      {\ensuremath{\rho L_{xy}}\xspace}

\newcommand{\piz}      {\ensuremath{\pi^{0}}\xspace}

\newcommand{\pp}       {\ensuremath{p+p}\xspace}
\newcommand{\AuAu}     {\ensuremath{\rm Au+Au}\xspace}

\newcommand{\Lepsi}    {\ensuremath{L_\epsilon}\xspace}

\newcommand{\lepsi}    {\ensuremath{L_\epsilon}\xspace}

\newcommand{\meanTAA}  {\ensuremath{\left<T_{{\rm AA}}\right>}\xspace}
\newcommand{\meanNcoll} {\ensuremath{\left<N_{{\rm coll}}\right>}\xspace}
\newcommand{\Ncoll}    {\ensuremath{N_{{\rm coll}}}\xspace}

\newcommand{\Npart}    {\ensuremath{N_{{\rm part}}}\xspace}
\newcommand{\npart}    {\ensuremath{N_{{\rm part}}}\xspace}

\newcommand{\rhopart}  {\ensuremath{\rho_{{\rm part}}}\xspace}
\newcommand{\rhocent}  {\ensuremath{\rho_{{\rm cent}}}\xspace}

\newcommand{\rhoLxy}   {\ensuremath{\rho {L_{xy}}}\xspace}
\newcommand{\rhoLxyDiv}{\ensuremath{\rhoLxy/\rhocent}\xspace}

\newcommand{\snn}      {\ensuremath{\sqrt{s_{_{NN}}}}\xspace}

\newcommand{\Emcal}  {EMCal\xspace}

\newcommand{\pythia} {{\sc pythia}\xspace}

\def\mean#1{\left<#1\right>}

\begin{document}

\title{High-\pT \piz Production with Respect to the Reaction Plane \\
in \AuAu Collisions at $\snn = 200$~GeV}

\newcommand{\abilene}{Abilene Christian University, Abilene, TX 79699, U.S.}
\newcommand{\banaras}{Department of Physics, Banaras Hindu University, Varanasi 221005, India}
\newcommand{\bnlphys}{Brookhaven National Laboratory, Upton, NY 11973-5000, U.S.}
\newcommand{\caucr}{University of California - Riverside, Riverside, CA 92521, U.S.}
\newcommand{\cns}{Center for Nuclear Study, Graduate School of Science, University of Tokyo, 7-3-1 Hongo, Bunkyo, Tokyo 113-0033, Japan}
\newcommand{\colorado}{University of Colorado, Boulder, CO 80309, U.S.}
\newcommand{\columbia}{Columbia University, New York, NY 10027 and Nevis Laboratories, Irvington, NY 10533, U.S.}
\newcommand{\dapnia}{Dapnia, CEA Saclay, F-91191, Gif-sur-Yvette, France}
\newcommand{\debrecen}{Debrecen University, H-4010 Debrecen, Egyetem t{\'e}r 1, Hungary}
\newcommand{\elte}{ELTE, E{\"o}tv{\"o}s Lor{\'a}nd University, H - 1117 Budapest, P{\'a}zm{\'a}ny P. s. 1/A, Hungary}
\newcommand{\fsu}{Florida State University, Tallahassee, FL 32306, U.S.}
\newcommand{\gsu}{Georgia State University, Atlanta, GA 30303, U.S.}
\newcommand{\hiroshima}{Hiroshima University, Kagamiyama, Higashi-Hiroshima 739-8526, Japan}
\newcommand{\ihepprot}{IHEP Protvino, State Research Center of Russian Federation, Institute for High Energy Physics, Protvino, 142281, Russia}
\newcommand{\illuiuc}{University of Illinois at Urbana-Champaign, Urbana, IL 61801, U.S.}
\newcommand{\isu}{Iowa State University, Ames, IA 50011, U.S.}
\newcommand{\jinrdubna}{Joint Institute for Nuclear Research, 141980 Dubna, Moscow Region, Russia}
\newcommand{\kaeri}{KAERI, Cyclotron Application Laboratory, Seoul, Korea}
\newcommand{\kek}{KEK, High Energy Accelerator Research Organization, Tsukuba, Ibaraki 305-0801, Japan}
\newcommand{\kfki}{KFKI Research Institute for Particle and Nuclear Physics of the Hungarian Academy of Sciences (MTA KFKI RMKI), H-1525 Budapest 114, POBox 49, Budapest, Hungary}
\newcommand{\korea}{Korea University, Seoul, 136-701, Korea}
\newcommand{\kurchatov}{Russian Research Center ``Kurchatov Institute", Moscow, Russia}
\newcommand{\kyoto}{Kyoto University, Kyoto 606-8502, Japan}
\newcommand{\labllr}{Laboratoire Leprince-Ringuet, Ecole Polytechnique, CNRS-IN2P3, Route de Saclay, F-91128, Palaiseau, France}
\newcommand{\lawllnl}{Lawrence Livermore National Laboratory, Livermore, CA 94550, U.S.}
\newcommand{\losalamos}{Los Alamos National Laboratory, Los Alamos, NM 87545, U.S.}
\newcommand{\lpc}{LPC, Universit{\'e} Blaise Pascal, CNRS-IN2P3, Clermont-Fd, 63177 Aubiere Cedex, France}
\newcommand{\lund}{Department of Physics, Lund University, Box 118, SE-221 00 Lund, Sweden}
\newcommand{\muenster}{Institut f\"ur Kernphysik, University of Muenster, D-48149 Muenster, Germany}
\newcommand{\myongji}{Myongji University, Yongin, Kyonggido 449-728, Korea}
\newcommand{\nagasaki}{Nagasaki Institute of Applied Science, Nagasaki-shi, Nagasaki 851-0193, Japan}
\newcommand{\newmex}{University of New Mexico, Albuquerque, NM 87131, U.S. }
\newcommand{\nmsu}{New Mexico State University, Las Cruces, NM 88003, U.S.}
\newcommand{\ornl}{Oak Ridge National Laboratory, Oak Ridge, TN 37831, U.S.}
\newcommand{\orsay}{IPN-Orsay, Universite Paris Sud, CNRS-IN2P3, BP1, F-91406, Orsay, France}
\newcommand{\pnpi}{PNPI, Petersburg Nuclear Physics Institute, Gatchina, Leningrad region, 188300, Russia}
\newcommand{\riken}{RIKEN Nishina Center for Accelerator-Based Science, Wako, Saitama 351-0198, JAPAN}
\newcommand{\rikjrbrc}{RIKEN BNL Research Center, Brookhaven National Laboratory, Upton, NY 11973-5000, U.S.}
\newcommand{\rikkyo}{Physics Department, Rikkyo University, 3-34-1 Nishi-Ikebukuro, Toshima, Tokyo 171-8501, Japan}
\newcommand{\saispbstu}{Saint Petersburg State Polytechnic University, St. Petersburg, Russia}
\newcommand{\saopaulo}{Universidade de S{\~a}o Paulo, Instituto de F\'{\i}sica, Caixa Postal 66318, S{\~a}o Paulo CEP05315-970, Brazil}
\newcommand{\seoulnat}{System Electronics Laboratory, Seoul National University, Seoul, Korea}
\newcommand{\stonybrkc}{Chemistry Department, Stony Brook University, Stony Brook, SUNY, NY 11794-3400, U.S.}
\newcommand{\stonycrkp}{Department of Physics and Astronomy, Stony Brook University, SUNY, Stony Brook, NY 11794, U.S.}
\newcommand{\subatech}{SUBATECH (Ecole des Mines de Nantes, CNRS-IN2P3, Universit{\'e} de Nantes) BP 20722 - 44307, Nantes, France}
\newcommand{\tenn}{University of Tennessee, Knoxville, TN 37996, U.S.}
\newcommand{\titech}{Department of Physics, Tokyo Institute of Technology, Oh-okayama, Meguro, Tokyo 152-8551, Japan}
\newcommand{\tsukuba}{Institute of Physics, University of Tsukuba, Tsukuba, Ibaraki 305, Japan}
\newcommand{\vandy}{Vanderbilt University, Nashville, TN 37235, U.S.}
\newcommand{\waseda}{Waseda University, Advanced Research Institute for Science and Engineering, 17 Kikui-cho, Shinjuku-ku, Tokyo 162-0044, Japan}
\newcommand{\weizmann}{Weizmann Institute, Rehovot 76100, Israel}
\newcommand{\yonsei}{Yonsei University, IPAP, Seoul 120-749, Korea}
\affiliation{\abilene}
\affiliation{\banaras}
\affiliation{\bnlphys}
\affiliation{\caucr}
\affiliation{\cns}
\affiliation{\colorado}
\affiliation{\columbia}
\affiliation{\dapnia}
\affiliation{\debrecen}
\affiliation{\elte}
\affiliation{\fsu}
\affiliation{\gsu}
\affiliation{\hiroshima}
\affiliation{\ihepprot}
\affiliation{\illuiuc}
\affiliation{\isu}
\affiliation{\jinrdubna}
\affiliation{\kaeri}
\affiliation{\kek}
\affiliation{\kfki}
\affiliation{\korea}
\affiliation{\kurchatov}
\affiliation{\kyoto}
\affiliation{\labllr}
\affiliation{\lawllnl}
\affiliation{\losalamos}
\affiliation{\lpc}
\affiliation{\lund}
\affiliation{\muenster}
\affiliation{\myongji}
\affiliation{\nagasaki}
\affiliation{\newmex}
\affiliation{\nmsu}
\affiliation{\ornl}
\affiliation{\orsay}
\affiliation{\pnpi}
\affiliation{\riken}
\affiliation{\rikjrbrc}
\affiliation{\rikkyo}
\affiliation{\saispbstu}
\affiliation{\saopaulo}
\affiliation{\seoulnat}
\affiliation{\stonybrkc}
\affiliation{\stonycrkp}
\affiliation{\subatech}
\affiliation{\tenn}
\affiliation{\titech}
\affiliation{\tsukuba}
\affiliation{\vandy}
\affiliation{\waseda}
\affiliation{\weizmann}
\affiliation{\yonsei}
\author{S.~Afanasiev} \affiliation{\jinrdubna}
\author{C.~Aidala} \affiliation{\columbia}
\author{N.N.~Ajitanand} \affiliation{\stonybrkc}
\author{Y.~Akiba} \affiliation{\riken} \affiliation{\rikjrbrc}
\author{J.~Alexander} \affiliation{\stonybrkc}
\author{A.~Al-Jamel} \affiliation{\nmsu}
\author{K.~Aoki} \affiliation{\kyoto} \affiliation{\riken}
\author{L.~Aphecetche} \affiliation{\subatech}
\author{R.~Armendariz} \affiliation{\nmsu}
\author{S.H.~Aronson} \affiliation{\bnlphys}
% ...
\collaboration{PHENIX Collaboration} \noaffiliation

\date{\today}

\begin{abstract}

Measurements of the azimuthal
anisotropy of high-$\pT$ neutral pion ($\pi^0$) production in
\mbox{Au+Au} collisions at $\sqrt{s}_{NN} = 200$~GeV by the PHENIX experiment 
are presented. The data included in this paper were collected during the 
2004 RHIC running period and represent approximately an order of
magnitude increase in the number of analyzed events relative to previously
published results. Azimuthal angle distributions of $\pi^0$s detected in the PHENIX electromagnetic
calorimeters are measured relative to the reaction plane determined event-by-event 
using the forward and backward beam-beam counters. Amplitudes of the second Fourier 
component ($v_2$) of the angular distributions are presented as a function of $\pi^0$
transverse momentum ($\pT$) for different bins in collision centrality. Measured
reaction plane dependent \piz yields are used to determine the azimuthal dependence 
of the \piz suppression as a function of \pt, $\RAA(\Delta\phi,\pt)$.
A jet-quenching motivated geometric analysis is presented that
attempts to simultaneously describe the centrality dependence and
reaction plane angle dependence of the \piz suppression in terms of
the path lengths of hypothetical parent partons in the medium. 
This set of results allows for a detailed examination of the influence of geometry 
in the collision region, and of the interplay between collective flow and 
jet-quenching effects along the azimuthal axis.

\end{abstract}

\pacs{21.65.Qr,25.75.-q,25.75.Dw}
\maketitle

\section{Introduction}
\label{sec:intro}

Over the past few years, experiments at the Relativistic Heavy Ion
Collider (RHIC) have established that a dense partonic medium is
formed in \AuAu collisions at
$\sqrt{s_{NN}}$=200~GeV~\cite{Adcox:2004mh,Adams:2005dq,Back:2004je,Arsene:2004fa}.
This medium thermalizes very quickly~\cite{Adcox:2004mh,
Mrowczynski:1993qm, Arnold:2004ti, Rebhan:2004ur, Dumitru:2005gp,
Schenke:2006xu, Scherer:2008zz, Xu:2008zi}, is extremely opaque to the
passage of high-\pT particles~\cite{Adcox:2001jp,Adler:2002xw}, and
the strong coupling of matter in the medium produces a system for
which the ratio of shear viscosity to entropy ($\eta/s$) 
approaches zero~\cite{Adler:2003kt,Adams:2003am,Romatschke:2007mq,Dusling:2007gi,
Song:2007ux}. Much of the current focus is on the extraction of key
transport and thermodynamic characteristics of the matter produced in
these collisions.  Measurements of high-\pT parton propagation in the
medium as well as medium-induced modification of the fragmentation parton
spectrum and its products provide a critical tool for probing
medium properties.

One of the most striking early results from RHIC was the observation
of strongly suppressed production of high-\pt particles
in central Au+Au events
compared to appropriately scaled \pp
collisions~\cite{Adcox:2001jp,Adler:2002xw}.  High-\pT partons are
formed from hard scattering between the initial colliding
partons, and these partons fragment into two or more jets of hadrons.
When propagating through a dense volume of deconfined
matter, these high-\pT partons are expected to scatter from color
charges in the medium,
losing energy through a combination of gluon bremsstrahlung radiation
and collisional energy transfer to partons in the medium.
These radiated gluons
eventually fragment into hadrons at lower \pT, resulting in a
depletion of the observed yields of hadrons at higher \pT.

A useful way to quantify the suppression of high-\pT hadrons is the
nuclear modification factor ($\RAA$) where the \pp cross section is
scaled by the thickness function $\left<T_{\rm AA}\right>$ of the two Au nuclei
\[
\RAA(\pt) = \frac{(1/N^{\rm evt}_{\rm AA})d^2N^{\piz}_{\rm AA}/d\pt
  dy}{\left<T_{\rm AA}\right>\times d^2\sigma^{\piz}_{pp}/d\pt dy}.
\]
PHENIX has measured a \piz \RAA close to unity in both peripheral
\AuAu collisions and
\d{\rm A}u collisions~\cite{Adare:2008qa,Adler:2003ii}, consistent with the
expectation that these collisions would not produce an extended, dense
medium. As the collisions become more central, \RAA
decreases to about 0.2, indicating a stronger parton energy loss.
Furthermore, the measured \piz \RAA is nearly constant as a function of \pT, for \pT
$\gtrsim 5$~\GeVc up to the highest currently accessibly \pT,
$20$~GeVc~\cite{Adare:2008qa}. 

These data can be well reproduced by models that calculate the energy
lost by the hard scattered partons as they traverse the dense medium.
The amount of energy-loss depends on the density of
the medium~\cite{Baier:2002tc}, so measurements of high-\pt hadron suppression provide constraints on the transport
coefficient $\mean{\hat{q}^2}$, a measure of mean transverse momentum
squared $\mean{k_{\rm T}^2}$ transferred by the medium to a high-energy parton.
However, multiple models with different
physical assumptions can reproduce the measured
$\RAA(\pT)$~\cite{Majumder:2007iu,Vitev:2008jh}.  The different models
vary widely in how they include the crucial interference terms between
multiple-scattering centers as well as the interplay between
inelastic, elastic and flavor-changing processes during the parton's
passage.

To discriminate between these models we need to increase our
experimental control of the path length, since the amount of energy
lost by a high-\pT parton strongly increases with the distance
traveled through the medium.  A quadratic dependence on the path length
is predicted for a static medium if the dominant energy-loss mechanism
is the bremsstrahlung radiation of gluons surviving the
destructive interference caused by multiple
scattering~\cite{Majumder:2007iu,Vitev:2008jh}. For an expanding
plasma the quadratic increase should be moderated to a linear
dependence~\cite{Gyulassy:2001nm}.

The centrality dependence of $\RAA(\pt)$ offers a probe of the path-length
dependence of partonic energy loss.  However, we can better test the
path-length dependence by studying the azimuthal variation of the high-\pT
suppression at a fixed centrality.  Since the collision zone has a nearly
elliptical shape in the transverse plane due to the non-central overlap of the
colliding Au nuclei, partons that travel along the short axis of the nuclear
overlap region lose less energy and should therefore be less suppressed. The
key observable is then the two-dimensional modification factor
$\RAA(\dphi,\pT)$, where \dphi is the angle of emission with respect to the
event plane.  The azimuthal dependence of the spectra can be also
parameterized by a Fourier expansion, where up to second order
$\frac{dN}{d\dphi} =
N_0[1+2\vtwo\cos(2\dphi)]$, with $\vtwo$ being called elliptic flow coefficient.  
While both quantities characterize azimuthal asymmetries, historically
and conceptually they have different roots.  The notion of elliptic
flow is primarily tied to lower $\pT$ phenomena (``soft physics''),
the domain where particle production is proportional to the number of
participating nucleons (\Npart), and positive $\vtwo$ arises from
the {\it boost} to the mean $\pT$ in the direction where the pressure
gradient is highest (along the reaction plane).  Conversely, $\RAA(\pT)$ and 
$\RAA(\dphi,\pT)$ are commonly used to describe high $\pT$ behavior (hard
scattering, which scales with the number of binary collisions
\Ncoll). When $\RAA$ deviates from unity at high \pT, it becomes a
valuable probe of the {\it loss} of energy/momentum in a particular direction.
However, there is no clear separation between soft and hard
regions, and both $\RAA$ and $\vtwo$ are well-defined in the entire
momentum range, so in this sense $\vtwo$ is sensitive to
differential energy loss at high \pT.

PHENIX has measured high-\pt \vtwo for \piz particles from Au+Au
collisions~\cite{Adler:2006bw}. The energy-loss models that reproduce 
$\RAA(\pT)$ diverge in their predictions of the
azimuthal anisotropy at high \pT. They generally under-predict the 
observed azimuthal variation of $\RAA(\dphi)$, or equivalently, are unable 
to describe the \pT dependence of \vtwo over the full range of \pT where 
one would naively expect them to be applicable~\cite{PhysRevC.69.034908,Renk:2006sx,majumder:041902}.  
These models include the hydrodynamical evolution of the medium, and therefore the
high-\pT probe loses energy in a medium that is becoming spatially isotropic 
with time. 
Several early papers noticed that the
measured $\vtwo$ values were larger than what one would expect from a
completely opaque almond-shape collision
zone~\cite{Shuryak:2001me,Drees:2003zh}. Other early energy-loss
calculations came close to reproducing the measured
$\RAA(\dphi)$~\cite{Dainese:2004te,Cole2006225}, but in these the plasma 
expansion 
was not taken into account, which resulted in unrealistically strong 
azimuthal anisotropy.  
Another calculation~\cite{Zhang:2007ja} has reproduced
$\RAA(\dphi)$, but in this model the Au nuclei were parameterized as 
hard-spheres instead of using a more realistic Woods-Saxon density profile, and this mechanism
artificially increases the azimuthal dependence of the energy density.

One potential resolution of the problem with energy loss calculations
not reproducing the measured azimuthal dependence of yields is a recent
calculation that allowed the
high-\pT parton to resonantly scatter with the
medium~\cite{Liao:2008dk} (and references therein), increasing the energy lost by a parton
at plasma densities that correspond to temperatures near the critical temperature. 
This produces a sharper dependence of the energy-loss on the spatial 
variation of the medium's energy density and hence the
model is able to simultaneously reproduce both $\RAA(\pT)$ and
$\RAA(\dphi)$. A critical check will be to examine whether the
same parameters work for the full range of collision centralities.

In order to discriminate among all the models that 
attempt to reproduce $\RAA(\dphi,\pT)$, the experimental challenge 
is to extend the range and increase the precision of observations which 
can be used to test different energy-loss models. In this paper we extend the range
of published data on $\RAA(\dphi)$~\cite{Adler:2006bw} by a) reaching higher \pT, and 
thereby moving to a \pT region that is completely
dominated by the fragmentation of hard partons and reducing the
possible contribution of particles from
recombination~\cite{Fries:2003kq}, b) using finer bins in centrality, thus
achieving less averaging of the path length, and c) reducing the
statistical and systematic uncertainties to further constrain models.
We present in this article measurements using data collected during the 2004 RHIC running period.  
These data represent a high-statistics sample
of \AuAu collisions (approximately 50 times that of the 2002 RHIC
running period) and
therefore extend our ability to measure \RAAphi and \vTwo to much
higher \pT with better precision.

\section{Experimental Details}
\label{sec:exp}

The data presented in this paper were taken by the PHENIX experiment~\cite{Adler:2003zu}
in 2004 (RHIC Run-4), and represent the analysis of 821M minimum bias
Au+Au collisions at $\sqrt{s_{NN}}=200$~GeV.  The detectors involved in
this analysis are the beam-beam counters~\cite{Allen:2003zt} (BBC; triggering, centrality
and reaction plane determination), the zero-degree calorimeter~\cite{Adler:2000bd}
(ZDC, centrality determination) and the electromagnetic calorimeter~\cite{Aphecetche:2003zr}
(EMCal, $\pi^0$ measurement).  

The BBCs are two groups of 64 hexagonal quartz \v{C}erenkov radiator
counters with photomultiplier readout surrounding the beampipe 144 cm
up- and downstream (``North'' and ``South'') from the center of the
nominal collision diamond, covering the $3 < |\eta| <3.9$
pseudorapidity range and the full azimuth.  Coincidence of signals in
at least two photomultiplier tubes in both BBCs served as a minimum
bias trigger and according to simulations it captured 92\% of all
inelastic collisions.  The size of the total signal in the BBCs
increases monotonically with collision centrality at this
$\sqrt{s_{NN}}$.  The collision vertex $z$ was calculated from the
difference between the fastest timing signals in the North and South BBCs,
respectively, with $\sigma<2.0$~cm resolution.  Only events with
$|z|<30$~cm were analyzed.

The ZDCs are small tungsten/scintillator hadron calorimeters with
quartz fiber lightguides and photomultiplier readout, located between
the beampipes at 18~m North and South from the collision point.  They
measure non-interacting ``spectator'' neutrons in a cone of about
2~mrad, and their signal is double-valued as a function of centrality
(it is low in very central and very peripheral collisions but large at
mid-centrality).  The correlation of ZDC {\it vs} BBC signals resolves
this ambiguity and allows a precise measurement of the true centrality
for all but the most peripheral collisions.

The reaction plane (spanned by the beam direction and the impact
vector of the colliding nuclei) is determined event-by-event from the
azimuthal charge distribution in the BBCs, after taking into account
small nonuniformities (in the response of individual radiators,
PMTs, electronics, etc.{}), using the assumption that over a large
number of events the $\phi$ distribution of per-event reaction planes
should be uniform.  Due to the large rapidity gap between the central
arm ($|\eta|<0.35$) where $\pi^0$s are measured, and the BBCs where
the reaction plane is established, we assume that the reaction plane
is unbiased and free from auto-correlations.  However, the relatively
coarse granularity of BBCs affects the resolution.  Note that in this
analysis precise knowledge of the reaction plane resolution, which
depends strongly on centrality, is crucial.  This will be discussed in
detail in the next Section.

Neutral pions are measured by reconstructing their decay photons 
($\pi^0 \rightarrow \gamma\gamma$) in the EMCal.  The EMCal consists
of 8 sectors at midrapidity ($|\eta|<0.35$), covering a total of
$2\times90^\circ$ in azimuth.  Six sectors are lead/scintillator
(PbSc) sampling calorimeter with photomultiplier readout and
$5.5\times5.5$~cm$^2$ granularity, two sectors are lead/glass (PbGl)
\v{C}erenkov counters with $4\times4$~cm$^2$ granularity and 
photomultiplier readout.  The two detectors are 18$X_0$ and 16$X_0$
radiation lengths deep, respectively, both ensuring essentially full
containment of electromagnetic showers in the relevant energy range.
The {\it in situ} energy resolution is well reproduced by
simulation both in PbSc and PbGl: the \piz peak positions and the
widths both agree with the data to better than $1\,$~MeV over the entire momentum
range.  Therefore, the error on the energy (and momentum) scale is
less than 1\%.
Timing resolution $\sigma_t$ is $\sim 450$~ps and $\sim 650$~ps 
for the PbSc and PbGl, respectively, allowing the
rejection of neutrons and antineutrons up to a few \GeVc transverse
momentum, which would otherwise be a major source of neutral showers
up to a few GeV energy.  
At sufficiently high transverse momenta, decay photons from
a nearly symmetric ($E_{\gamma_1} \approx E_{\gamma_2}$) decay may
produce showers in the calorimeter that start to merge into one
reconstructed cluster.  In the PbSc this effect is first visible
around \pT~$\sim10$~\GeVc, at the upper end of the \pT region
considered in this paper.  Due to its higher granularity and smaller
Moli\`ere-radius the PbGl is immune to this ``merging'' problem up to
\pT~$\sim$15~\GeVc.
The hadronic response, timing properties and other sources 
of systematic errors are very different for the two calorimeter types.
Therefore, when extracting the $\phi$-integrated \RAA, which
serves as absolute normalization, the PbSc and PbGl were analyzed separately
and the results combined to decrease the total systematic 
uncertainty.

\section{Data Analysis}
\label{sec:data_ana}

\subsection{Centrality}

As mentioned, the minimum-bias trigger in the Run-4 PHENIX configuration is supplied
by the BBCs, and
the correlation of the
charge deposited in the BBCs with energy deposited in the ZDCs
provided a determination of the centrality of the
collision.  The elliptic flow measurement presented in this paper is
measured in seven bins of the centrality range 0-92\%, 
with lowest corresponding to the most central: 0-5\%, 5-10\%,
10-20\%, 20-30\%, 30-40\%, 40-50\%, and 50-60\%.  In addition, the
combined ranges 0-20\%, 20-40\%, 40-60\%, and minimum bias bins are
included.  For the yields with respect to the reaction plane, the
centralities presented are 0-10\%, 10-20\%, 20-30\%, 30-40\%, 40-50\%,
50-60\%.  Finally, the \RAA versus nuclear path length result excludes the
most central bin due to its smaller intrinsic ellipticity 
(the average path length is insensitive to $\dphi$).

\subsection{Reaction plane determination}
\label{sec:reactionplane}
The technique used to determine the reaction plane on an
event-by-event basis is the same used in previous PHENIX
analyses~\cite{Adler:2003kt,Adler:2005rg,Afanasiev:2007tv}.
The quartz radiators of each counter are arranged in approximately
concentric circles around the beam axis.  The light collected in the
photo-multiplier tubes (PMTs) allows for an estimate of the number of
charged particles passing through the detector.

The number of charged particles at a given PMT position, $N_i$, is
weighted in a manner to reduce the bias of the inner rings and used to
measure the orientation of the reaction plane from the formula
\begin{equation}
\tan(2\Psi) = \frac{\sum_i w_i N_i \sin(2\phi_i)-\left<\sum_i w_i N_i
\sin(2\phi_i)\right>}{\sum_i w_i N_i \cos(2\phi_i)-\left<\sum_i w_i N_i
\cos(2\phi_i)\right>},
\end{equation}
where $\phi_i$ is the nominal azimuth of the radiator.  Subtraction of
the average centroid removes biases due to various detector effects.
A final flattening technique is used to remove the residual
non-uniformities in the distribution of angles.

To estimate the resolution of the reaction plane measurement, we use the
sub-event technique~\cite{Ollitrault:1997di}.  The approach consists
of dividing the event up into two sub-events roughly equal in size.  The two
individual BBCs provide a natural sub-event division, so we analyze the
distribution of event-by-event differences between the reaction plane 
angles measured in the north and south counters, $\Delta\Psi =
\Psi_{N}-\Psi_{S}$.  In the presence of pure flow, this distribution 
takes the form \cite{Ollitrault:1997di}:

\begin{figure}[tb]
\includegraphics[width=1.0\linewidth]{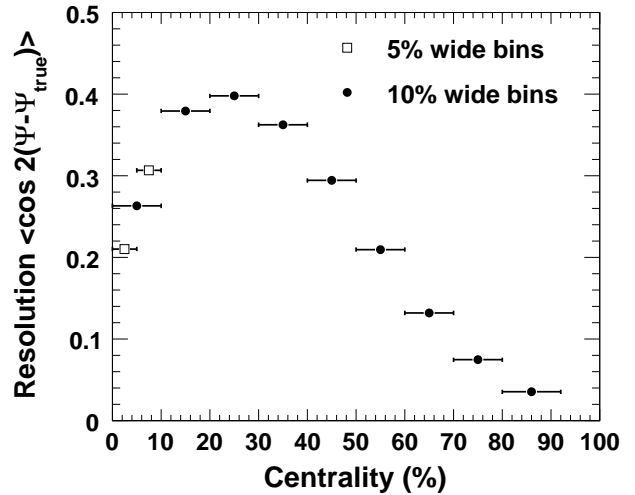}
\caption{Reaction plane resolution correction as a function of centrality.}
\label{fig:rpres}
\end{figure}

\clearpage

\begin{widetext}
\begin{equation}
\frac{dN}{d\Delta\Psi} =
\frac{e^{-\chi^2}}{2}\left\{\frac{2}{\pi}(1+\chi^2)+z\left[I_0(z)+\rm{L}_0(z)\right] +
  \chi^2\left[I_1(z)+\rm{L}_1(z)\right]\right\} \label{eq:Olli}.
\label{eq:ollitrault}
\end{equation}
where $z=\chi^2 \cos{(2\Delta\Psi)}$ and the functions $I_n$ and
$\rm{L}_n$ are the modified Bessel functions of the first kind and
modified Struve functions, respectively. The parameter $\chi$
describes the dispersion of the flow $\vec{Q}$ vector and thus
determines the correction required for the reaction plane resolution. 
Since $\Delta\Psi$ represents the whole-event difference distribution and we
are dealing with sub-events with roughly half the multiplicity of the event, we
replace $\chi \rightarrow \chi/\sqrt{2}$ in Eq.~\ref{eq:ollitrault}
and fit this function to the measured $\Delta\Psi$ distribution to
extract $\chi$. The resulting value is then used to evaluate
the resolution of the event-plane of $n^{th}$
order~\cite{Ollitrault:1997di}:
\begin{equation}
\langle\cos n\left(\Psi-\Psi_{RP}\right)\rangle = \frac{\sqrt{\pi}}{2}\chi
  e^{-\frac{\chi^2}{2}} \left[I_{\frac{n-1}{2}}\left(\frac{\chi^2}{2}\right) +
  I_{\frac{n+1}{2}}\left(\frac{\chi^2}{2}\right)\right] \label{eq:rp_res}
\end{equation}
\end{widetext}
where the true reaction plane orientation is denoted by $\Psi_{RP}$
and the observed orientation by $\Psi$.  Figure~\ref{fig:rpres} shows
the resolution correction obtained using the above-described procedure
as a function of centrality.  Both 5\% and 10\% wide bins are shown
for comparison. 

Eq.~\ref{eq:Olli} is derived under the assumption that the azimuthal
distributions are free of non-flow effects.  Due to the large rapidity
gap between the BBCs and the central arm region, it is expected that
particles observed in the BBCs have no correlation with those measured
in the central arm detectors. \pythia~\cite{Sjostrand:2006za} studies
have been used to confirm that jets observed in the central arm have
negligible effect on the reaction plane measurement from the
BBCs\cite{Adare:2008cqb}.

\subsection {Neutral pion measurement}
\label{sec:yield}

Measurement of neutral pions has played a critical role in the study of
high-\pt phenomena at RHIC, and especially by
PHENIX~\cite{Adcox:2001jp,Adler:2006bw,Adare:2008qa}. 
The two-particle decay channel $\pi^0 \rightarrow \gamma + \gamma$ provides a
clean signal of identified hadrons out to the highest \pt regions.  

\Emcal showers are found by clustering contiguous towers with energy
above a threshold energy (10 MeV) and requiring at least 50 MeV in the
tower with highest energy deposit.  The impact position is calculated
from the positions of the participating towers weighted by the
logarithm of deposited energy.  The energy of the cluster is corrected for
non-perpendicular incidence -- the angle being derived by assuming a
straight path between the actual vertex and the calculated impact
point -- as well as nonlinearities~\cite{Adler:2006bw}.  In
high-multiplicity events such as central \AuAu collisions, there is
an increasing probability for clusters to overlap (one tower accumulates energy
from more than one particle), which can distort an energy
measurement from a simple sum over contiguous towers.  
To mitigate this effect, the EMCal clustering algorithm also provides
a quantity called {\it ecore}, which is determined by
extrapolating the ``core'' energy represented by the central four or five
towers in the cluster, assuming an electromagnetic shower profile.  The
energy- and impact angle- dependent
shower profile is a model developed from and checked against beam test 
data.  In this
way, {\it  ecore} provides a more realistic measurement of the shower 
energy, 
less prone to contributions from accidental overlaps (particles
hitting close enough they deposit energy in the same towers) than a simple energy sum of participating towers
would be.
We use this 
{\it ecore} for the energy of reconstructed clusters in this analysis.

The invariant mass of a photon pair $\gamma_i,\gamma_j$ 
as measured in the EMCal 
is calculated from the energy of the clusters and their measured position:
\begin{equation}
  m_{\gamma_i\gamma_j} = \sqrt{(P_{\gamma_i}^2+P_{\gamma_j}^2)} = \sqrt{2E_iE_j\cos(1-\theta_{ij})}
\end{equation}
where $\theta_{ij}$ is the opening angle between the two photons
and $m_{\gamma_i\gamma_j}$ is equal to the \piz mass for 
photons from the decay of the same \piz.  Since the photons from the
\piz are not tagged, such pairs have to be formed from each photon
pair in the event where the pair momentum falls in a particular \pT
bin, and some of these pairs might accidentally reproduce the \piz
mass as well (combinatorial background), particularly at lower \pT
and higher centralities (multiplicities).
Since \piz{}s cannot be uniquely identified, raw \piz yields are
extracted statistically, by subtracting the combinatorial background
from the invariant mass distribution.

A well-known technique to reduce the combinatorial background is to
place a cut on the energy asymmetry of the pair, as defined by:
\begin{equation}
\alpha = \frac{|E_1 - E_2|}{E_1+E_2} = \beta|\cos\theta^*|.
\end{equation}
Because the angular distribution $d\sigma/d\cos\theta^*$ of the pairs in the
rest frame of the \piz is uniform, the asymmetry distribution should
be flat.  
However, due to the steeply falling photon
spectrum, fake (non-correlated) pairs which still give the
proper \piz mass are strongly peaked towards $\alpha=1$. A pair of 
clusters in the \Emcal is considered a neutral pion candidate 
only if the pair's asymmetry is less than 0.8.  In addition, the two photons 
are required to be separated by at least 8~cm for the combination 
to be considered as a \piz candidate.

There remains a non-trivial background contribution which passes these cuts:
pairs of photons from different \piz{}s, or, more generally, from
pairs of uncorrelated clusters which pass the photon identification
cuts and accidentally give an invariant mass near the true \piz
mass.  This 
remaining combinatorial background is estimated and subtracted
using the event mixing method.  The procedure
involves forming pairs from different events, which will by definition be
uncorrelated.  Each photon candidate is combined with all the photon
candidates  in previous events stored in memory.  
In order to replicate the background from
uncorrelated pairs within the same event as 
closely as possible in
the mixed events, mixing is performed within bins of
centrality, vertex $z$ position, and reaction plane orientation.  
Since all events analyzed are minimum-bias, no special steps are
needed to avoid the distortions of the mixed-event background by 
the trigger requirement.  All cuts applied to
the combinations of same event pairs are also applied to mixed-event pairs.
The number of events buffered determines the statistics of
the event-mixed distributions, chosen as a tradeoff between desired statistical
accuracy and computational resources.  The data presented in this article are mixed with
five previous events (in each centrality, vertex, and reaction plane bin).  

\begin{figure*}
\includegraphics[width=0.9\linewidth]{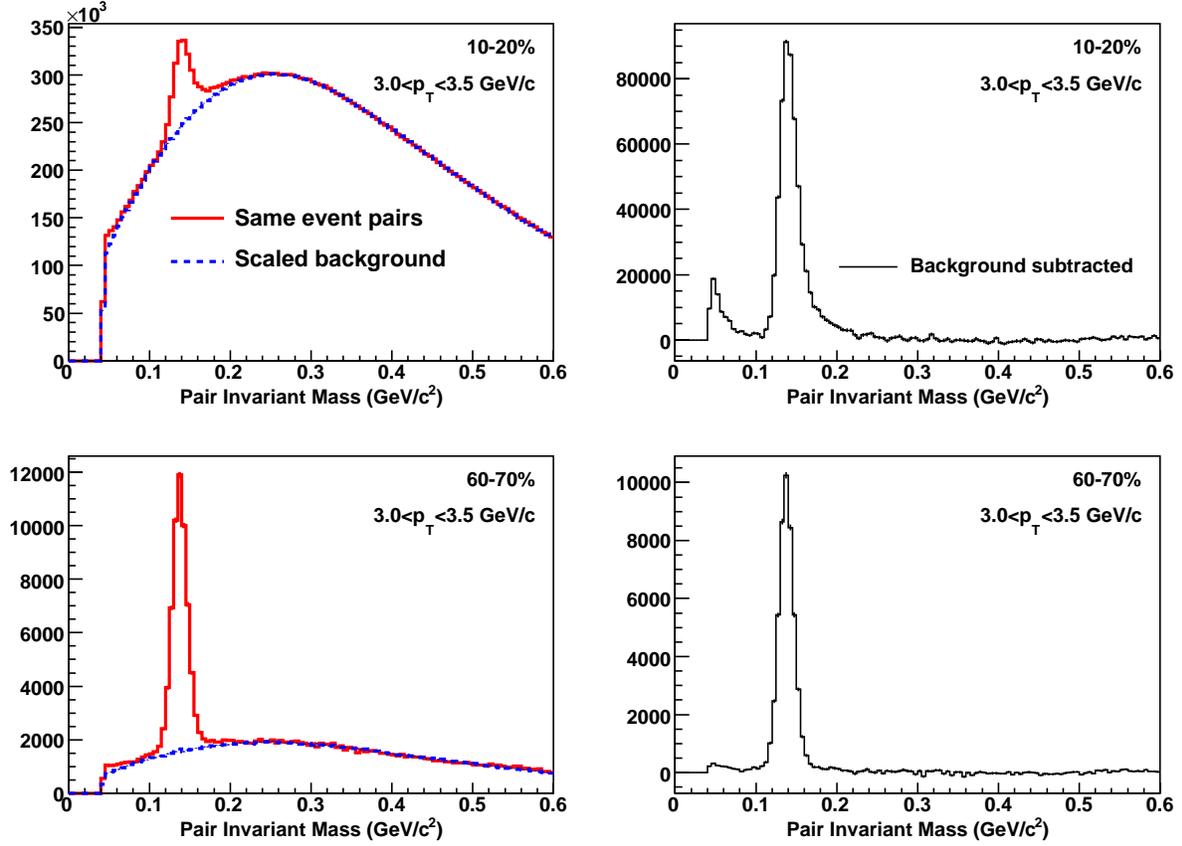}
\caption{Invariant mass distributions at moderate \pt and two
  different centralities.  Left panels: same event and normalized
  mixed event distributions.  Right panel: the subtracted
  distributions, which are then fitted with the sum of a first degree
  polynomial and a Gaussian. }
\label{fig:pi0mass}
\end{figure*}

For a given \pt bin, the mixed-event mass distribution is normalized to the
same-event distribution in a region away from the \piz mass peak.  The
normalization region is 0.25--0.45~GeV/$c^2$ for $\pT < 6.0$~GeV/$c$ and
0.21--0.45~GeV/$c^2$ otherwise.  
Fig.~\ref{fig:pi0mass} shows an example of this subtraction
process for two \pt ranges in two centrality bins.

The scaled background distribution is then subtracted from the same-event pair
distribution.  The subtracted result thus represents a sample of real \piz{}s.
The peak is fit to a Gaussian to determine its width and mean position.  The
raw yield of \piz{}s is determined by integrating the counts in a window of
$\pm 2\sigma$ around the mean.  The width and mean are recorded and
parameterized as a function of \pt and centrality based upon this
$\phi$-integrated, large sample.
The positions and widths from this parameterization are then used
when we extract the (much smaller) raw yields in bins of angle
$\Delta\phi$ with respect to the reaction plane. The
maximum variation of the yields (multiplicities) with $\Delta\phi$ is
only about a factor of 2, and therefore the means and
widths are not expected to change substantially. 
Furthermore, the statistics are
much poorer in the $\Delta\phi$ bins, which would make individual \piz
fits unreliable.

There is a residual background in the invariant mass distributions even after
the mixed-event distribution has been removed, especially at lower \pT (below
$\sim 2$~GeV/$c$).  This is due to correlations that event mixing cannot
reproduce, like the ``sub-event structure'' due to the presence of jets or
multiple, close-by showers from an annihilating anti-neutron, or imperfections
of the reconstruction algorithm, such as cluster merging, cluster splitting,
and a host of other contributions.  Much of the residual background is
excluded by starting the fit at 0.09~\GeVc. What is left is accounted for by
including a first-order polynomial in the fits to the (already
background-subtracted) invariant mass distribution, and subtracting its
integral from the raw \piz yield (see Fig.~\ref{fig:pi0mass}).  In the more
central events, the peak deviates slightly from gaussian on the high mass
side, due to overlapping clusters.  The use of {\it ecore} mitigates this
effect, and the systematic uncertainty on yield extraction arising from the
remaining asymmetry has been estimated to be 3-4\%~\cite{Adare:2008qa}.

\subsection {Elliptic flow measurement}
\label{sec:v2}

To obtain the azimuthal angle dependence of \piz production, we
measure raw \piz yields in a given \pt bin as a function of the \piz
angle with respect to the reaction plane orientation in six
equally-spaced bins of $\Delta\phi = \phi(\piz)-\Psi_{RP}$ covering
the range $0 < \dphi < \pi/2$.  The \piz yields are measured in each
\dphi bin using the same procedure described 
in~\ref{sec:yield} for the
reaction-plane inclusive measurement except that the mass fits are not
performed in each \dphi bin. Instead, the peak integration window is
set $\pm 2\sigma$ around the mean where the width and mean are taken
from the inclusive analysis. The resulting raw \piz angular
distribution $dN/d\Delta\phi$ can then be fit to determine the
strength of the modulation in the yield.  Because the PHENIX BBCs have
uniform azimuthal coverage, the \piz measurements have uniform
acceptance in \dphi when averaged over a large event sample,
despite the limited azimuthal acceptance of the PHENIX electromagnetic
calorimeters.

\begin{figure*}
\includegraphics[width=0.9\linewidth]{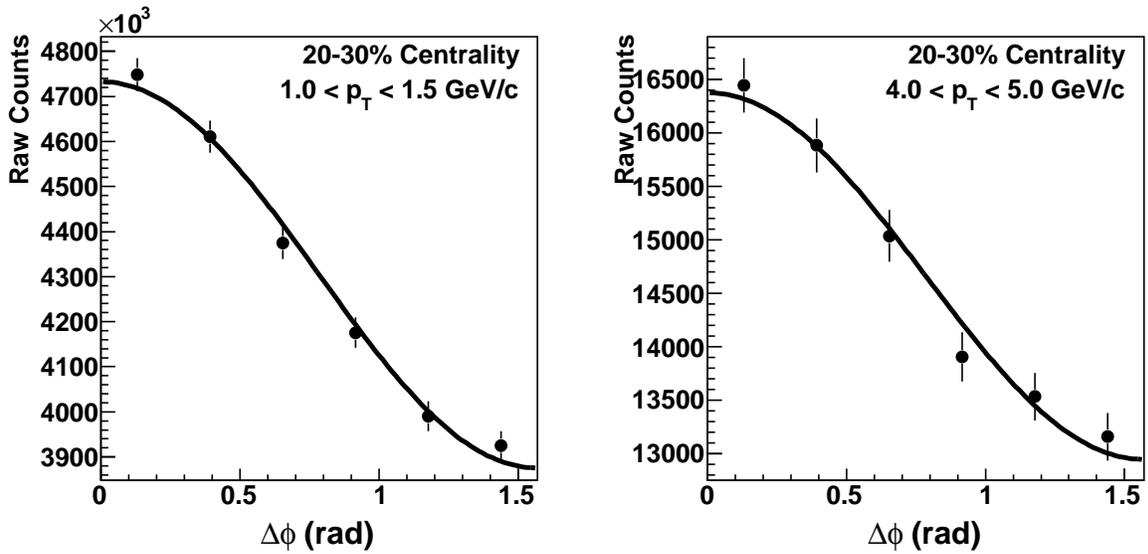}
\caption{Example of analytic fitting of raw $dN/d\dphi$ distributions}
\label{fig:dndphi_example}
\end{figure*}

Assuming elliptic flow is the dominant source of \dphi variation in
the \piz yields, we perform a fit to the angular distributions of the form
\begin{equation}
\frac{dN}{d\dphi} = N_0 (1 + 2 \vtwo^{\rm meas} \cos 2 \dphi).
\label{eq:v2fit}
\end{equation}
We use an analytic linear $\chi^2$ fitting procedure that matches the
integral of Eq.~\ref{eq:v2fit} over each of the \dphi bins to the
measured \piz yield  within the corresponding bin. In the definition of
the $\chi^2$ function we account for non-zero covariances between the
yields in the different \dphi bins resulting from the limited
acceptance of the calorimeters.
These covariances have been evaluated separately for each \pt and
centrality bin. 
Examples of the raw $dN/d\Delta\phi$ distributions and the results of
the $\chi^2$ fits are shown in Fig.~\ref{fig:dndphi_example}. The
resulting \vtworaw values are then corrected upward to account
for reaction plane resolution using correction factors described in
Section~\ref{sec:reactionplane}. 

\clearpage

\subsection {\RAA(\dphi) measurement}

The nuclear modification factor \RAA has played a critical role in
understanding energy loss mechanisms.  \RAA is defined as
\begin{equation}
\RAA(\pt) = \frac{(1/N^{\rm evt}_{\rm AA})d^2N^{\piz}_{\rm AA}/d\pt
  dy}{\left<T_{\rm AA}\right>\times d^2\sigma^{\piz}_{pp}/d\pt dy}
\end{equation}
where $\left<T_{\rm AA}\right>$ is the mean Glauber overlap function for
the centrality being analyzed:
\begin{equation}
\left<T_{\rm AA}\right> \equiv
\frac{\int{T_{\rm AA}(\mathbf{b})d\mathbf{b}}}
     {\int{(1-e^{-\sigma^{inel}_{pp}T_{\rm AA}(\mathbf{b})})d\mathbf{b}}},
\end{equation}
from which the mean number of binary nucleon-nucleon collisions can be
calculated, $\meanNcoll = \sigma^{inel}_{pp}\meanTAA$.

For each \pT bin,
we can calculate the ratio
\begin{equation}
R(\dphi_i,\pt) = \frac{N(\dphi_i,\pt)}{\sum_{i=1}^{6} N(\dphi_i,\pt)}
\end{equation}
where $N(\dphi_i,\pt)$ is the number of \piz{}s observed in the given
$(\dphi_i,\pt)$ bin.  
Since the BBC is azimuthally symmetric 
the PHENIX acceptance has no \dphi
dependence, there should be no azimuthal dependence to efficiency and
acceptance corrections.  As a result,
\begin{equation}
\RAA(\dphi_i,\pt) = R(\dphi_i,\pt) \times \RAA(\pt).
\label{eq:raaphidef}
\end{equation}
Thus, we can use measured inclusive $\RAA(\pt)$ to convert
$R(\dphi_i,\pt)$ to $\RAA(\dphi_i,\pt)$.  Since the detector
efficiency and acceptance corrections are already contained in
$\RAA(\pt)$, there is no need to apply them to $R(\dphi_i,\pt)$.

Prior to calculating $\RAA(\dphi_i,\pt)$ we correct the ratios
$R(\dphi_i,\pt)$ for the finite reaction plane resolution using an approximate unfolding
technique. For a pure flow \dphi distribution, we can express the
influence of the resolution broadening on the measured \dphi distribution
\begin{equation}
R^{\rm meas}(\dphi_i,\pt) =
R^{\rm true}(\dphi_i,\pt)
\left[\frac{1+2\vtworaw\cos(2\dphi)}{1+2\vtwocorr\cos(2\dphi)}\right],
\end{equation}
where according to the results of Section~\ref{sec:reactionplane}
$\vtworaw = \vtwocorr/\langle\cos 2(\Psi-\Psi_{RP})\rangle$. Then, if
the measured \dphi distribution resulted from pure elliptic
flow, it could be corrected back to the true distribution by
\begin{equation}
R^{\rm corr}(\dphi_i,\pt) =
R^{\rm meas}(\dphi_i,\pt)
\left[\frac{1+2\vtwocorr\cos(2\dphi)}{1+2\vtworaw\cos(2\dphi)}\right].
\label{eq:dndphicorr}
\end{equation}
As shown above, the general features of the measured \piz \dphi
distributions are well-described by pure $\cos{(2\dphi)}$
modulation. However, we wish to preserve in our measurements of 
the azimuthal dependence of the \piz production the full shape of
the measured \dphi distribution,  including possible small
non-elliptic contributions. For this 
purpose, the correction described in Eq.~\ref{eq:dndphicorr}
applied to the data represents an approximation to a full unfolding
procedure that becomes exact when the distribution is purely
$\cos{(2\dphi)}$ in form. We have checked for a few cases that a full unfolding
procedure applied to the measured $dN/\dphi$ distributions using
singular value decomposition regulation of the response matrix
reproduces the correction in Eq.~\ref{eq:dndphicorr}. From the
corrected ratios, $R^{\rm corr}(\dphi_i,\pt)$, we use
Eq.~\ref{eq:raaphidef} to obtain $\RAA(\dphi_i,\pt)$.

\section{Results}

\subsection{Elliptic flow coefficient}

\begin{figure*}
\includegraphics[width=0.8\linewidth]{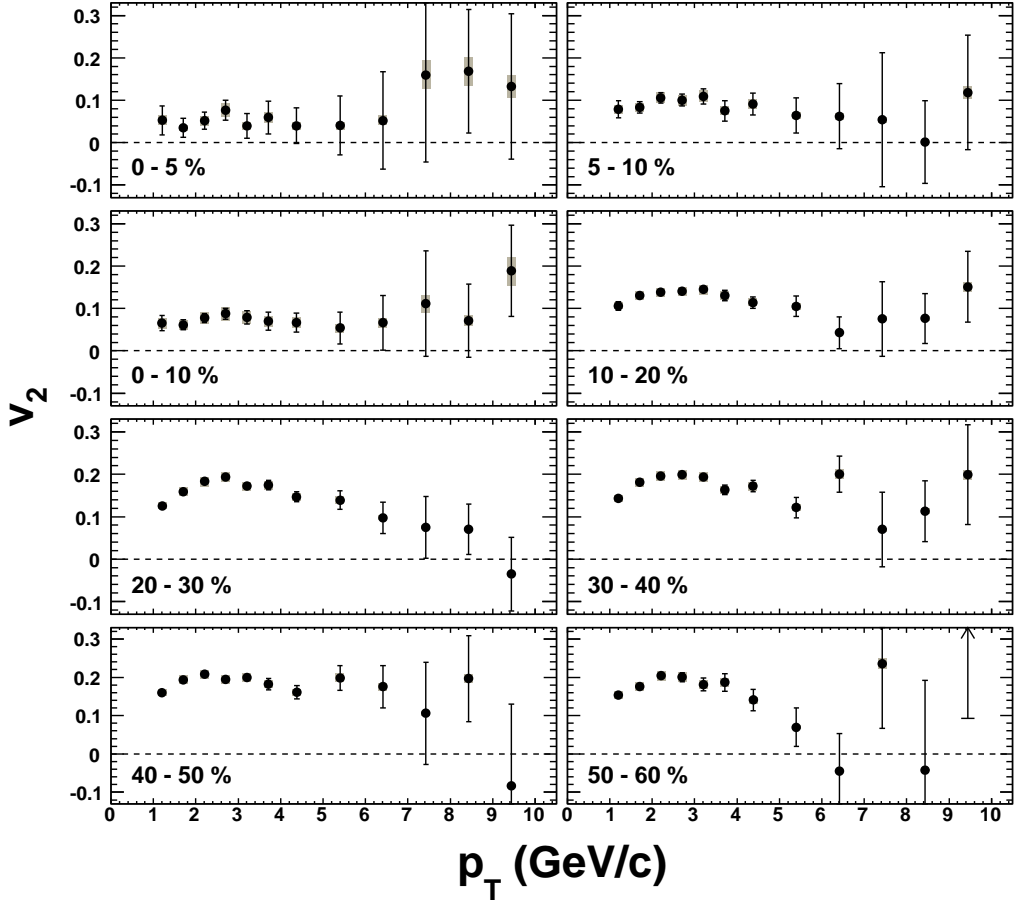}
\caption{\piz \vtwo versus \pT for centralities 0-5\%, 5-10\%, 0-10\%, 10-20\%,
  20-30\%, 30-40\%, 40-50\%, and 50-60\%.  The arrow in the 50-60\%
  panel shows the lower limit of the uncertainty on the data point,
  which lies outside the bounds of the plot.}  
\label{fig:pi0v2_nofits}
\end{figure*}

\begin{figure*}
\includegraphics[width=0.8\linewidth]{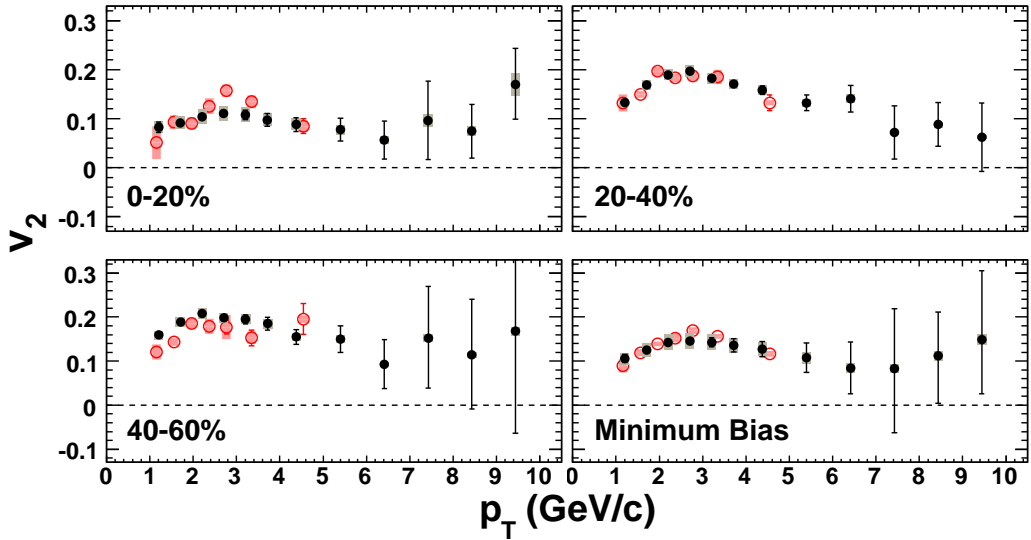}
\caption{(Color online) \piz $\vtwo$ versus \pT for centralities 0-20\%, 20-40\%, 40-60\%, and
  minimum bias.  The closed (black) circles are data presented in this paper, and the
  open (red) circles are previously published
  results~\cite{Adler:2005rg}.}
\label{fig:pi0v2_comb_nofits}
\end{figure*}

The results of the \vtwo measurements using the methods described in
Sec.~\ref{sec:v2} are presented in Fig.~\ref{fig:pi0v2_nofits} as a function
of \pt for different centrality bins. The data points in the figure
are plotted at the mean \piz \pt in bins of width $\Delta \pt =
0.5$~\GeVc for $\pt < 4$~\GeVc and \mbox{$\Delta \pt = 1$~\GeVc} for \mbox{$\pt
> 4$~\GeVc.} The error bars shown on the \vtwo data points were
obtained by multiplying the raw \vtwo fit errors (see
Section~\ref{sec:v2}) by the same reaction plane resolution
correction factor applied to the \vtwo values themselves. The 
error bars, then, represent uncorrelated statistical errors on the
measured \vtwo values arising from statistical errors on the
$dN/d\dphi$ data points used in the fits (these errors would be
categorized as Type A uncertainties in the framework described
in~\cite{Adare:2008cg} or \pt-uncorrelated). Systematic errors on the 
\vtwo measurements due to the reaction plane determination procedure
and from systematic uncertainties in the reaction plane
resolution correction are represented in
Figs.~\ref{fig:pi0v2_nofits},\ref{fig:pi0v2_comb_nofits} by filled boxes, which
for most data points are similar in size or smaller than the
data points (these uncertainties are classified as Type
B~\cite{Adare:2008cg} or \pt-correlated).

Figure~\ref{fig:pi0v2_comb_nofits} shows $\vtwo(\pt)$ for four centrality ranges, 
obtained by combining data from the centrality bins shown in
Fig.~\ref{fig:pi0v2_nofits}. The corrected $dN/d\dphi$ distributions from
individual centrality bins were summed over a given centrality range
and then fit to obtain the corrected \vtwo values shown in
Fig.~\ref{fig:pi0v2_comb_nofits}. The reaction plane resolution correction
produces correlated errors in the corrected $dN/d\dphi$ distributions
for each original centrality bin, and these correlated errors persist
in the summed $dN/d\dphi$ distribution. These correlated errors are not
included in the statistical errors for the fit to the combined
$dN/d\dphi$ distribution, but their impact on the final \vtwo value is
estimated separately by evaluating the changes in the \vtwo fit
parameter that result from adding to the summed $dN/d\dphi$ values $\pm
1~\sigma$ of the correlated errors. Since this estimated uncertainty
results from the statistical uncertainties on the \vtwo values for the
original centrality bins, we include the $1\sigma$ bounds obtained
from this procedure in the statistical error on the \vtwo values for
the combined centrality bins.  Systematic errors for the combined bins
are plotted similarly to those in Fig.~\ref{fig:pi0v2_nofits}.

The results presented here nearly double the \pt range of previous
PHENIX \piz \vtwo measurements from RHIC 

\clearpage

\noindent Run-2
\cite{Adler:2005rg}. Those measurements are shown for comparison
purposes in Fig.~\ref{fig:pi0v2_comb_nofits}. Good agreement is seen between
the Run-4 measurements presented here and the Run-2 results except in the
40-60\% centrality bin where the new $v_2$ measurements are
systematically higher by $\sim 30\%$. This difference is attributed
to improved reaction plane resolution corrections for the 40-60\%
centrality bin resulting from the combining of {\em corrected}
$dN/\dphi$ distributions from smaller centrality bins. This summing
procedure better handles the rapid variation of reaction plane
resolution with centrality in mid-central to peripheral
collisons. Furthermore, we have cross-checked the procedure using 5\%
bins, verifying the combined result reproduces the data analyzed in
wider bins.  The
previous Run~2 analysis did not have a sufficiently large data sample to allow
the use of separate 40-50\% and 50-60\% bins, and therefore the reaction plane
resolution correction was necessarily less accurate.
The measured \vtwo values presented in Fig.~\ref{fig:pi0v2_nofits} and
Fig.~\ref{fig:pi0v2_comb_nofits} are also consistent with previously
published PHENIX charged pion \vtwo measurements \cite{Adler:2003kt,Adler:2005rg}.

The results in Figures~\ref{fig:pi0v2_nofits} and~\ref{fig:pi0v2_comb_nofits} show
a rapid increase of \vtwo with increasing \pt at low \pt, a maximum in
the range $2<\pt<3$~\GeVc, and then at higher \pt a decrease of \vtwo
with increasing \pt. An increase in \vtwo at low \pt is
well-established \cite{Adare:2006ti,Adams:2004bi,Back:2004mh} and is
understood to result from the collective elliptic flow of the medium
generated by the initial spatial anisotropy of the collision
zone. Hydrodynamical models have been successful in quantitatively
describing the pion $\vtwo(\pt)$ in the region $\pt <
1.5$~\GeVc. However, it has also been well-established that the pion
$\vtwo(\pt)$ deviates from the hydrodynamic prediction above
1.5~\GeVc, a result that is understood to imply the contribution of
hard processes, distortions of the spectrum due to recombination at
freeze-out, and/or effects from dispersive hadronic evolution after
freeze-out. Thus, a change in the variation of \vtwo with \pt near
$\pt \sim 2$~GeV/c is not unexpected. If the large \vtwo values at
lower \pt are interpreted as resulting from soft, collective
mechanismsm, then a decrease in \vtwo for $\pt >3$~GeV/c suggested by
the data in Fig.~\ref{fig:pi0v2_nofits} would naturally reflect an
increasing contribution of hard processes with smaller \vtwo.

\begin{figure*}[tbh]
\includegraphics[width=0.8\linewidth]{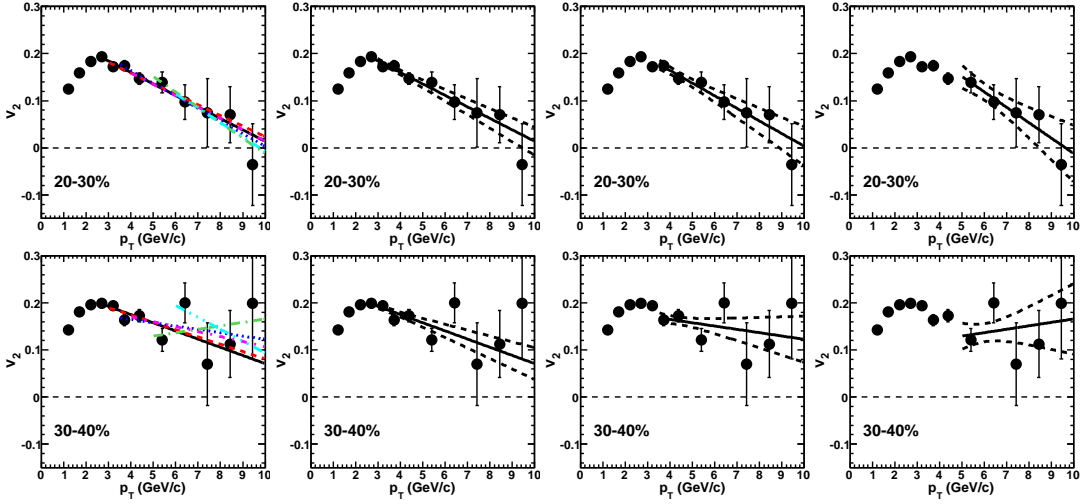}
\caption{(Color online) \vtwo versus \pt for 20-30\% and 30-40\% centralities,
  with fits of the high-\pt data to a first order polynomial.  From
  right to left, the first panel shows the series of fits, with each
  fit starting with a successively higher \pt.  The second, third, and
  fourth panels show selected fits with uncertainty bands based on the
  1-$\sigma$ variation of the fit parameters, including their covariance. 
}  
\label{fig:v2_pol1}
\end{figure*}

To statistically test the significance of the decrease of \vtwo with
\pt, we show in Fig.~\ref{fig:v2_pol1} the results of linear
fits to the high-\pt \vtwo values for the 20-30\%
and 30-40\% centralities.  The panels on the left-hand side of
Fig.~\ref{fig:v2_pol1} display a series of fits beginning at higher
\pt values, the first fit starting at the \pt near the maximum \vtwo,
$\pt = 2.5$~\GeVc.  The right-hand panels show the 1-$\sigma$ limits
of the functions for the three fits in the series, calculated
from the 1-$\sigma$ variation of the two fit parameters (and including the
covariance between them). The results of the fits indicate that the
decrease of \vtwo with \pt at higher \pt is statistically significant,
though the data points for $\pt > 5$~GeV/c are not sufficient by
themselves to establish a trend. We can state, however, that the
points for $\pt > 5$ GeV/c are consistent with the linear decrease
obtained including the lower \pt points. A question we would like to
answer, then, is whether the data show any indications of devation
from a monotonic decrease in $\vtwo(\pt)$ indicating the transition to
a quenching-dominated azimuthal variation.

\begin{figure*}
\includegraphics[width=0.9\linewidth]{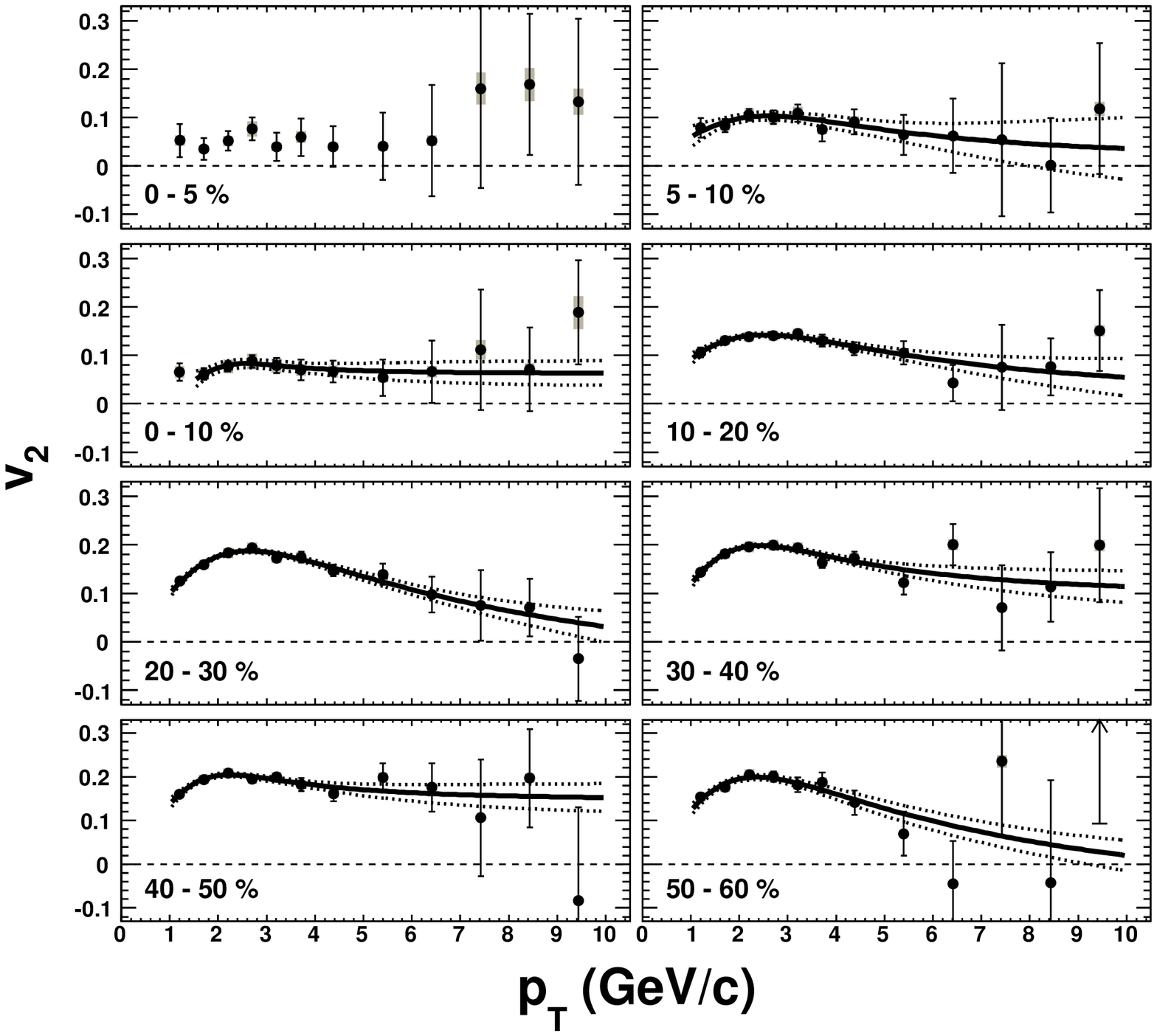}
\caption{\piz \vtwo versus \pT for centralities 0-5\%, 5-10\%, 0-10\%,
  10-20\%, 20-30\%, 30-40\%, 40-50\%, and 50-60\%.  The arrow in the
  50-60\% panel shows the lower limit of the uncertainty on the data
  point, which lies outside the bounds of the plot.  The solid lines
  represent the fit to the data, Eq.~\ref{eq:v2fitfunc}.  The dashed
  lines represent the 1~$\sigma$ deviations of the fit function.  See
  text for more details.}
\label{fig:pi0v2}
\end{figure*}

\begin{figure*}
\includegraphics[width=0.9\linewidth]{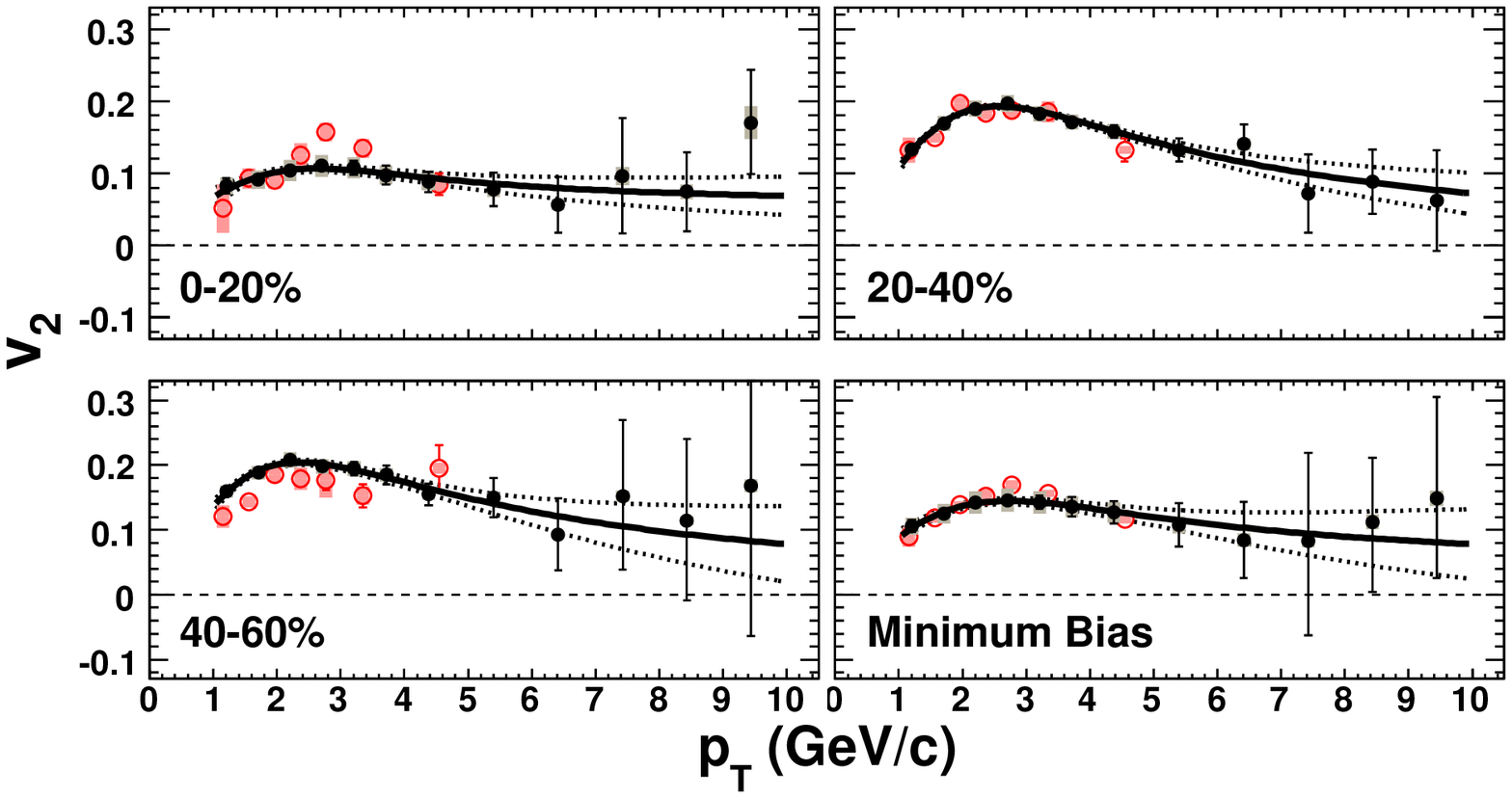}
\caption{(Color online) \piz $\vtwo$ versus \pT for centralities
  0-20\%, 20-40\%, 40-60\%, and minimum bias.  The closed (black)
  circles are data presented in this paper, and the open (red) circles
  are previously published results~\cite{Adler:2005rg}.  The solid and
  dashed lines as in Fig.~\ref{fig:pi0v2}.}
\label{fig:pi0v2_comb}
\end{figure*}

A complete understanding of $\vtwo(\pt)$ over the measured \pt range
therefore requires the treatment of the transition from soft to hard
dominated physics. According to the above discussion, in the \pt range
where \vtwo is maximum, particle production is likely not dominated by
hard processes and the reduction of \vtwo with increasing \pt
indicates increasing hard-scattering contributions (or decreasing soft
contamination). Motivated by this general argument, we have attempted
to describe the results in  Figs.~\ref{fig:pi0v2_nofits} and
~\ref{fig:pi0v2_comb_nofits} by a functional form 
\begin{equation}
\vtwo(\pt) =
\left(\frac{\left(\pt/\lambda\right)^m}{1+\left(\pt/\lambda\right)^m}\right)\left(a+\frac{1}{\pt^n}\right).
\label{eq:v2fitfunc}
\end{equation}
The first term is intended to describe a rapidly rising and saturating
soft \vtwo resulting from collective motion while the second term
represents a rapidly falling soft/hard ratio. The additive constant in
the second term represents an asymptotic \vtwo that could describe a
constant or slowly varying azimuthal-dependent quenching. 
We show in
Figs.~\ref{fig:pi0v2}-\ref{fig:pi0v2_comb} the optimum fits to the full set of $\vtwo(\pt)$
values in the different centrality bins and the result of $1\sigma$
variation of the fit parameters taking into account the complete
covariance matrix from the fits.

The fits to the data show that the measured \pt dependence of the \piz
\vtwo is qualitatively compatible with a description of the of
low and intermediate \pt region in terms of a collective flow
modulation diluted by a decreasing relative soft contribution with
increasing \pt. Assuming this picture, it is then important to
determine at what \pt the contamination from the soft production no
longer dominates the measured \dphi variation of \piz yield. For most
of the centrality bins, the fits suggest that \vtwo decreases over
most of the measured \pt range albeit with a decreasing slope at
higher \pt. The
central bins are compatible with \vtwo continuing to decrease beyond
the measured \pt range although the $1\sigma$ uncertainty bands also
accommodate \vtwo saturating within the measured range. The more
peripheral bins (30-40\% and 40-50\%) suggest that the \vtwo has
reached a nearly \pt independent value by $\sim 5$~GeV/c. The 50-60\%
centrality bin has sufficient fluctuations that little can be inferred
from the \pt dependence of \vtwo in that centrality bin. In all of the
centrality bins, the data are consistent with a smooth reduction of
$\vtwo(\pt)$ from a maximum to a non-zero value at high \pt with that
value increasing in more peripheral collisions as would be expected
from quenching in an increasingly anisotropic medium. While the
functional form in Eq.~\ref{eq:v2fitfunc} can describe the \pt
variation of \vtwo within the range of the current data and within the
statistical fluctuations of the data points, it is possible that this
description will fail at higher \pt with improved statistics. In fact,
a statistically significant deviation of 
$\vtwo(\pt)$ from the form in
Eq.~\ref{eq:v2fitfunc} might provide the most direct evidence of the
dominance of quenching effects in determining \vtwo.

\newpage

\subsection{Nuclear modification factor with respect to the reaction
  plane}
\label{sec:raa_dphi}

\begin{figure*}
\includegraphics[width=0.8\linewidth]{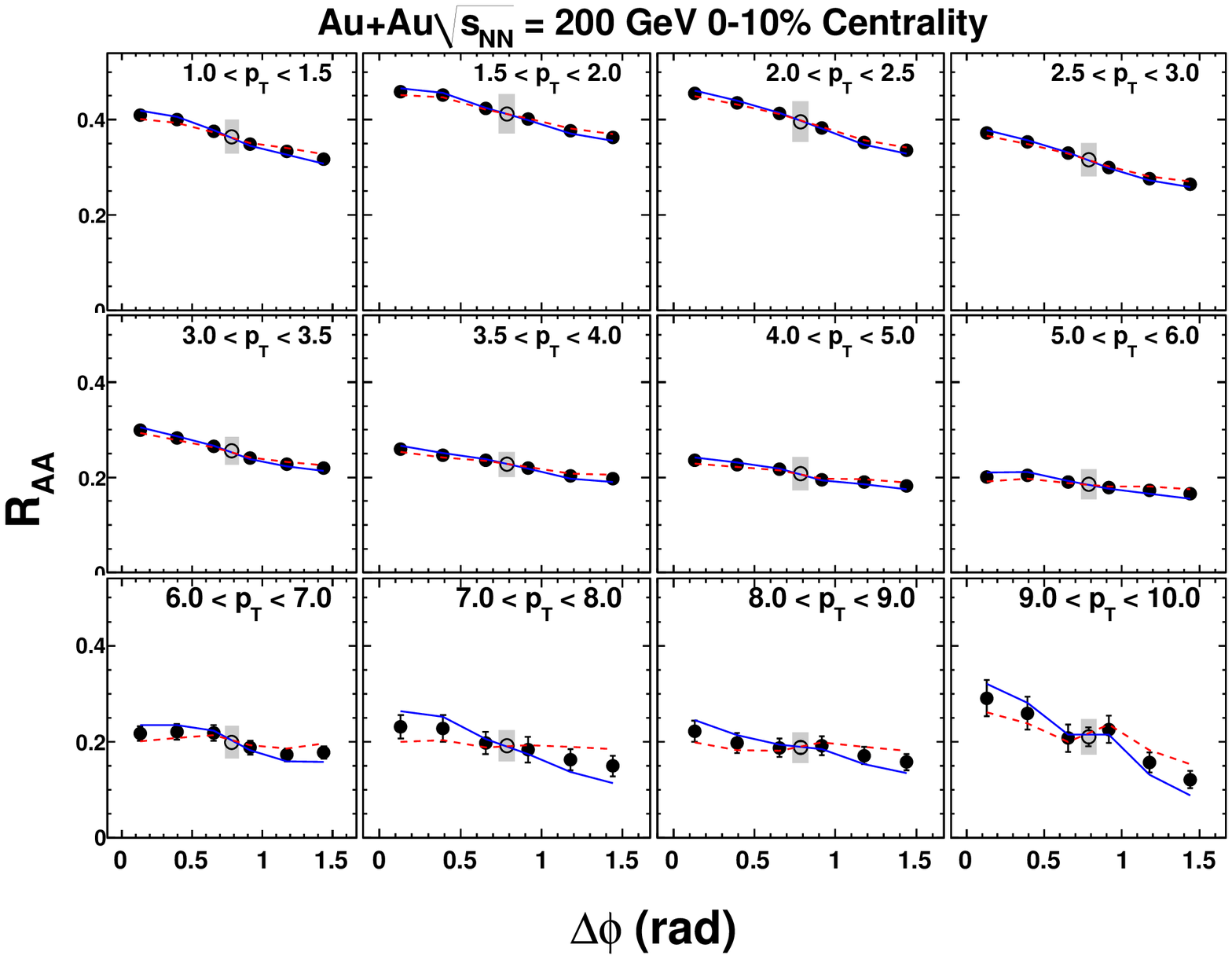}
\caption{(Color online) \piz \RAA versus angle of emission with
  respect to the reaction plane for 0-10\% centrality.  The error bars
  denote the statistical errors, while the solid (blue) line and
  dashed (red) line represent the systematic error due to the
  resolution correction factor.  The inclusive \RAA measurement is
  shown with the open circle, for which the error bar shows the
  statistical error and the box shows the systematic error.  }
\label{fig:pi0RaaDphi_00_10}
\end{figure*}

\begin{figure*}
\includegraphics[width=0.8\linewidth]{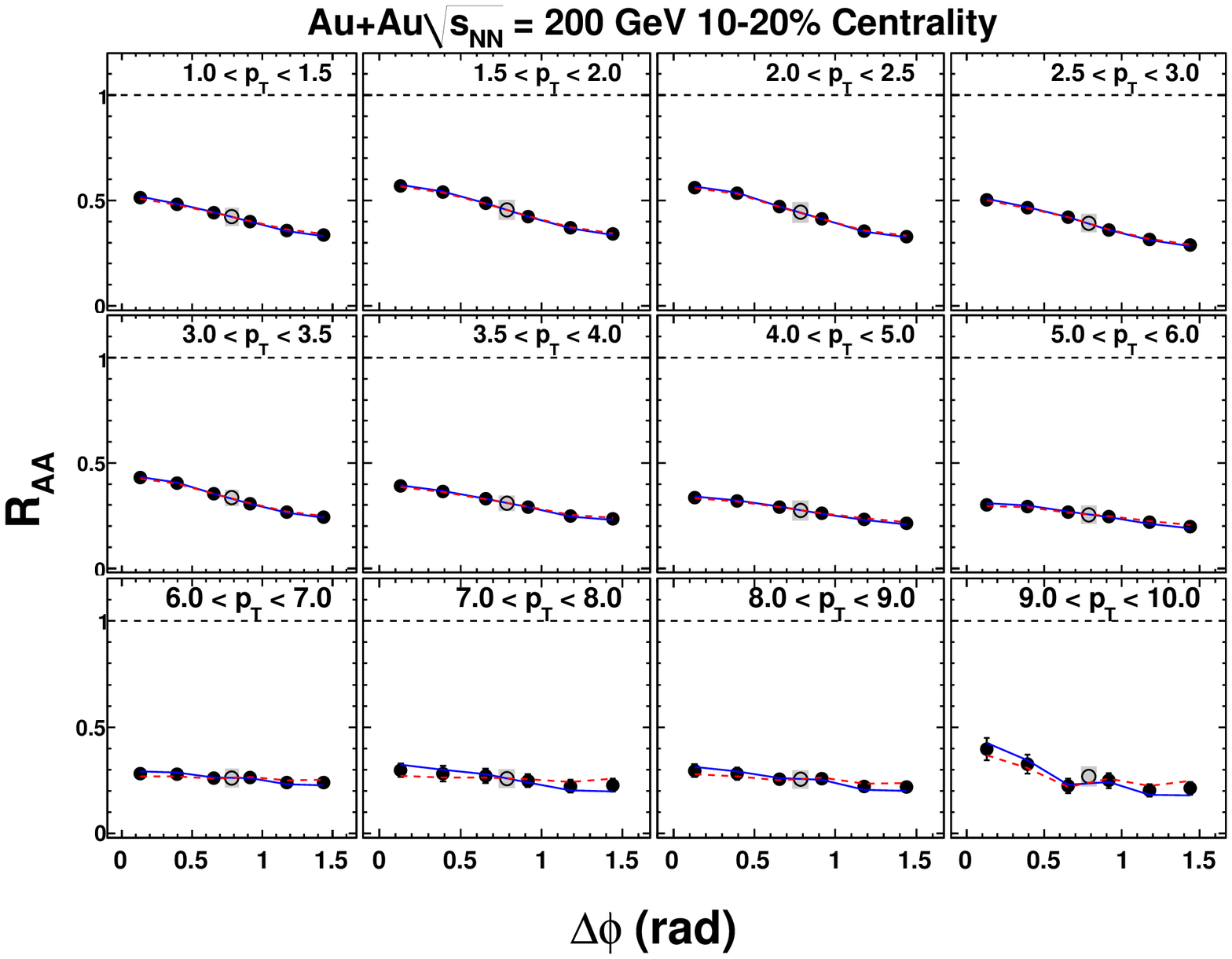}
\caption{\piz \RAA versus angle of emission with respect to the reaction plane
  for 10-20\% centrality.  Colors/data points as in
  Fig.~\ref{fig:pi0RaaDphi_00_10}.}
\label{fig:pi0RaaDphi_10_20}
\end{figure*}

\begin{figure*}
\includegraphics[width=0.8\linewidth]{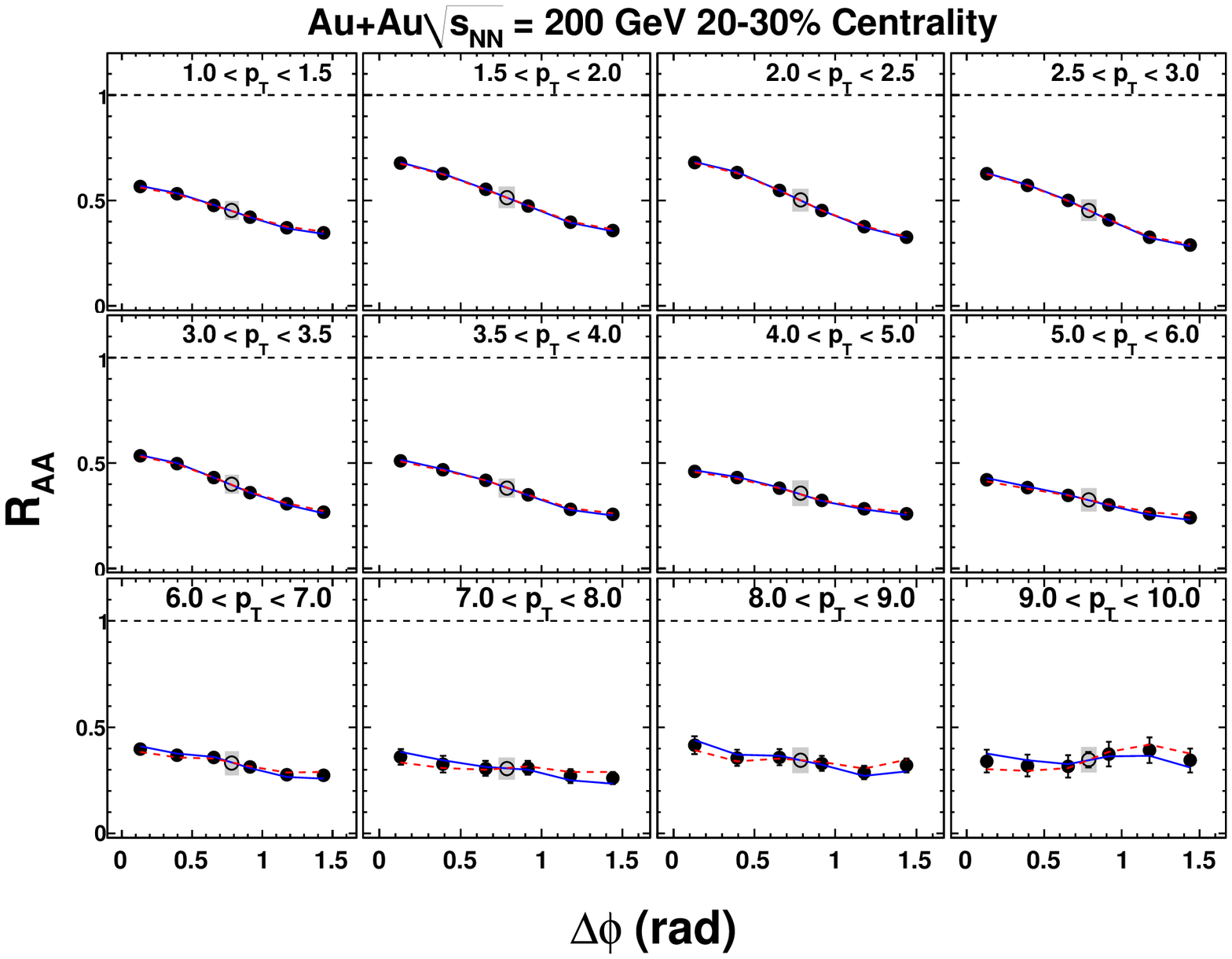}
\caption{\piz \RAA versus angle of emission with respect to the
  reaction plane for 20-30\% centrality.  Colors/data points as in
  Fig.~\ref{fig:pi0RaaDphi_00_10}.}
\label{fig:pi0RaaDphi_20_30}
\end{figure*}

\begin{figure*}
\includegraphics[width=0.8\linewidth]{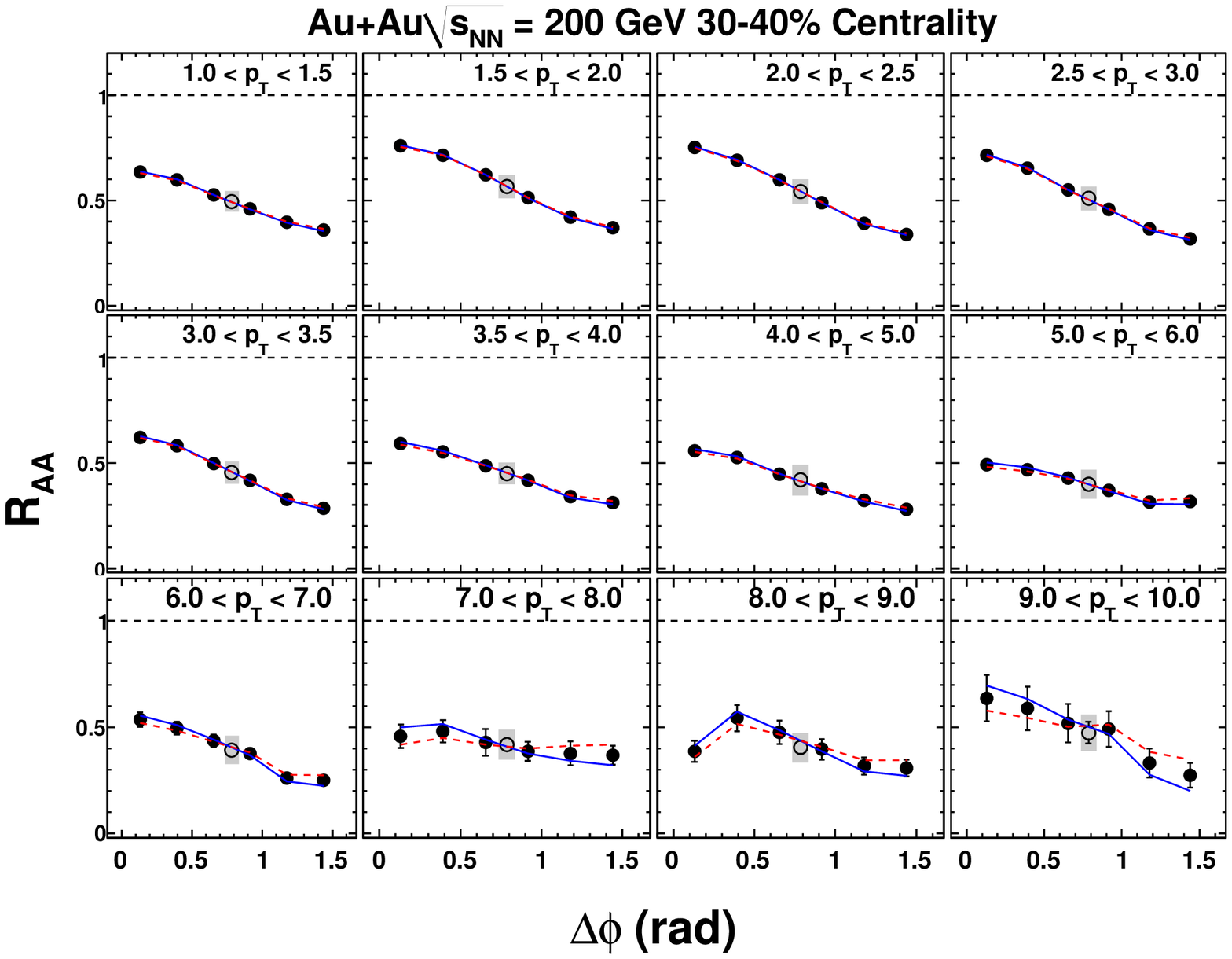}
\caption{\piz \RAA versus angle of emission with respect to the
  reaction plane for 30-40\% centrality.  Colors/data points as in
  Fig.~\ref{fig:pi0RaaDphi_00_10}.}
\label{fig:pi0RaaDphi_30_40}
\end{figure*}

\begin{figure*}
\includegraphics[width=0.8\linewidth]{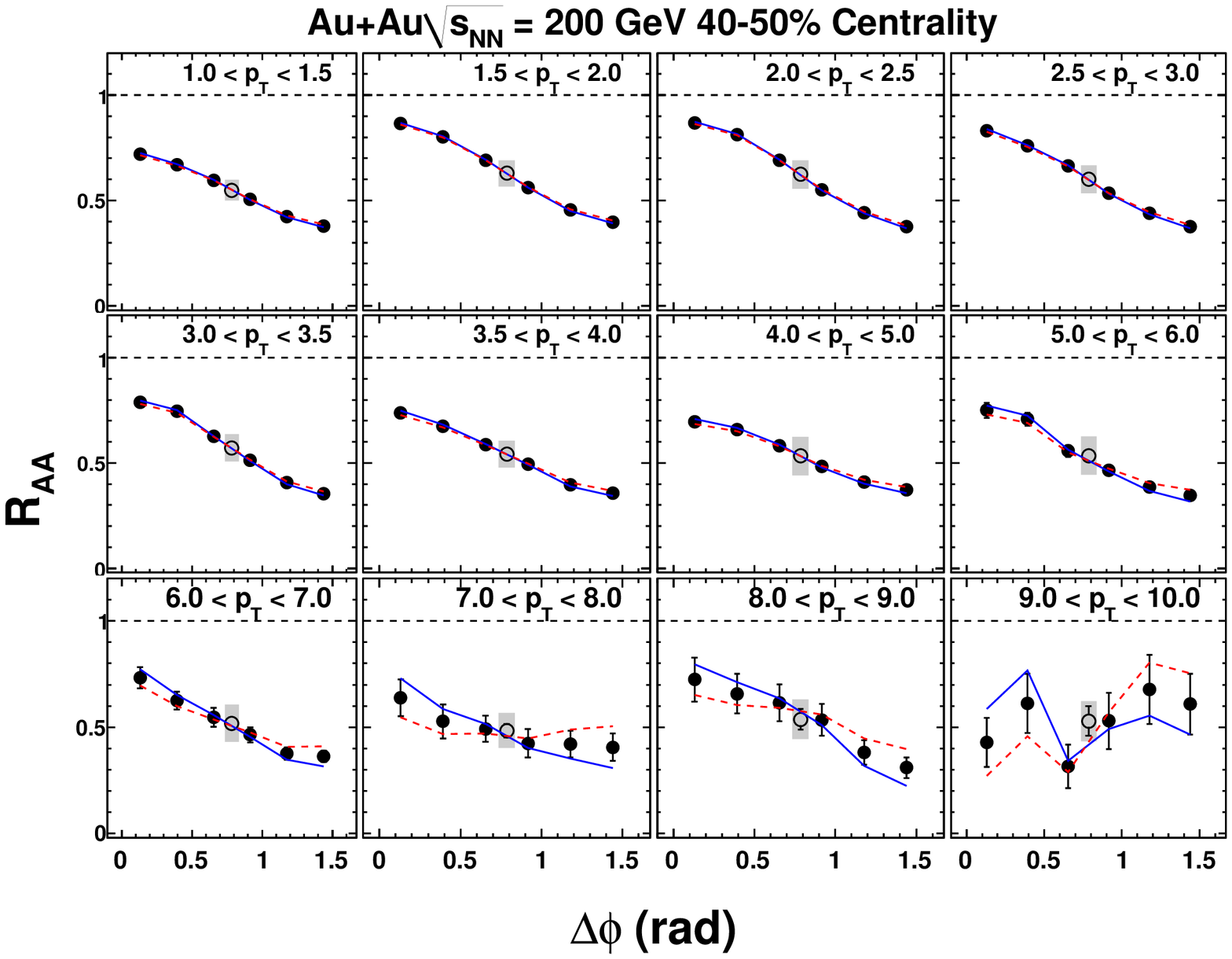}
\caption{\piz \RAA versus angle of emission with respect to the
  reaction plane for 40-50\% centrality.  Colors/data points as in
  Fig.~\ref{fig:pi0RaaDphi_00_10}.}
\label{fig:pi0RaaDphi_40_50}
\end{figure*}

\begin{figure*}
\includegraphics[width=0.8\linewidth]{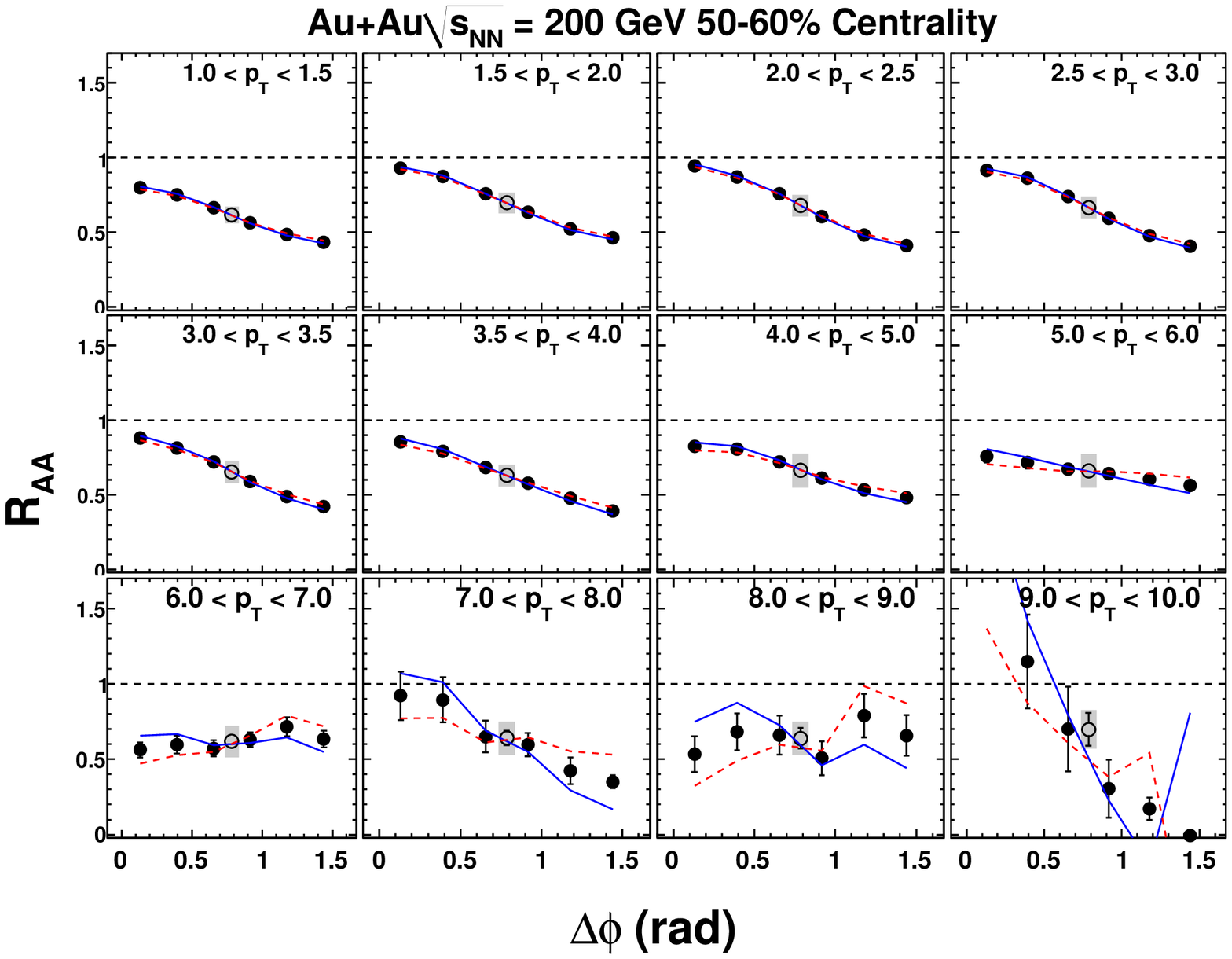}
\caption{\piz \RAA versus angle of emission with respect to the
  reaction plane for 50-60\% centrality.  Colors/data points as in
  Fig.~\ref{fig:pi0RaaDphi_00_10}.}
\label{fig:pi0RaaDphi_50_60}
\end{figure*}
The nuclear modification factor as a function of \dphi for six
centrality bins is shown in
Figs.~\ref{fig:pi0RaaDphi_00_10}-\ref{fig:pi0RaaDphi_50_60}.  The
closed circles represent the \dphi-dependent measurements described in
this paper while the open circles positioned at $\dphi=\pi/4$
represent the inclusive \RAA measurement~\cite{Adare:2008qa}.  In both
cases statistical uncertainties (i.e.\ Type A) are represented by the
error bars.  For the inclusive \RAA measurement, the total systematic
uncertainties (or Type C~\cite{Adare:2008cg}) are shown by the boxes.
The upper and lower $1\sigma$ ranges of the correlated statistical
uncertainties (i.e.\ Type B) on the $\RAA(\dphi)$ measurements resulting
from the reaction plane resolution correction are shown by the (blue)
solid and (red) dashed lines. For all bins except the 0-10\%
centrality bin a dotted line is plotted at $\RAA=1$ for reference. We
note that by construction, the average $\RAA(\dphi)$ from the reaction
plane dependent measurements must be equal to the inclusive \RAA.

The results in the
Figs.~\ref{fig:pi0RaaDphi_00_10}-\ref{fig:pi0RaaDphi_50_60} show that
the in-plane \piz\ suppression is generally weaker and varies more
rapidly with \pt\ than the suppression for \piz{}s produced at larger
angles. As the collisions become more peripheral (for example, 50-60\%), the
small suppression seen in the inclusive \RAA almost vanishes for
\piz{}s emitted close to the reaction plane.  In a previous
analysis~\cite{Adler:2006bw}, it was observed that the in-plane \RAA
even exceeded unity for peripheral collisions; these data exhibit
no such enhancement.  However, the results presented in this article agree
within systematic errors with previously reported data.

\begin{figure*}
\includegraphics[width=0.85\linewidth]{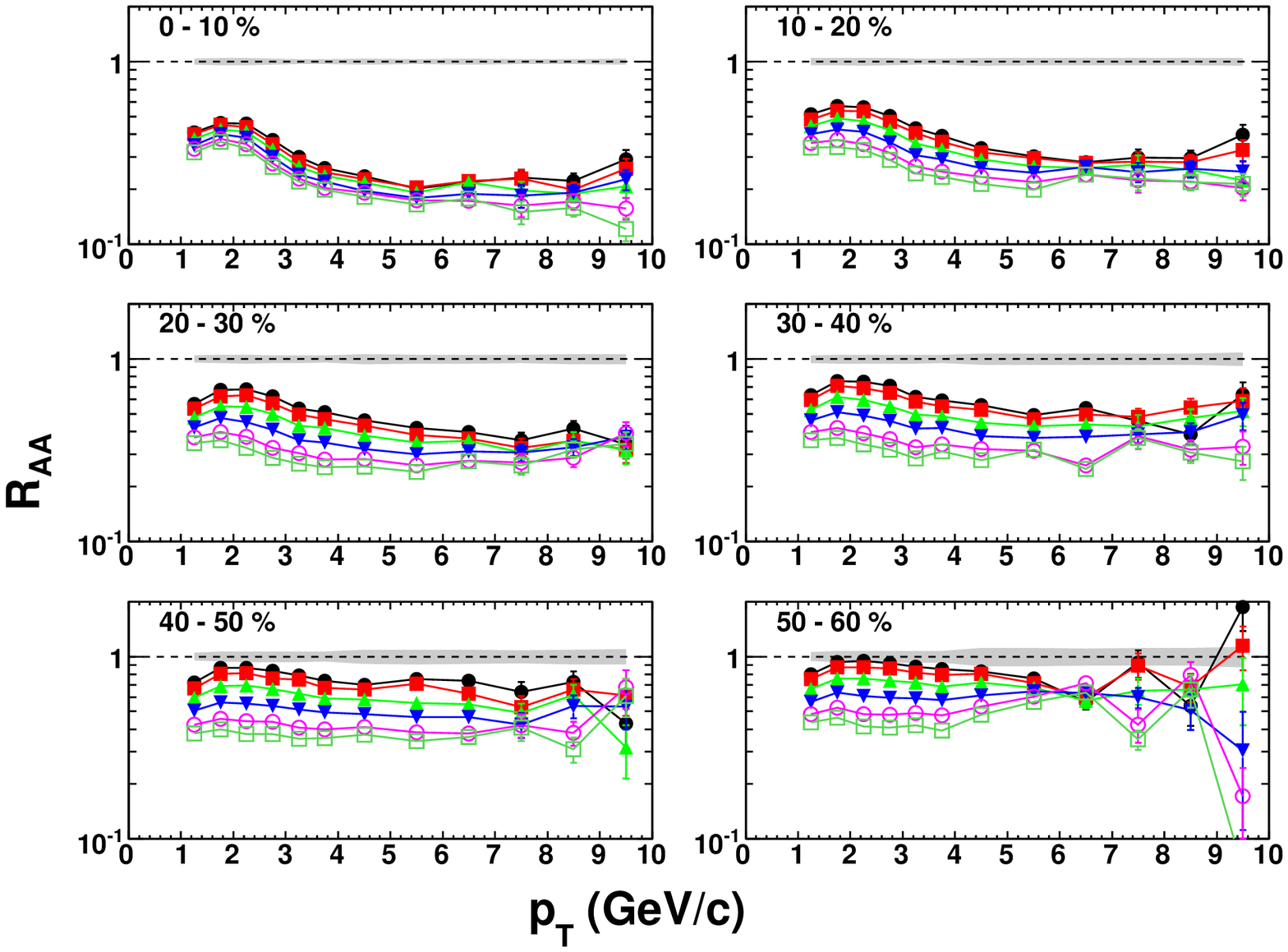}
\caption{(Color online) Semi-log plots of $\RAA(\pt)$ for each \dphi
  bin, in different centrality ranges. The \dphi bins are represented
  as follows: closed (black) circles, 0-15$^\circ$; closed (red)
  squares, 15-30$^\circ$; closed (light green) triangles,
  30-45$^\circ$; closed inverted (blue) triangles, 45-60$^\circ$;
  open (magenta) circles, 60-75$^\circ$; and open (dark green) squares,
  75-90$^\circ$.  The systematic error in the inclusive \RAA is
  represented by the grey bands.  Errors due to the correction factor
  have been omitted for clarity.}
\label{fig:pi0RaaPtDphi}
\end{figure*}
\begin{figure*}
\includegraphics[width=0.85\linewidth]{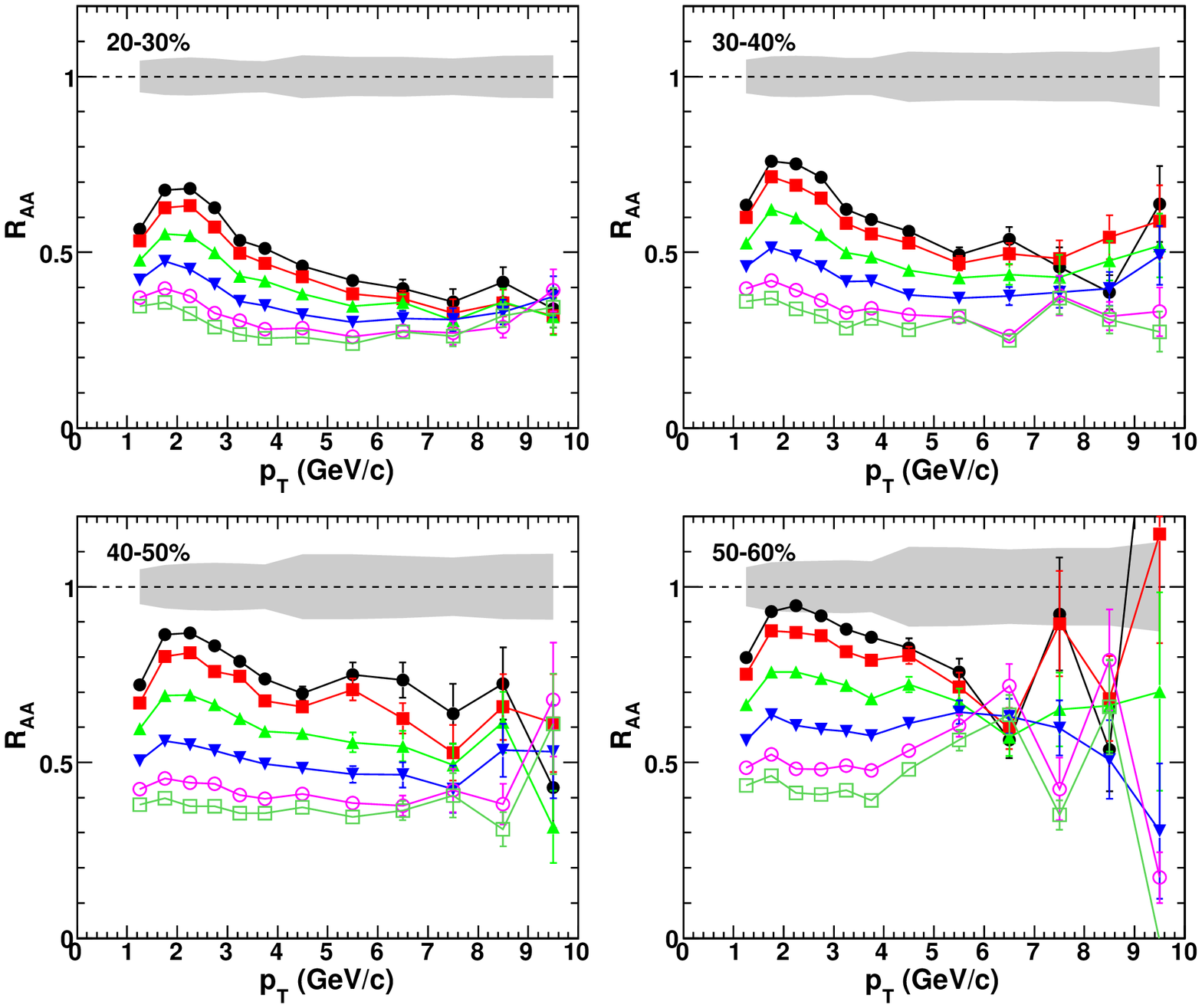}
\caption{(Color online) $\RAA(\pt)$ for different $\dphi$ bins in the
  20-30\%, 30-40\%, 40-50\%, and 50-60\% centrality ranges. Colors/data points as in Fig.~\ref{fig:pi0RaaPtDphi}.}
\label{fig:pi0RaaPtDphi_subset}
\end{figure*}

The \RAAphi results are combined in Fig.~\ref{fig:pi0RaaPtDphi} that shows the
\pt dependence of the \RAA in each of the six \dphi bins included in
this analysis. We use a semi-log scale so that the reduction of the
\dphi-integrated \RAA in more central collisions does not confuse the
interpretation of the results.  For clarity, the results from the
20-30\%, 30-40\%, 40-50\%, and 50-60\% centrality bins are shown on linear plots in 
Fig.~\ref{fig:pi0RaaPtDphi_subset}. 

The $\RAA(\pt)$ results exhibit a peak near 2~\GeVc, which
becomes more prominent for more central collisions. The peak is
strongest in the 0-10\% bin where there is little modulation of the 
\dphi\ distributions at low or high \pt, so the peak cannot be
directly attributed to elliptic flow. The peak in \RAA\ near 2~\GeVc
is much weaker in the more peripheral (40-50\% and 50-60\%) centrality
bins, particularly for \piz{}s produced at larger \dphi, and the
primary variation seen in these peripheral bins with increasing \dphi
is a reduction in \RAA that is only weakly \pt dependent.

For the intermediate
centrality bins (10-20\% through 30-40\%) the peaking in \RAA\ is seen
in all \dphi bins, but is much stronger in the in-plane bins. For
these intermediate centralities and for \pt values above the peak in
\RAA ($\pt \gtrsim 3$~\GeVc), the \RAA for \piz{}s produced at angles
normal to the reaction plane is nearly constant with \pt while the \RAA
for \piz{}s produced at small angles from the reaction plane 
decreases rapidly with increasing \pt. The near constancy of the
out-of-plane \RAA together with the rapid reduction in in-plane \RAA
indicates that in the intermediate centrality bins, the \vtwo and
inclusive \RAA decrease simultaneously with increasing \pt such that 
$\RAA(\pi/2, \pt) = \RAA(\pt) (1 - 2v_2)$ is approximately
constant. We will argue below that a correlation between \RAA and
\vtwo may naturally result from the underlying physics responsible for
the azimuthal variation of the particle yields. However, we observe
that a simultaneous reduction in integrated \RAA\ and \vtwo suggested
by the more central $\RAA(\pt)$ data would be contrary to naive energy
loss expectations since smaller \RAA\ would imply stronger quenching
in the medium which would, in turn, imply larger variation between
in-plane and out-of-plane quenching. 

\begin{figure*}
\includegraphics[width=1.0\linewidth]{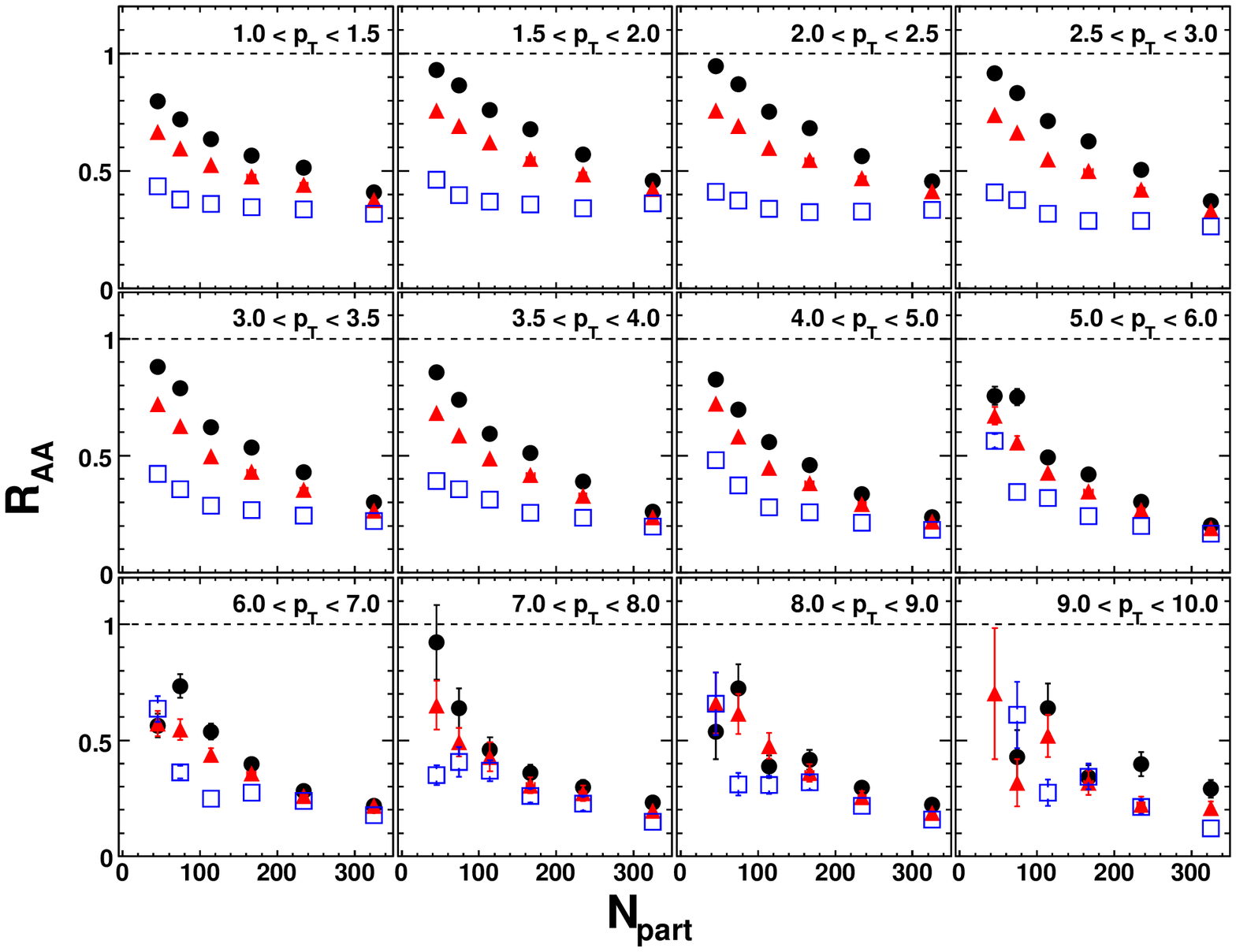}
\caption{(Color online) \piz $\RAA(\Npart)$ in reaction-plane bins at
  fixed \pt. The three bins are as follows: closed (black) circles are
  $\RAA(0<\dphi<15^\circ)$ (in plane), the closed (red) triangles
  are the $\RAA(30<\dphi<45^\circ)$, and open (blue) squares are
  $\RAA(85<\dphi<90^\circ)$ (out of plane).}
\label{fig:pi0raa_npart}
\end{figure*}

\begin{figure*}
\includegraphics[width=1.0\linewidth]{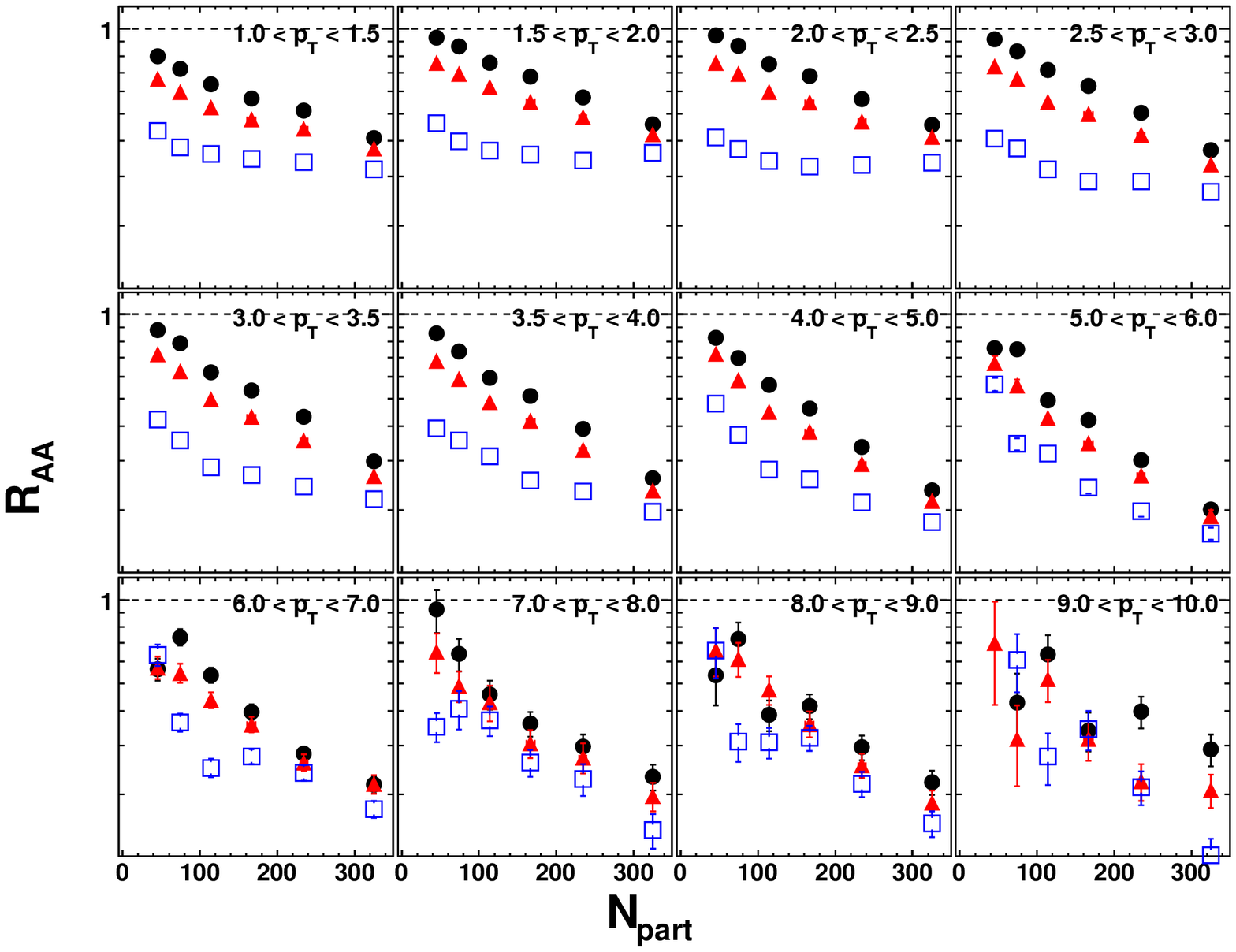}
\caption{(Color online) \piz $\RAA(\Npart)$ in reaction-plane bins at
  fixed \pt, with log scale on the $y$ axis. Colors/data as in Fig.~\ref{fig:pi0raa_npart}.}
\label{fig:pi0raa_npart_logy}
\end{figure*}

A similar implicit correlation between integrated \RAA\ and \vtwo is
seen in the centrality dependence of the \RAAphi 
results. These are re-plotted in Fig.~\ref{fig:pi0raa_npart} as a
function of \Npart for three \dphi bins -- the bins closest to and
further from the reaction plane and one of the intermediate bins.  For
$\Npart>100$, the out-of-plane \RAA values are nearly independent of
centrality while the in-plane \RAA values decrease rapidly with
increasing centrality. This result would have a natural geometric
interpretation for \piz production dominated by hard scattering and
jet quenching. The length of the medium normal to the reaction plane
varies only slowly with centrality except in the most peripheral
collisions. Then, if the \piz suppression is determined primarily by
the path length of its parent parton in the medium, the \piz \RAA
would be nearly constant. Following the same argument, the yield for
pions in the direction of the reaction plane would be much less
suppressed in non-central collisions due to the short path lengths of
the parent partons in the medium. However, with increasing centrality
and decreasing anisotropy of the collision zone, the in-plane parton
path lengths would grow to match those in the out-of-plane
direction. Thus, if the \piz\ suppression depended primarily on path
length, the in-plane \RAA\ would naturally drop to match the
out-of-plane values reproducing the behavior of Fig.~\ref{fig:pi0raa_npart}.
In order to better see the difference between the in- and out-of-plane
behaviors, these data are also plotted on Fig.~\ref{fig:pi0raa_npart_logy} with a
semi-log scale.

One difficulty with this geometric interpretation of the \RAAphi
results given above is that the trend in the data that it is supposed
to explain persists down to low \pt, where the \vtwo values are too
large to be accounted for via perturbative or formation time based 
energy loss scenarios \cite{Shuryak:2001me,Pantuev:2005jt,
Drees:2003zh,PhysRevC.69.034908,Renk:2006sx,majumder:041902}. That fact
coupled with the pronounced peaking in \RAAphi near 2~\GeVc suggests
that physics other than hard scattering and jet quenching must be
invoked to explain the \piz yields at intermediate \pt. However, the
fact that the out-of-plane yields show less 
pronounced peaking near 2~\GeVc, that they vary little as a
function of \pt above 3~\GeVc, and that they vary little with
centrality for $\Npart > 100$  could be interpreted to imply that the
\piz suppression at angles normal to the reaction plane more directly
represents the effects of quenching of hard quarks and gluons while
the yield of \piz{}s produced more closely aligned with the reaction
plane is enhanced by the collective motion of the system. That
additional enhancement could either be due to soft hadrons being
boosted to larger \pt values by the collective elliptic flow or could
result from weaker quenching for partons moving in the direction of
the flow field \cite{Armesto:2004vz,Renk:2006sx}. 
Simultaneous description of the in-plane and out-of-plane behavior is
a sensitive test of energy loss models.

\begin{figure}
\includegraphics[width=1.0\linewidth]{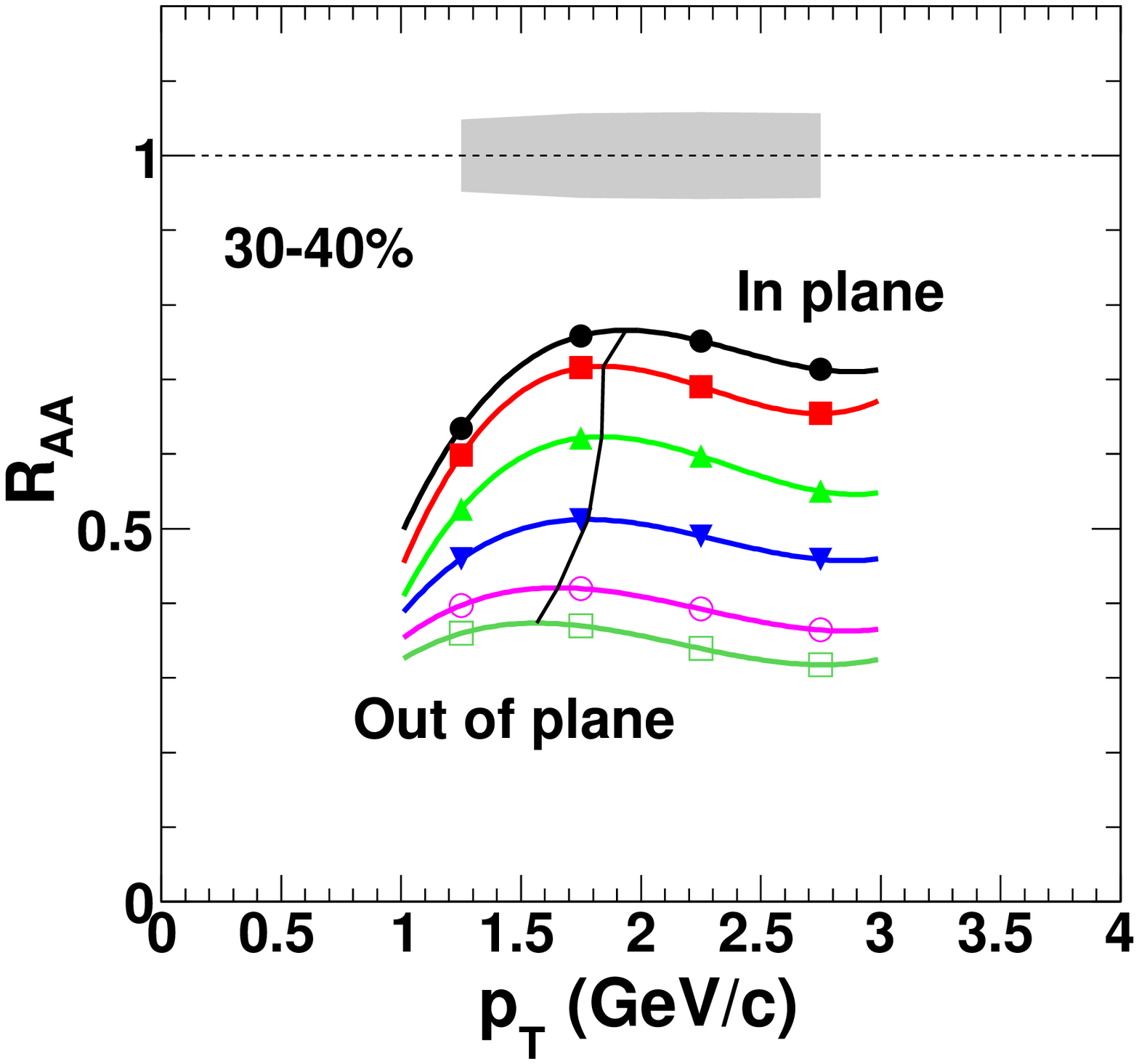}
\caption{(Color online) Illustration of the shift in the peak positions in $\RAA(\pt)$
for the 30-40\% centrality bin.  Colors/data points as in Fig.~\ref{fig:pi0RaaPtDphi}.}
\label{fig:raapt_peakshift}
\end{figure}

The \vtwo values presented in
Figs.~\ref{fig:pi0v2}-\ref{fig:pi0v2_comb} also peak near 2~\GeVc, but
the locations of the maxima in \vtwo are shifted to higher \pt than
the maxima in $\RAA(\dphi)$. This suggests that the two effects
may not directly related, but we observe that the maxima in the
$\RAA(\pt)$ distributions in
Figs.~\ref{fig:pi0RaaPtDphi}-\ref{fig:pi0RaaPtDphi_subset} shift to larger
\pt for smaller \dphi. To better illustrate the shift
of the maxima in $\RAA(\pt)$ we show in Fig.~\ref{fig:raapt_peakshift}
the $\RAA(\pt)$ values for the different \dphi bins and indicate the
variation of the peak position obtained using polynomial fits to the
first four \pt bins. For the 30-40\% centrality bin, the maximum in
$\RAA(\pt)$ for $\dphi<\pi/12$ is shifted by 0.4~\GeVc relative to the
$5\pi/12 < \dphi < \pi/2$ bin. This shift in the peak \RAApt with
\dphi can produce a $\vtwo(\pt)$ that peaks higher in \pt than the
inclusive \RAA.

The observed shift in the peak of \RAApt with \dphi illustrates an
important property of collective motion of the medium. The collective
motion does not simply superimpose azimuthal variation on the
particles produced at a given \pt, it provides a \dphi 
dependent shift and/or broadening in the transverse momentum spectrum
of the produced particles. The resulting distortion will be the
smallest for particles produced at angles normal to the reaction
plane and will be largest for particles produced in the plane. Any
collective shift of soft particles to higher \pt will increase the
measured  $\RAA(\dphi,\pt)$ for small \dphi relative to large \dphi
values producing a simultaneous increase in both the \dphi-integrated
\RAA\ and the \vtwo. With increasing \pt, an expected decrease in the
soft contamination would naturally explain the simultaneous reduction
in \vtwo and \dphi-integrated \RAA 
 evident in the 10-20\% and 20-30\% bins where the separation
between the $\RAA(\pt)$ curves for different $\dphi$ bins decreases
while the average \RAA also decreases. We will return to investigate
this correlation again below.

The 40-50\% and 50-60\% centrality bins in Fig.~\ref{fig:pi0RaaPtDphi}
show little of the peaking near 2~\GeVc, especially in \dphi bins not
aligned with the reaction plane. Nonetheless the \vtwo values for the
more peripheral bins reach the same large maximum values, $\vtwo \sim
0.2$, at intermediate \pt as the \vtwo values for more central bins
where the peak in \RAApt is more prominent. Thus, while the peaking in
\RAApt is less prominent in the more peripheral bins, the relative
difference between the in-plane and out-of-plane \piz yields in the
40-50\% centrality bin is comparable to that in the 20-30\% centrality
bin. However, it is possible that the large \dphi dependence in the
more peripheral bins and the apparent persistence of that variation to
high \pt in the 40-50\% centrality bin more directly reflects the
larger spatial anisotropy of the collision zone in more peripheral
collisions. The question of whether the \piz suppression measurements
presented here can be understood on the basis of geometry and jet
quenching will be more fully explored in the following section. 

\begin{figure}
\includegraphics[width=1.0\linewidth]{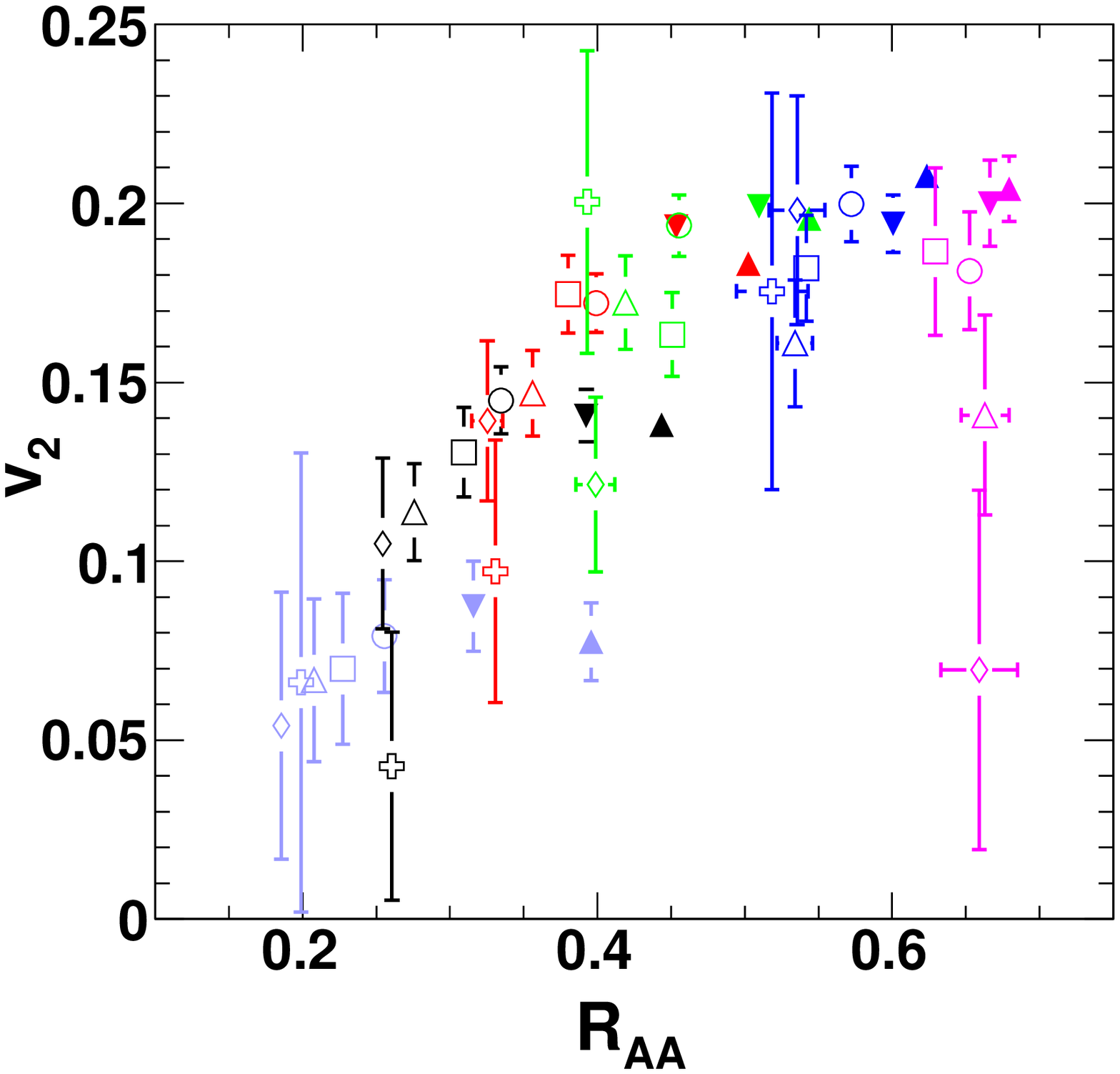}
\caption{(Color online) \piz \vtwo vs.\ inclusive $\RAA$.  The points denote
  bins in \pt as follows: triangles $2<\pt<2.5$~\GeVc, inverted triangles
  $2.5<\pt<3$~\GeVc, circles, $3<\pt<3.5$~\GeVc; squares, $3.5<\pt<4$~\GeVc;
  open triangles, $4.0<\pt<4.5$~\GeVc; diamonds, $4.5<\pt<5$~\GeVc; crosses,
  $5<\pt<6$~\GeVc.  Centrality bins are indicated by the colors: light blue,
  0-10\%; black, 10-20\%; red, 20-30\%; green, 30-40\%; blue, 40-50\%; magenta
  50-60\%.}
\label{fig:v2_vs_raa}
\end{figure}

We have observed above that the \pt and centrality dependence of \RAA
indicate a correlation between inclusive \RAA and \vtwo such that the
out-of-plane \piz yields vary only slowly with \pt or centrality while
the in-plane yields approach the out-of-plane yields with increasing
\pt or increasing \Npart. Such a correlation between these two
seemingly unrelated quantities merely indicates that the yields or
\RAA\ of \piz{}'s measured in-plane and out-of-plane more directly 
reflect the underlying physics responsible for the azimuthal variation
than the \dphi-integrated yield or \RAA and the amplitude of the \dphi
modulation, \vtwo. Indeed, we have argued
above that at higher \pt the centrality dependence of $\RAA(\dphi)$
may reflect the geometry of jet quenching. At  
intermediate \pt, the \RAApt results suggest contamination of the
in-plane yields by soft production and a simultaneous decrease in \RAA
and \vtwo with increasing \pt as the relative contribution of
collective soft processes to \piz production decreases. To more
directly demonstrate the correlation 
that forms the basis of these arguments we show in
Fig.~\ref{fig:v2_vs_raa} a plot of \vtwo versus the inclusive \RAA for
centralities from 0 to 60\%.  Data are displayed for $\pt>2$~\GeVc.
The intermediate centrality bins show a correlated increase of 
\vtwo and \RAA consistent with the discussion above and a possible
saturation of \vtwo for larger \RAA values. In fact, the trends for
different centrality bins appear to be in general agreement. However, 
their exact relationship and establishing or excluding a causal
connection requires further investigation. 

\subsection{Nuclear modification factor dependence on path length}

The centrality of a collision fixes the geometry of the overlap region between
the nuclei, and fixing the angle of emission of the particles further
constrains the path length through the medium.  We can use this feature to
study the dependence of the nuclear modification factor on the path length
traversed by the partons.  We investigate the path length dependence by
expanding on several methods previously described
in~\cite{Adler:2006bw}.  We start with three estimators of the path
length that are purely geometric, and one that includes the color
density of the medium in its calculation:
\begin{enumerate}
\item We start by modeling the overlap region as an ellipse
  defined by
  \begin{equation}
    \frac{x^2}{b^2}+\frac{y^2}{a^2} = 1
  \end{equation}
  where the minor axis $b$ is oriented in the $x$ direction and is
  parallel to the reaction plane.  The axes $a$ and $b$ are fixed by
  the intersection in the transverse plane of two hard spheres with $R
  = 6.8$~fm. In terms of the spatial eccentricity $\epsilon =
  \left<y^2-x^2\right>/\left<y^2+x^2\right>$ (often used with Glauber
  calculations), we can express the distance from the origin to the
  edge of the ellipse at a given angle:
  \begin{equation}
    L_{\epsilon}(\dphi) =
    \frac{b\sqrt{1+\epsilon}}{\sqrt{1+\epsilon\cos2\dphi}}.
      \label{eq:lepsi}
  \end{equation}
  Since this length starts at the origin, and does not take into
  account color density, the expression provides a very simple estimator
  with which we can evaluate the dependence of the \RAA on path
  length.  We will refer to the hard sphere result as
  $L_{\epsilon,hs}$.
\item Instead of an ellipse strictly defined by the transverse profile of two
  hard spheres, we model the collision region as an effective ellipse whose
  dimensions are determined by equating the RMS radius and eccentricity to the
  corresponding quantities calculated from the transverse distribution of
  participant density based on standard Glauber calculations.  This length,
  \lepsi, is evaluated using the same expression as Eq.~\ref{eq:lepsi}, with
  $b=\sqrt{x^2}$.  Both quantities are determined using the PHENIX Glauber
  model~\cite{Miller:2007ri}.
\item For a more realistic approach, we evaluate the distance along the
  parton's path weighted by the participant density,
  \begin{equation}
    \rhoLxy = \int_0^\infty dl
    \rho_{{\rm part}}(x_0+l\cos\dphi,y_0+l\sin\dphi),
    \label{eq:rhoLxy}
  \end{equation}
  where $(x_0,y_0)$ is the hard-scattering position and \dphi is the
  angle of the jet with respect to the $x$ axis.  The jet production
  point is sampled from a Monte Carlo using a weighted $T_{\rm AA}(x,y)$
  distribution and a uniform $\dphi$ distribution.  The participant
  density, assumed to be proportional to the color density, is
  calculated from the Glauber model.  The $\rhopart$ density in
  Eq.~\ref{eq:rhoLxy} is modeled using a 1D Bjorken expansion,
  \[\rho_c(\tau) \propto
  \left(\frac{\tau^2/\tau_0^2}{1+\tau^2/\tau^2_0}\right)
  \left(\frac{\tau_0}{\tau}\right).\] Thus \rhoLxy roughly represents
  LPM energy loss~\cite{Aurenche:2000gf}.  Note that $\rho_c$
  approaches the same $1/\tau$ dependence as the standard $\rho
  \propto \tau_0/\tau$ but differs from by a factor of 2 at $\tau =
  \tau_0$ (additionally this form is regular at $\tau = 0$).
\item Finally, we modify \rhoLxy by normalizing it by the value of the
  participant density at the center of the collision region, $\rhocent
  = \rhopart(0,0)$.  As a result, $\rhoLxy/\rhocent$ is an estimator
  based on geometry alone, but also accounts for the effect of the
  density distribution both on the jet production point as well as the
  path from that point to the edge of the medium.
\end{enumerate}

\begin{figure*}
\includegraphics[width=0.9\linewidth]{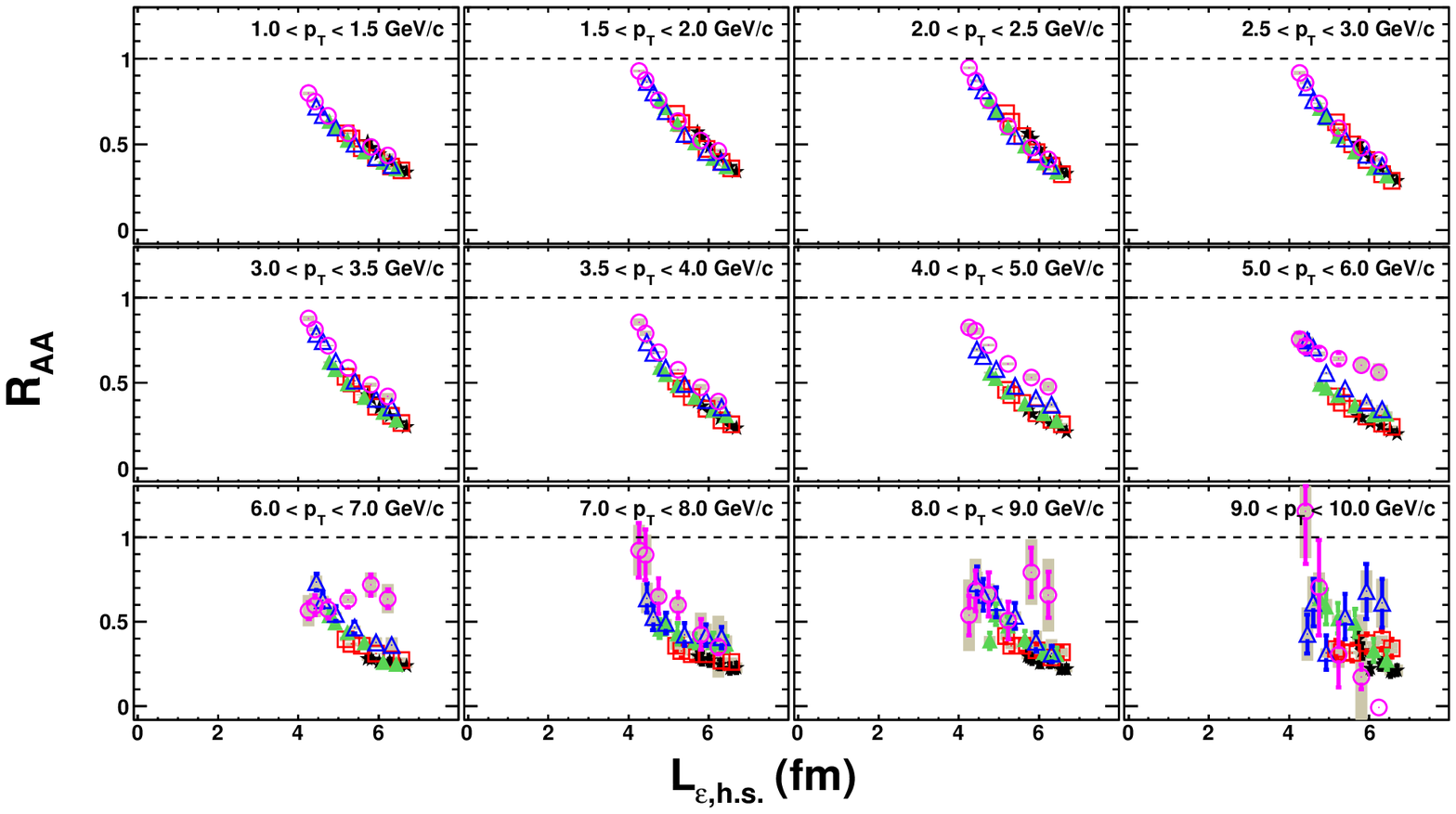}
\caption{(Color online) \piz \RAA versus \lepsi based on the hard
  sphere calculation.  Each panel corresponds to a \pt bin.  Each data
  point corresponds to a particular centrality bin and \dphi value.
  The centralities are represented as follows: (black) stars, 10-20\%;
  open (red) squares, 20-30\%; (green) triangles, 30-40\%, open (blue) 
  triangles, 40-50\%; open (magenta) circles, 50-60\%.  The height of the
  boxes represent the systematic errors on \RAA for the corresponding
  \lepsi.}
\label{fig:pi0raa_lepsihard}
\end{figure*}

\begin{figure*}
\includegraphics[width=0.9\linewidth]{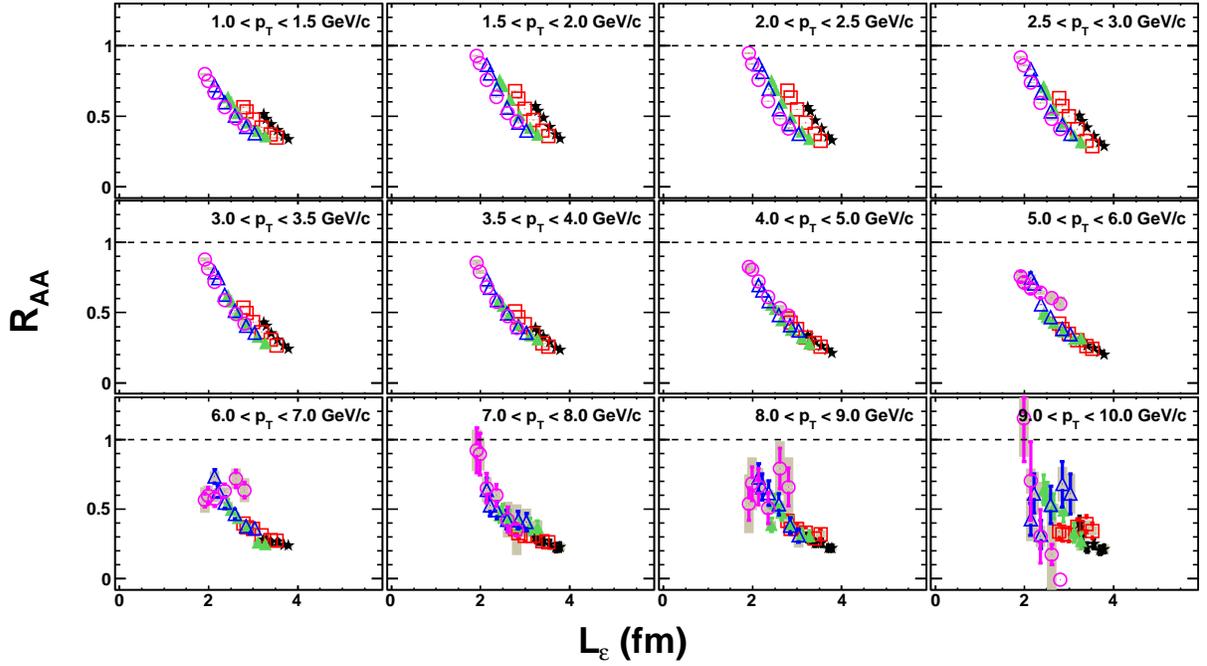}
\caption{(Color online) \piz \RAA versus \lepsi based on the RMS
  radius. Colors/data points as in Fig.~\ref{fig:pi0raa_lepsihard}.}
\label{fig:pi0raa_lepsi}
\end{figure*}

\begin{figure*}
\includegraphics[width=0.9\linewidth]{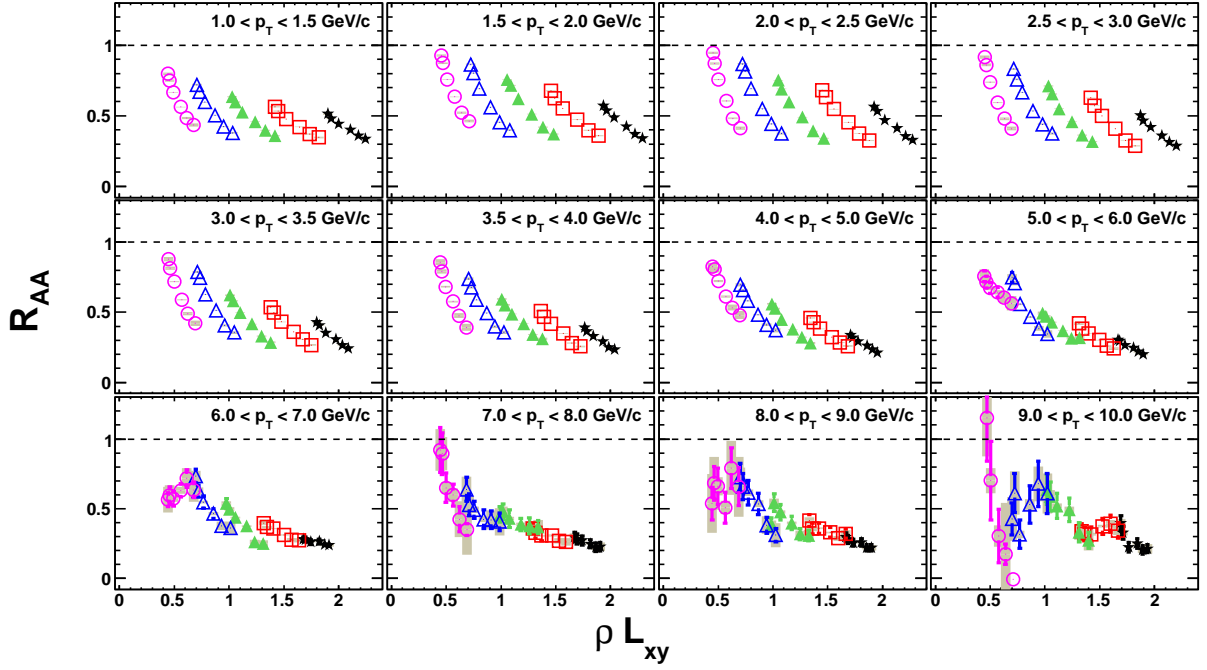}
\caption{(Color online) \piz \RAA versus \leff.  The units of \leff
  are participant $\times$ fm.  Colors/data points as in
  Fig.~\ref{fig:pi0raa_lepsihard}.}
\label{fig:pi0raa_leff}
\end{figure*}

\begin{figure*}
\includegraphics[width=0.9\linewidth]{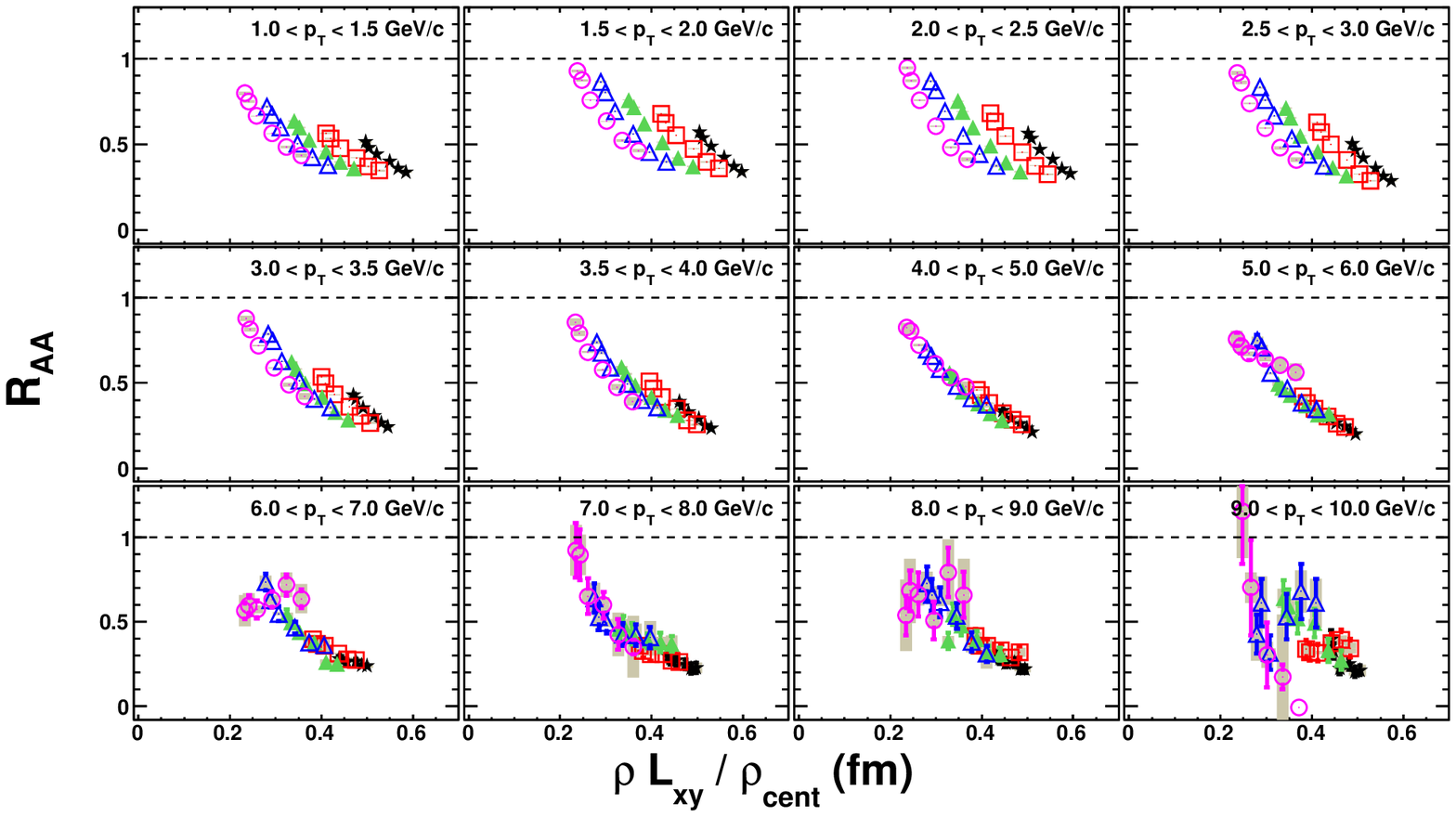}
\caption{(Color online) \piz \RAA versus $\leff/\rhocent$.  The
  units of $\leff/\rhocent$ are fm.  Colors/data points as in
  Fig.~\ref{fig:pi0raa_lepsihard}.}
\label{fig:pi0raa_rhoLxydiv}
\end{figure*}

Figs.~\ref{fig:pi0raa_lepsihard}-\ref{fig:pi0raa_rhoLxydiv} show the
\RAA dependence on $L_{\epsilon,hs}$, \Lepsi, \rhoLxy, and \rhoLxyDiv
respectively.  The results shown in this paper cover the \pt range
$\pt = 1$-$10$~\GeVc, not only extending the measurement presented
previously but allowing a much finer binning in \pt.  The statistical
errors in the \RAA measurements are represented by error bars (see
Section~\ref{sec:raa_dphi}).  The systematic errors shown in these
data are on the \RAA values only, and are indicated by the filled
boxes.  The major contribution to the systematic error in the \lepsi
values is the determination of $\npart$, and is at the 10-20\% level.

Both the $L_{\epsilon,hs}$ and \Lepsi behavior show an interesting
degree of scaling.  This result is all the more unexpected because of
the overly simple geometric picture they represent.  Despite the
simplistic picture they present, they both exhibit striking
universality: the hard sphere $\RAA(L_{\epsilon,hs})$ scales well in
the low \pt region (as high as $\pt \approx 4~\GeVc$) while
$\RAA(\lepsi)$ scales well to higher \pt, at least one bin in \pt
beyond $L_{\epsilon,hs}$ (though one might argue qualitatively this
trend continues even higher when the most peripheral centrality is
excluded).  The more precise \pt dependence available in this data set
reveals a slight deviation from the universality with \lepsi that was
previously reported~\cite{Adler:2006bw}.

By contrast, we expect the \rhoLxy estimator to provide a somewhat
more intuitive and concrete picture, as it represents a more realistic
approach to the geometry and medium.  Since we expect radiative energy
loss to play a greater role at high \pt, this should be the estimator
that would provide the best scaling.  In fact, at the higher \pt
range, a universality does emerge, though not until $\pt \approx
6$~\GeVc.  Below that value, the measured \RAA points lie on parallel,
but separate, curves.  When \rhoLxy is normalized to the central
density, data again exhibit a more universal dependence over a wider
\pt range than what is seen with \rhoLxy alone, as shown in
Fig.~\ref{fig:pi0raa_rhoLxydiv}.

When considered together these results offer a rich picture. At low to
moderate \pt simple geometry may play a larger role in determining the
final level of \RAA than conventionally thought. At higher \pt the
scaling motivated by energy loss (\rhoLxy) describes the data well. We
note that there are three possible (and not necessarily exclusive)
interpretations: 1) at low to moderate \pt the combined effects of the
boost due to expansion and fragmentation are sensitive primarily to
the difference in lengths traveled by the partons, and only weakly
dependent on other parameters 2) we need to restrict the analysis of
the \piz \RAA to $\pt > 5$~\GeVc to be in the range where
fragmentation followed by energy loss dominates, or 3) the assumption
that energy loss does not depend linearly on color density~\cite{Liao:2008dk}
is incorrect and leads a departure from scaling with \rhoLxy at low to moderate \pt.

\section{Summary and Conclusions}
\label{sec:concl}

We have presented measurements for the azimuthal anisotropy of neutral pions
in \AuAu collisions at $\snn=200$~GeV.  These measurements include the
\vtwo and $\RAA(\dphi)$ of \piz{}s as a function of transverse momentum 
and centrality.
The \vtwo has been measured from $\pt \approx 1$--$10~\GeVc$ in eight
centrality bins and four combined centrality bins.  The \RAA has been measured
in the same \pt range in six centrality bins.  In addition, the \RAA
dependence on effective path length through the collision region has been
presented for five centralities.  

The general trend seen in the $\vtwo(\pt)$ data is an initial increase in
\vtwo with increasing \pt, peaking of the \vtwo in the region of
$2<\pt<3$~\GeVc, followed by a decrease in the \vtwo.  We have argued that
such a trend implies a transition from particle production dominated by soft
processes to a \pt region dominated by hard processes.  In order to quantify
the \pt and centrality evolution of \vtwo, we have fit its \pt dependence to
an empirical expression that allows for such a transition.  While the
statistical precision of the high-\pT data limits the conclusions we can draw
from the fits, it is clear that the data support the assumption of a
decreased dominance of soft processes transitioning to an increased 
dominance of hard
process with increasing \pT.  The more peripheral bins suggest that a
\pt-independent \vtwo may be reached by $\pt \sim 5$~\GeVc.

The differential probes represented by $\RAA(\dphi)$ provide a more
sophisticated handle on the role of geometry in the collision region.  For
example, we see that in mid-centrality collisions the suppression of pions
out-of-plane is approximately the same as the suppression in more central
collisions. The data suggest that the interplay between the two main effects, 
namely collective flow and jet quenching, may take place not only along the 
expected transition in \pT from soft to hard physics, but perhaps also 
azimuthally, with the quenching effects being dominant along the direction 
normal to the reaction plane. To further shed light on the transition from 
soft to hard \pt regions, we have fit the maxima of mid-central $\RAA(\pt)$ 
in each \dphi range.  Between the in and out of plane directions, we observe a
shift of $0.4$~\GeVc in position of the peak of the spectrum.

In order to further clarify the centrality evolution of the azimuthal
dependence of the \piz suppression, we have presented the \RAA as
a function of \npart in fixed \pt bins, for three directions: along the
reaction plane, normal to the reaction plane, and midway between.  For $\npart
\gtrsim 100$, the \RAA along the normal to the reaction plane is almost 
constant, a
trend seen in most \pt bins.  By contrast, the \RAA nearest to the reaction
plane drops by almost a factor of two, converging on the out-of-plane value at the
highest \npart.  This important feature may provide the most compelling
argument for geometry as the source of suppression.  Since the path length in
the direction normal to the reaction plane varies slowly with centrality, we
would expect the \RAA to be nearly flat.  Conversely, the in-plane path length
will be sensitive to the degree of overlap and strongly influence the observed
\RAA.  Thus it would lead to a small suppression in peripheral collisions
while eventually converging on the same value as seen normal to the plane as
the anisotropy vanishes in more central collisions.  This effect is further
borne out in the correlation observed between \vtwo and the inclusive
\RAA. However, we have also argued that contamination from soft production
could produce a similar behavior and we have no independent indication of how
far in \pt soft contamination might extend. Nonetheless, under both
interpretations we can conclude that the \RAA\ for pions produced along the
normal to the reaction plane more directly reflects the physics of
quenching. We also conclude that the correlation between \RAA and \vtwo makes
separate treatment of these quantities disadvantageous.

We have examined \RAA as function of the average path length of the parent
parton through the overlap region in the collision, through the estimators
$L_{\epsilon,hs}$, \lepsi, \rhoLxyDiv, and \rhoLxy.  Each of the first three
quantities represents a progressively more sophisticated estimator for the
distance traveled by the parton, with \rhoLxy at the end providing a proxy for
LPM energy loss.
Comparison of the scaling with these three measures of 
lengths seems to suggest
that the pion suppression at low- to moderate-\pt is mostly dependent
upon simply the geometric length. The estimator that should in
principle be the most realistic one, \rhoLxy, exhibits good
universality at the highest \pT values, suggesting that energy loss
comparisons should be restricted to the \pt range $\pt>5$~\GeVc. The
importance of simple geometry at low- to moderate-\pt is further
supported when \rhoLxy is normalized by the participant density at the
center.  This normalization effectively makes \rhoLxy a length.  These
geometric descriptions offer a description of the suppression both at
low and high \pt regions, clearly showing a transition between the
ranges.  The features seen in the \RAA as function of path length tie
in consistently with the observations of a transition in the behavior
of the measured \vtwo.  These two observables, \vtwo and $\RAA(\dphi)$,
analyzed together provide a valuable set of probes for understanding
the processes governing the suppression of yields in \AuAu collisions.

\begin{acknowledgments}

We thank the staff of the Collider-Accelerator and Physics
Departments at Brookhaven National Laboratory and the staff of
the other PHENIX participating institutions for their vital
contributions.  We acknowledge support from the 
Office of Nuclear Physics in the
Office of Science of the Department of Energy, the
National Science Foundation, Abilene Christian University
Research Council, Research Foundation of SUNY, and Dean of the
College of Arts and Sciences, Vanderbilt University (U.S.A),
Ministry of Education, Culture, Sports, Science, and Technology
and the Japan Society for the Promotion of Science (Japan),
Conselho Nacional de Desenvolvimento Cient\'{\i}fico e
Tecnol{\'o}gico and Funda\c c{\~a}o de Amparo {\`a} Pesquisa do
Estado de S{\~a}o Paulo (Brazil),
Natural Science Foundation of China (People's Republic of China),
Centre National de la Recherche Scientifique, Commissariat
{\`a} l'{\'E}nergie Atomique, and Institut National de Physique
Nucl{\'e}aire et de Physique des Particules (France),
Ministry of Industry, Science and Tekhnologies,
Bundesministerium f\"ur Bildung und Forschung, Deutscher
Akademischer Austausch Dienst, and Alexander von Humboldt Stiftung (Germany),
Hungarian National Science Fund, OTKA (Hungary), 
Department of Atomic Energy (India), 
Israel Science Foundation (Israel), 
Korea Research Foundation 
and Korea Science and Engineering Foundation (Korea),
Ministry of Education and Science, Rassia Academy of Sciences,
Federal Agency of Atomic Energy (Russia),
VR and the Wallenberg Foundation (Sweden), 
the U.S. Civilian Research and Development Foundation for the
Independent States of the Former Soviet Union, the US-Hungarian
NSF-OTKA-MTA, and the US-Israel Binational Science Foundation.

\end{acknowledgments}

\clearpage

\appendix

\appendix

\section{Data Tables}

\begin{table}[htb]
\caption{\piz \vTwo for 0--5\% and 5--10\% centrality.  All errors are absolute.}
\begin{ruledtabular}\begin{tabular}{cccccc}
Centrality & $\left<p_{\rm T}\right> {\rm Gev}/c$ & $v_2$ & \multicolumn{2}{c}{Stat Err} & Syst Err \\
\hline
&1.21 & 0.052 & +0.034 & -0.034 & 0.011 \\
&1.70 & 0.035 & +0.022 & -0.022 & 0.007 \\
&2.20 & 0.051 & +0.020 & -0.020 & 0.010 \\
&2.70 & 0.076 & +0.023 & -0.023 & 0.016 \\
&3.21 & 0.039 & +0.029 & -0.029 & 0.008 \\
&3.71 & 0.059 & +0.039 & -0.039 & 0.012 \\
0--5\%
&4.37 & 0.040 & +0.042 & -0.042 & 0.008 \\
&5.40 & 0.040 & +0.070 & -0.070 & 0.008 \\
&6.41 & 0.052 & +0.115 & -0.115 & 0.011 \\
&7.43 & 0.160 & +0.206 & -0.206 & 0.033 \\
&8.43 & 0.168 & +0.146 & -0.146 & 0.034 \\
&9.44 & 0.132 & +0.171 & -0.171 & 0.027 \\
\hline
&1.21 & 0.079 & +0.020 & -0.020 & 0.010 \\
&1.71 & 0.083 & +0.013 & -0.013 & 0.010 \\
&2.20 & 0.106 & +0.012 & -0.012 & 0.013 \\
&2.70 & 0.100 & +0.014 & -0.014 & 0.012 \\
&3.21 & 0.109 & +0.018 & -0.018 & 0.014 \\
&3.71 & 0.075 & +0.024 & -0.024 & 0.009 \\
5--10\%
&4.38 & 0.091 & +0.026 & -0.026 & 0.011 \\
&5.40 & 0.064 & +0.041 & -0.041 & 0.008 \\
&6.41 & 0.062 & +0.077 & -0.077 & 0.008 \\
&7.42 & 0.054 & +0.158 & -0.158 & 0.007 \\
&8.43 & 0.002 & +0.097 & -0.097 & 0.0002 \\
&9.44 & 0.118 & +0.136 & -0.136 & 0.015 \\
\end{tabular}\end{ruledtabular}
\end{table}

\begin{table*}[tbh]
\caption{\piz \vTwo for other centralities.  All errors are absolute.}
\begin{ruledtabular}\begin{tabular}{cccccc|cccccc}
Centrality & $\left<p_{\rm T}\right> {\rm Gev}/c$ & $v_2$ & \multicolumn{2}{c}{Stat Err} & Syst Err &
Centrality & $\left<p_{\rm T}\right> {\rm Gev}/c$ & $v_2$ & \multicolumn{2}{c}{Stat Err} & Syst Err \\
\hline
&1.20 & 0.066 & +0.018 & -0.018 & 0.012  & &1.21 & 0.153 & +0.010 & -0.010 & 0.009 \\
&1.71 & 0.061 & +0.012 & -0.012 & 0.011  & &1.71 & 0.175 & +0.008 & -0.008 & 0.010 \\
&2.20 & 0.078 & +0.011 & -0.011 & 0.014  & &2.21 & 0.204 & +0.009 & -0.009 & 0.012 \\
&2.70 & 0.087 & +0.013 & -0.013 & 0.016  & &2.71 & 0.200 & +0.012 & -0.012 & 0.012 \\
&3.21 & 0.079 & +0.016 & -0.016 & 0.014  & &3.21 & 0.181 & +0.016 & -0.016 & 0.011 \\
&3.71 & 0.070 & +0.021 & -0.021 & 0.013  & &3.72 & 0.187 & +0.023 & -0.023 & 0.011 \\
0--10\% 
&4.37 & 0.067 & +0.023 & -0.023 & 0.012  & 
50--60\% &4.38 & 0.141 & +0.028 & -0.028 & 0.008 \\
&5.40 & 0.054 & +0.037 & -0.037 & 0.010  & &5.40 & 0.070 & +0.050 & -0.050 & 0.004 \\
&6.41 & 0.066 & +0.064 & -0.064 & 0.012  & &6.41 & -0.044 & +0.097 & -0.097 & 0.003 \\
&7.42 & 0.112 & +0.124 & -0.124 & 0.020  & &7.42 & 0.235 & +0.168 & -0.168 & 0.014 \\
&8.43 & 0.071 & +0.087 & -0.087 & 0.013  & &8.43 & -0.042 & +0.234 & -0.234 & 0.002 \\
&9.44 & 0.189 & +0.108 & -0.108 & 0.034  & &9.44 & 0.520 & +0.428 & -0.428 & 0.031 \\
\hline 
&1.20 & 0.106 & +0.010 & -0.010 & 0.008  & &1.20 & 0.083 & +0.011 & -0.011 & 0.011 \\
&1.71 & 0.131 & +0.006 & -0.006 & 0.010  & &1.71 & 0.091 & +0.007 & -0.007 & 0.012 \\
&2.20 & 0.138 & +0.006 & -0.006 & 0.010  & &2.20 & 0.104 & +0.007 & -0.007 & 0.014 \\
&2.70 & 0.141 & +0.007 & -0.007 & 0.010  & &2.70 & 0.110 & +0.008 & -0.008 & 0.015 \\
&3.21 & 0.145 & +0.009 & -0.009 & 0.011  & &3.21 & 0.107 & +0.010 & -0.010 & 0.015 \\
&3.71 & 0.130 & +0.012 & -0.012 & 0.010  & &3.71 & 0.097 & +0.013 & -0.013 & 0.013 \\
10--20\% 
&4.37 & 0.114 & +0.014 & -0.014 & 0.008  & 
0--20\% &4.37 & 0.088 & +0.014 & -0.014 & 0.012 \\
&5.40 & 0.105 & +0.024 & -0.024 & 0.008  & &5.40 & 0.077 & +0.023 & -0.023 & 0.010 \\
&6.41 & 0.043 & +0.037 & -0.037 & 0.003  & &6.41 & 0.056 & +0.039 & -0.039 & 0.007 \\
&7.42 & 0.075 & +0.088 & -0.088 & 0.006  & &7.42 & 0.096 & +0.081 & -0.079 & 0.013 \\
&8.43 & 0.076 & +0.059 & -0.059 & 0.006  & &8.43 & 0.074 & +0.055 & -0.054 & 0.010 \\
&9.44 & 0.151 & +0.083 & -0.083 & 0.011  & &9.44 & 0.170 & +0.074 & -0.072 & 0.023 \\
\hline 
&1.21 & 0.125 & +0.008 & -0.008 & 0.008  & &1.21 & 0.133 & +0.005 & -0.005 & 0.008 \\
&1.71 & 0.159 & +0.005 & -0.005 & 0.010  & &1.71 & 0.169 & +0.004 & -0.004 & 0.010 \\
&2.20 & 0.183 & +0.005 & -0.005 & 0.011  & &2.20 & 0.189 & +0.004 & -0.004 & 0.011 \\
&2.70 & 0.193 & +0.006 & -0.006 & 0.012  & &2.71 & 0.197 & +0.005 & -0.005 & 0.012 \\
&3.21 & 0.172 & +0.008 & -0.008 & 0.010  & &3.21 & 0.182 & +0.006 & -0.006 & 0.011 \\
&3.71 & 0.175 & +0.011 & -0.011 & 0.010  & &3.71 & 0.171 & +0.008 & -0.008 & 0.010 \\
20--30\% 
&4.38 & 0.147 & +0.012 & -0.012 & 0.009  & 
20--40\% &4.38 & 0.158 & +0.009 & -0.009 & 0.009 \\
&5.40 & 0.139 & +0.022 & -0.022 & 0.008  & &5.40 & 0.132 & +0.016 & -0.016 & 0.008 \\
&6.41 & 0.097 & +0.037 & -0.037 & 0.006  & &6.41 & 0.140 & +0.027 & -0.027 & 0.008 \\
&7.42 & 0.074 & +0.073 & -0.073 & 0.004  & &7.42 & 0.071 & +0.055 & -0.054 & 0.004 \\
&8.43 & 0.070 & +0.059 & -0.059 & 0.004  & &8.43 & 0.088 & +0.045 & -0.044 & 0.005 \\
&9.44 & -0.035 & +0.087 & -0.087 & 0.002  & &9.44 & 0.062 & +0.070 & -0.069 & 0.004 \\
\hline 
&1.21 & 0.143 & +0.007 & -0.007 & 0.008  & &1.21 & 0.159 & +0.007 & -0.007 & 0.009 \\
&1.71 & 0.181 & +0.005 & -0.005 & 0.010  & &1.71 & 0.188 & +0.005 & -0.005 & 0.010 \\
&2.20 & 0.196 & +0.005 & -0.005 & 0.011  & &2.21 & 0.208 & +0.006 & -0.006 & 0.012 \\
&2.71 & 0.199 & +0.007 & -0.007 & 0.011  & &2.71 & 0.198 & +0.008 & -0.008 & 0.011 \\
&3.21 & 0.194 & +0.009 & -0.009 & 0.011  & &3.21 & 0.195 & +0.010 & -0.010 & 0.011 \\
&3.71 & 0.163 & +0.012 & -0.012 & 0.009  & &3.72 & 0.185 & +0.014 & -0.014 & 0.010 \\
30--40\% 
&4.38 & 0.172 & +0.013 & -0.013 & 0.010  &
40--60\% & 4.38 & 0.155 & +0.017 & -0.017 & 0.009 \\
&5.40 & 0.121 & +0.024 & -0.024 & 0.007  & &5.40 & 0.149 & +0.031 & -0.030 & 0.008 \\
&6.41 & 0.200 & +0.042 & -0.042 & 0.011  & &6.41 & 0.093 & +0.056 & -0.055 & 0.005 \\
&7.43 & 0.070 & +0.088 & -0.088 & 0.004  & &7.42 & 0.152 & +0.118 & -0.113 & 0.008 \\
&8.43 & 0.113 & +0.071 & -0.071 & 0.006  & &8.43 & 0.114 & +0.126 & -0.124 & 0.006 \\
&9.44 & 0.199 & +0.118 & -0.118 & 0.011  & &9.44 & 0.168 & +0.266 & -0.231 & 0.009 \\
\hline
&1.20 & 0.160 & +0.008 & -0.008 & 0.009  & &1.20 & 0.106 & +0.009 & -0.009 & 0.013 \\
&1.71 & 0.193 & +0.006 & -0.006 & 0.010  & &1.71 & 0.125 & +0.006 & -0.006 & 0.015 \\
&2.21 & 0.208 & +0.006 & -0.006 & 0.011  & &2.20 & 0.142 & +0.006 & -0.006 & 0.017 \\
&2.71 & 0.194 & +0.008 & -0.008 & 0.010  & &2.71 & 0.146 & +0.008 & -0.008 & 0.018 \\
&3.21 & 0.200 & +0.011 & -0.011 & 0.011  & &3.21 & 0.143 & +0.011 & -0.011 & 0.017 \\
&3.72 & 0.182 & +0.015 & -0.015 & 0.010  & &3.71 & 0.136 & +0.015 & -0.015 & 0.016 \\
40--50\% 
&4.38 & 0.161 & +0.018 & -0.018 & 0.009  & 
0--92\% &4.38 & 0.127 & +0.017 & -0.017 & 0.015 \\
&5.40 & 0.198 & +0.032 & -0.032 & 0.011  & &5.40 & 0.108 & +0.033 & -0.033 & 0.013 \\
&6.41 & 0.175 & +0.055 & -0.055 & 0.009  & &6.41 & 0.084 & +0.059 & -0.059 & 0.010 \\
&7.42 & 0.106 & +0.133 & -0.133 & 0.006  & &7.42 & 0.083 & +0.136 & -0.145 & 0.009 \\
&8.43 & 0.196 & +0.112 & -0.112 & 0.011  & &8.43 & 0.113 & +0.099 & -0.109 & 0.011 \\
&9.44 & -0.084 & +0.214 & -0.214 & 0.005 & &9.44 & 0.149 & +0.156 & -0.123 & 0.012 \\
\end{tabular}\end{ruledtabular}
\end{table*}

\begin{table*}[tbh]
\caption{$R_{\rm AA}$ vs. path length for (upper) $1.0<p_{\rm T}<1.5$ and (lower) $1.5<p_{\rm T}<2.0~{\rm GeV}/c$.
\label{tab:raa_vs_leff_dataA}
} 
\begin{ruledtabular}\begin{tabular}{cclllll}
Centrality & $\dphi$ & $\lepsi$ & $\leff$ & $R_{\rm AA}$ & Stat Error(abs) & Sys Error(abs) \\
\hline
& 0--15 & 3.23 & 1.90 & 0.514 & 0.009 & 0.005 \\
& 15-30 & 3.29 & 1.93 & 0.481 & 0.009 & 0.003 \\
& 30--45 & 3.41 & 2.00 & 0.443 & 0.008 & 0.001 \\
10--20\%
& 45-60 & 3.55 & 2.11 & 0.400 & 0.007 & 0.001 \\
& 60--75 & 3.69 & 2.18 & 0.358 & 0.007 & 0.004 \\
& 75-90 & 3.78 & 2.24 & 0.337 & 0.006 & 0.006 \\
\hline 
& 0--15 & 2.78 & 1.42 & 0.565 & 0.009 & 0.004 \\
& 15-30 & 2.85 & 1.45 & 0.533 & 0.008 & 0.003 \\
& 30--45 & 2.99 & 1.52 & 0.477 & 0.008 & 0.001 \\
20--30\% 
& 45-60 & 3.19 & 1.64 & 0.421 & 0.007 & 0.001 \\
& 60--75 & 3.39 & 1.74 & 0.371 & 0.006 & 0.003 \\
& 75-90 & 3.52 & 1.82 & 0.346 & 0.006 & 0.005 \\
\hline 
& 0--15 & 2.43 & 1.03 & 0.634 & 0.009 & 0.004 \\
& 15-30 & 2.50 & 1.06 & 0.599 & 0.009 & 0.003 \\
& 30--45 & 2.66 & 1.12 & 0.526 & 0.008 & 0.001 \\
30--40\% 
& 45-60 & 2.87 & 1.24 & 0.459 & 0.007 & 0.001 \\
& 60--75 & 3.11 & 1.33 & 0.397 & 0.006 & 0.004 \\
& 75-90 & 3.27 & 1.42 & 0.360 & 0.005 & 0.005 \\
\hline 
& 0--15 & 2.14 & 0.70 & 0.721 & 0.009 & 0.005 \\
& 15-30 & 2.21 & 0.73 & 0.669 & 0.009 & 0.004 \\
& 30--45 & 2.37 & 0.78 & 0.596 & 0.008 & 0.002 \\
40--50\% 
& 45-60 & 2.59 & 0.87 & 0.505 & 0.007 & 0.002 \\
& 60--75 & 2.84 & 0.95 & 0.424 & 0.006 & 0.005 \\
& 75-90 & 3.03 & 1.03 & 0.379 & 0.005 & 0.007 \\
\hline 
& 0--15 & 1.92 & 0.44 & 0.798 & 0.010 & 0.009 \\
& 15-30 & 1.99 & 0.46 & 0.751 & 0.010 & 0.007 \\
& 30--45 & 2.14 & 0.49 & 0.665 & 0.008 & 0.002 \\
50--60\%
& 45-60 & 2.36 & 0.56 & 0.562 & 0.007 & 0.003 \\
& 60--75 & 2.61 & 0.61 & 0.485 & 0.006 & 0.007 \\
& 75-90 & 2.81 & 0.68 & 0.435 & 0.006 & 0.010 \\
\hline
Centrality & $\dphi$ & $\lepsi$ & $\leff$ & $R_{\rm AA}$ & Stat Error(abs) & Sys Error(abs) \\
\hline 
& 0--15 & 3.23 & 1.94 & 0.570 & 0.007 & 0.003 \\
& 15-30 & 3.29 & 1.96 & 0.539 & 0.007 & 0.002 \\
& 30--45 & 3.41 & 2.03 & 0.486 & 0.006 & 0.001 \\
10--20\% 
& 45-60 & 3.55 & 2.15 & 0.424 & 0.009 & 0.001 \\
& 60--75 & 3.69 & 2.22 & 0.370 & 0.005 & 0.003 \\
& 75-90 & 3.78 & 2.29 & 0.341 & 0.005 & 0.004 \\
\hline 
& 0--15 & 2.78 & 1.46 & 0.677 & 0.008 & 0.003 \\
& 15-30 & 2.85 & 1.49 & 0.626 & 0.008 & 0.002 \\
& 30--45 & 2.99 & 1.57 & 0.552 & 0.007 & 0.001 \\
20--30\%
& 45-60 & 3.19 & 1.70 & 0.475 & 0.006 & 0.001 \\
& 60--75 & 3.39 & 1.80 & 0.397 & 0.005 & 0.002 \\
& 75-90 & 3.52 & 1.89 & 0.358 & 0.004 & 0.003 \\
\hline 
& 0--15 & 2.43 & 1.06 & 0.758 & 0.009 & 0.003 \\
& 15-30 & 2.50 & 1.09 & 0.716 & 0.008 & 0.002 \\
& 30--45 & 2.66 & 1.16 & 0.622 & 0.007 & 0.001 \\
30--40\% 
& 45-60 & 2.87 & 1.28 & 0.512 & 0.006 & 0.001 \\
& 60--75 & 3.11 & 1.38 & 0.420 & 0.005 & 0.003 \\
& 75-90 & 3.27 & 1.48 & 0.370 & 0.004 & 0.004 \\
\hline 
& 0--15 & 2.14 & 0.72 & 0.865 & 0.010 & 0.004 \\
& 15-30 & 2.21 & 0.75 & 0.802 & 0.009 & 0.003 \\
& 30--45 & 2.37 & 0.80 & 0.691 & 0.008 & 0.001 \\
40--50\% 
& 45-60 & 2.59 & 0.90 & 0.561 & 0.006 & 0.001 \\
& 60--75 & 2.84 & 0.99 & 0.455 & 0.005 & 0.004 \\
& 75-90 & 3.03 & 1.08 & 0.398 & 0.005 & 0.006 \\
\hline 
& 0--15 & 1.92 & 0.45 & 0.929 & 0.010 & 0.008 \\
& 15-30 & 1.99 & 0.47 & 0.874 & 0.010 & 0.006 \\
& 30--45 & 2.14 & 0.51 & 0.757 & 0.009 & 0.002 \\
50--60\%
& 45-60 & 2.36 & 0.58 & 0.635 & 0.007 & 0.002 \\
& 60--75 & 2.61 & 0.64 & 0.523 & 0.006 & 0.007 \\
& 75-90 & 2.81 & 0.71 & 0.462 & 0.005 & 0.009 \\
\end{tabular}\end{ruledtabular}
\end{table*}

\begin{table*}[tbh]
\caption{$R_{\rm AA}$ vs. path length for (upper) $2.0<p_{\rm T}<2.5$ and (lower) $2.5<p_{\rm T}<3.0~{\rm GeV}/c$.
\label{tab:raa_vs_leff_dataC}
} 
\begin{ruledtabular}\begin{tabular}{cclllll}
Centrality & $\dphi$ & $\lepsi$ & $\leff$ & $R_{\rm AA}$ & Stat Error(abs) & Sys Error(abs) \\
\hline 
& 0--15 & 3.23 & 1.92 & 0.562 & 0.007 & 0.003 \\
& 15-30 & 3.29 & 1.95 & 0.534 & 0.007 & 0.002 \\
& 30--45 & 3.41 & 2.02 & 0.470 & 0.006 & 0.001 \\
10--20\% 
& 45-60 & 3.55 & 2.13 & 0.413 & 0.006 & 0.001 \\
& 60--75 & 3.69 & 2.21 & 0.354 & 0.005 & 0.003 \\
& 75-90 & 3.78 & 2.28 & 0.329 & 0.004 & 0.004 \\
\hline 
& 0--15 & 2.78 & 1.45 & 0.681 & 0.008 & 0.003 \\
& 15-30 & 2.85 & 1.48 & 0.633 & 0.008 & 0.002 \\
& 30--45 & 2.99 & 1.56 & 0.547 & 0.007 & 0.001 \\
20--30\%
& 45-60 & 3.19 & 1.69 & 0.452 & 0.006 & 0.001 \\
& 60--75 & 3.39 & 1.79 & 0.375 & 0.005 & 0.003 \\
& 75-90 & 3.52 & 1.88 & 0.325 & 0.004 & 0.004 \\
\hline 
& 0--15 & 2.43 & 1.05 & 0.752 & 0.009 & 0.003 \\
& 15-30 & 2.50 & 1.08 & 0.691 & 0.008 & 0.002 \\
& 30--45 & 2.66 & 1.15 & 0.597 & 0.007 & 0.001 \\
30--40\% 
& 45-60 & 2.87 & 1.27 & 0.490 & 0.006 & 0.001 \\
& 60--75 & 3.11 & 1.37 & 0.392 & 0.005 & 0.003 \\
& 75-90 & 3.27 & 1.46 & 0.339 & 0.004 & 0.004 \\
\hline 
& 0--15 & 2.14 & 0.72 & 0.869 & 0.010 & 0.005 \\
& 15-30 & 2.21 & 0.75 & 0.813 & 0.009 & 0.004 \\
& 30--45 & 2.37 & 0.80 & 0.691 & 0.008 & 0.001 \\
40--50\% 
& 45-60 & 2.59 & 0.90 & 0.550 & 0.006 & 0.001 \\
& 60--75 & 2.84 & 0.99 & 0.443 & 0.005 & 0.004 \\
& 75-90 & 3.03 & 1.08 & 0.375 & 0.004 & 0.006 \\
\hline 
& 0--15 & 1.92 & 0.45 & 0.946 & 0.011 & 0.009 \\
& 15-30 & 1.99 & 0.47 & 0.870 & 0.010 & 0.007 \\
& 30--45 & 2.14 & 0.50 & 0.757 & 0.009 & 0.003 \\
50--60\%
& 45-60 & 2.36 & 0.57 & 0.606 & 0.007 & 0.003 \\
& 60--75 & 2.61 & 0.63 & 0.482 & 0.006 & 0.007 \\
& 75-90 & 2.81 & 0.70 & 0.413 & 0.005 & 0.010 \\
\hline
Centrality & $\dphi$ & $\lepsi$ & $\leff$ & $R_{\rm AA}$ & Stat Error(abs) & Sys Error(abs) \\
\hline
& 0--15 & 3.23 & 1.87 & 0.504 & 0.007 & 0.003 \\
& 15-30 & 3.29 & 1.90 & 0.467 & 0.007 & 0.002 \\
& 30--45 & 3.41 & 1.96 & 0.421 & 0.006 & 0.001 \\
10--20\% 
& 45-60 & 3.55 & 2.07 & 0.360 & 0.005 & 0.001 \\
& 60--75 & 3.69 & 2.14 & 0.315 & 0.005 & 0.003 \\
& 75-90 & 3.78 & 2.20 & 0.288 & 0.004 & 0.004 \\
\hline 
& 0--15 & 2.78 & 1.42 & 0.626 & 0.009 & 0.003 \\
& 15-30 & 2.85 & 1.45 & 0.573 & 0.008 & 0.002 \\
& 30--45 & 2.99 & 1.52 & 0.499 & 0.007 & 0.001 \\
20--30\%
& 45-60 & 3.19 & 1.64 & 0.409 & 0.006 & 0.001 \\
& 60--75 & 3.39 & 1.74 & 0.326 & 0.005 & 0.003 \\
& 75-90 & 3.52 & 1.82 & 0.287 & 0.004 & 0.004 \\
\hline 
& 0--15 & 2.43 & 1.03 & 0.713 & 0.009 & 0.004 \\
& 15-30 & 2.50 & 1.07 & 0.654 & 0.009 & 0.003 \\
& 30--45 & 2.66 & 1.13 & 0.550 & 0.007 & 0.001 \\
30--40\% 
& 45-60 & 2.87 & 1.25 & 0.459 & 0.006 & 0.001 \\
& 60--75 & 3.11 & 1.34 & 0.364 & 0.005 & 0.003 \\
& 75-90 & 3.27 & 1.43 & 0.318 & 0.004 & 0.005 \\
\hline 
& 0--15 & 2.14 & 0.72 & 0.832 & 0.011 & 0.006 \\
& 15-30 & 2.21 & 0.74 & 0.758 & 0.010 & 0.004 \\
& 30--45 & 2.37 & 0.79 & 0.665 & 0.009 & 0.002 \\
40--50\% 
& 45-60 & 2.59 & 0.89 & 0.534 & 0.007 & 0.002 \\
& 60--75 & 2.84 & 0.98 & 0.439 & 0.006 & 0.005 \\
& 75-90 & 3.03 & 1.07 & 0.376 & 0.005 & 0.007 \\
\hline 
& 0--15 & 1.92 & 0.45 & 0.916 & 0.012 & 0.011 \\
& 15-30 & 1.99 & 0.47 & 0.861 & 0.011 & 0.008 \\
& 30--45 & 2.14 & 0.50 & 0.738 & 0.010 & 0.003 \\
50--60\%
& 45-60 & 2.36 & 0.57 & 0.594 & 0.008 & 0.003 \\
& 60--75 & 2.61 & 0.63 & 0.480 & 0.006 & 0.010 \\
& 75-90 & 2.81 & 0.70 & 0.409 & 0.006 & 0.013 \\
\end{tabular}\end{ruledtabular}
\end{table*}

\begin{table*}[tbh]
\caption{$R_{\rm AA}$ vs. path length for (upper) $3.0<p_{\rm T}<3.5$ and (lower) $3.5<p_{\rm T}<4.0~{\rm GeV}/c$.
\label{tab:raa_vs_leff_dataE}
} 
\begin{ruledtabular}\begin{tabular}{cclllll}
Centrality & $\dphi$ & $\lepsi$ & $\leff$ & $R_{\rm AA}$ & Stat Error(abs) & Sys Error(abs) \\
\hline
& 0--15 & 3.23 & 1.80 & 0.430 & 0.008 & 0.003 \\
& 15-30 & 3.29 & 1.82 & 0.405 & 0.007 & 0.003 \\
& 30--45 & 3.41 & 1.88 & 0.355 & 0.006 & 0.001 \\
10--20\% 
& 45-60 & 3.55 & 1.97 & 0.307 & 0.006 & 0.001 \\
& 60--75 & 3.69 & 2.04 & 0.266 & 0.005 & 0.003 \\
& 75-90 & 3.78 & 2.09 & 0.243 & 0.004 & 0.004 \\
\hline 
& 0--15 & 2.78 & 1.38 & 0.534 & 0.009 & 0.003 \\
& 15-30 & 2.85 & 1.41 & 0.497 & 0.008 & 0.003 \\
& 30--45 & 2.99 & 1.48 & 0.431 & 0.007 & 0.001 \\
20--30\%
& 45-60 & 3.19 & 1.59 & 0.360 & 0.006 & 0.001 \\
& 60--75 & 3.39 & 1.67 & 0.305 & 0.005 & 0.003 \\
& 75-90 & 3.52 & 1.75 & 0.267 & 0.005 & 0.004 \\
\hline 
& 0--15 & 2.43 & 1.01 & 0.622 & 0.010 & 0.004 \\
& 15-30 & 2.50 & 1.04 & 0.582 & 0.009 & 0.003 \\
& 30--45 & 2.66 & 1.10 & 0.498 & 0.008 & 0.001 \\
30--40\% 
& 45-60 & 2.87 & 1.21 & 0.417 & 0.007 & 0.001 \\
& 60--75 & 3.11 & 1.30 & 0.328 & 0.005 & 0.004 \\
& 75-90 & 3.27 & 1.38 & 0.285 & 0.005 & 0.006 \\
\hline 
& 0--15 & 2.14 & 0.71 & 0.788 & 0.012 & 0.007 \\
& 15-30 & 2.21 & 0.74 & 0.745 & 0.012 & 0.006 \\
& 30--45 & 2.37 & 0.79 & 0.625 & 0.010 & 0.002 \\
40--50\% 
& 45-60 & 2.59 & 0.88 & 0.514 & 0.008 & 0.002 \\
& 60--75 & 2.84 & 0.96 & 0.407 & 0.007 & 0.007 \\
& 75-90 & 3.03 & 1.05 & 0.355 & 0.006 & 0.009 \\
\hline 
& 0--15 & 1.92 & 0.45 & 0.880 & 0.015 & 0.015 \\
& 15-30 & 1.99 & 0.46 & 0.814 & 0.014 & 0.011 \\
& 30--45 & 2.14 & 0.50 & 0.720 & 0.012 & 0.004 \\
50--60\%
& 45-60 & 2.36 & 0.57 & 0.589 & 0.010 & 0.004 \\
& 60--75 & 2.61 & 0.63 & 0.490 & 0.008 & 0.013 \\
& 75-90 & 2.81 & 0.69 & 0.421 & 0.007 & 0.018 \\
\hline
Centrality & $\dphi$ & $\lepsi$ & $\leff$ & $R_{\rm AA}$ & Stat Error(abs) & Sys Error(abs) \\
\hline 
& 0--15 & 3.23 & 1.76 & 0.390 & 0.009 & 0.004 \\
& 15-30 & 3.29 & 1.78 & 0.364 & 0.008 & 0.003 \\
& 30--45 & 3.41 & 1.84 & 0.329 & 0.007 & 0.001 \\
10--20\% 
& 45-60 & 3.55 & 1.93 & 0.291 & 0.007 & 0.001 \\
& 60--75 & 3.69 & 1.99 & 0.249 & 0.006 & 0.004 \\
& 75-90 & 3.78 & 2.04 & 0.234 & 0.005 & 0.005 \\
\hline 
& 0--15 & 2.78 & 1.36 & 0.511 & 0.011 & 0.004 \\
& 15-30 & 2.85 & 1.39 & 0.468 & 0.010 & 0.003 \\
& 30--45 & 2.99 & 1.46 & 0.418 & 0.009 & 0.001 \\
20--30\%
& 45-60 & 3.19 & 1.57 & 0.348 & 0.007 & 0.001 \\
& 60--75 & 3.39 & 1.65 & 0.281 & 0.006 & 0.004 \\
& 75-90 & 3.52 & 1.72 & 0.256 & 0.005 & 0.006 \\
\hline 
& 0--15 & 2.43 & 1.01 & 0.593 & 0.012 & 0.006 \\
& 15-30 & 2.50 & 1.04 & 0.552 & 0.011 & 0.004 \\
& 30--45 & 2.66 & 1.10 & 0.487 & 0.010 & 0.002 \\
30--40\% 
& 45-60 & 2.87 & 1.21 & 0.418 & 0.008 & 0.002 \\
& 60--75 & 3.11 & 1.29 & 0.341 & 0.007 & 0.005 \\
& 75-90 & 3.27 & 1.38 & 0.311 & 0.007 & 0.007 \\
\hline 
& 0--15 & 2.14 & 0.70 & 0.738 & 0.015 & 0.010 \\
& 15-30 & 2.21 & 0.73 & 0.676 & 0.014 & 0.007 \\
& 30--45 & 2.37 & 0.78 & 0.588 & 0.012 & 0.003 \\
40--50\% 
& 45-60 & 2.59 & 0.87 & 0.495 & 0.010 & 0.003 \\
& 60--75 & 2.84 & 0.95 & 0.397 & 0.008 & 0.009 \\
& 75-90 & 3.03 & 1.03 & 0.356 & 0.008 & 0.012 \\
\hline 
& 0--15 & 1.92 & 0.44 & 0.856 & 0.019 & 0.021 \\
& 15-30 & 1.99 & 0.46 & 0.791 & 0.017 & 0.015 \\
& 30--45 & 2.14 & 0.50 & 0.682 & 0.015 & 0.006 \\
50--60\%
& 45-60 & 2.36 & 0.56 & 0.577 & 0.013 & 0.006 \\
& 60--75 & 2.61 & 0.62 & 0.477 & 0.011 & 0.018 \\
& 75-90 & 2.81 & 0.68 & 0.392 & 0.009 & 0.023 \\
\end{tabular}\end{ruledtabular}
\end{table*}

\begin{table*}[tbh]
\caption{$R_{\rm AA}$ vs. path length for (upper) $4.0<p_{\rm T}<5.0$ and (lower) $5.0<p_{\rm T}<6.0~{\rm GeV}/c$.
\label{tab:raa_vs_leff_dataG}
} 
\begin{ruledtabular}\begin{tabular}{cclllll}
Centrality & $\dphi$ & $\lepsi$ & $\leff$ & $R_{\rm AA}$ & Stat Error(abs) & Sys Error(abs) \\
\hline 
& 0--15 & 3.23 & 1.71 & 0.336 & 0.010 & 0.004 \\
& 15-30 & 3.29 & 1.73 & 0.319 & 0.009 & 0.003 \\
& 30--45 & 3.41 & 1.78 & 0.292 & 0.009 & 0.001 \\
10--20\% 
& 45-60 & 3.55 & 1.86 & 0.261 & 0.008 & 0.001 \\
& 60--75 & 3.69 & 1.91 & 0.234 & 0.007 & 0.004 \\
& 75-90 & 3.78 & 1.95 & 0.213 & 0.006 & 0.005 \\
\hline 
& 0--15 & 2.78 & 1.34 & 0.461 & 0.012 & 0.004 \\
& 15-30 & 2.85 & 1.37 & 0.430 & 0.012 & 0.003 \\
& 30--45 & 2.99 & 1.43 & 0.381 & 0.010 & 0.001 \\
20--30\%
& 45-60 & 3.19 & 1.54 & 0.322 & 0.009 & 0.001 \\
& 60--75 & 3.39 & 1.61 & 0.284 & 0.008 & 0.004 \\
& 75-90 & 3.52 & 1.68 & 0.258 & 0.007 & 0.006 \\
\hline 
& 0--15 & 2.43 & 0.99 & 0.559 & 0.015 & 0.006 \\
& 15-30 & 2.50 & 1.02 & 0.527 & 0.014 & 0.005 \\
& 30--45 & 2.66 & 1.08 & 0.448 & 0.012 & 0.002 \\
30--40\% 
& 45-60 & 2.87 & 1.18 & 0.379 & 0.010 & 0.002 \\
& 60--75 & 3.11 & 1.26 & 0.322 & 0.009 & 0.006 \\
& 75-90 & 3.27 & 1.34 & 0.279 & 0.008 & 0.008 \\
\hline 
& 0--15 & 2.14 & 0.70 & 0.697 & 0.020 & 0.012 \\
& 15-30 & 2.21 & 0.72 & 0.658 & 0.019 & 0.009 \\
& 30--45 & 2.37 & 0.77 & 0.583 & 0.016 & 0.003 \\
40--50\% 
& 45-60 & 2.59 & 0.87 & 0.483 & 0.014 & 0.004 \\
& 60--75 & 2.84 & 0.94 & 0.411 & 0.012 & 0.010 \\
& 75-90 & 3.03 & 1.02 & 0.372 & 0.011 & 0.014 \\
\hline 
& 0--15 & 1.92 & 0.45 & 0.826 & 0.026 & 0.026 \\
& 15-30 & 1.99 & 0.47 & 0.804 & 0.025 & 0.020 \\
& 30--45 & 2.14 & 0.50 & 0.722 & 0.022 & 0.008 \\
50--60\%
& 45-60 & 2.36 & 0.57 & 0.611 & 0.019 & 0.008 \\
& 60--75 & 2.61 & 0.63 & 0.534 & 0.017 & 0.022 \\
& 75-90 & 2.81 & 0.70 & 0.480 & 0.0155 & 0.030 \\
\hline
Centrality & $\dphi$ & $\lepsi$ & $\leff$ & $R_{\rm AA}$ & Stat Error(abs) & Sys Error(abs) \\
\hline
& 0--15 & 3.23 & 1.67 & 0.301 & 0.014 & 0.007 \\
& 15-30 & 3.29 & 1.68 & 0.294 & 0.013 & 0.005 \\
& 30--45 & 3.41 & 1.73 & 0.267 & 0.012 & 0.002 \\
10--20\% 
& 45-60 & 3.55 & 1.81 & 0.246 & 0.011 & 0.002 \\
& 60--75 & 3.69 & 1.85 & 0.218 & 0.010 & 0.006 \\
& 75-90 & 3.78 & 1.90 & 0.199 & 0.009 & 0.008 \\
\hline 
& 0--15 & 2.78 & 1.31 & 0.420 & 0.018 & 0.008 \\
& 15-30 & 2.85 & 1.34 & 0.382 & 0.017 & 0.006 \\
& 30--45 & 2.99 & 1.40 & 0.347 & 0.015 & 0.002 \\
20--30\%
& 45-60 & 3.19 & 1.50 & 0.301 & 0.013 & 0.002 \\
& 60--75 & 3.39 & 1.56 & 0.260 & 0.012 & 0.007 \\
& 75-90 & 3.52 & 1.63 & 0.240 & 0.011 & 0.010 \\
\hline 
& 0--15 & 2.43 & 0.98 & 0.492 & 0.022 & 0.011 \\
& 15-30 & 2.50 & 1.01 & 0.469 & 0.020 & 0.008 \\
& 30--45 & 2.66 & 1.06 & 0.428 & 0.019 & 0.003 \\
30--40\% 
& 45-60 & 2.87 & 1.17 & 0.370 & 0.017 & 0.003 \\
& 60--75 & 3.11 & 1.24 & 0.315 & 0.015 & 0.009 \\
& 75-90 & 3.27 & 1.32 & 0.318 & 0.014 & 0.014 \\
\hline 
& 0--15 & 2.14 & 0.70 & 0.750 & 0.035 & 0.021 \\
& 15-30 & 2.21 & 0.72 & 0.708 & 0.032 & 0.016 \\
& 30--45 & 2.37 & 0.77 & 0.557 & 0.028 & 0.006 \\
40--50\% 
& 45-60 & 2.59 & 0.87 & 0.466 & 0.023 & 0.006 \\
& 60--75 & 2.84 & 0.94 & 0.385 & 0.019 & 0.019 \\
& 75-90 & 3.03 & 1.02 & 0.345 & 0.017 & 0.027 \\
\hline 
& 0--15 & 1.92 & 0.45 & 0.757 & 0.039 & 0.050 \\
& 15-30 & 1.99 & 0.47 & 0.715 & 0.040 & 0.036 \\
& 30--45 & 2.14 & 0.50 & 0.672 & 0.038 & 0.013 \\
50--60\%
& 45-60 & 2.36 & 0.57 & 0.642 & 0.033 & 0.014 \\
& 60--75 & 2.61 & 0.63 & 0.605 & 0.032 & 0.038 \\
& 75-90 & 2.81 & 0.69 & 0.563 & 0.030 & 0.051 \\
\end{tabular}\end{ruledtabular}
\end{table*}

\begin{table*}[tbh]
\caption{$R_{\rm AA}$ vs. path length for (upper) $6.0<p_{\rm T}<7.0$ and (lower) $7.0<p_{\rm T}<8.0~{\rm GeV}/c$.
\label{tab:raa_vs_leff_dataI}
} 
\begin{ruledtabular}\begin{tabular}{cclllll}
Centrality & $\dphi$ & $\lepsi$ & $\leff$ & $R_{\rm AA}$ & Stat Error(abs) & Sys Error(abs) \\
\hline 
& 0--15 & 3.23 & 1.68 & 0.281 & 0.017 & 0.011 \\
& 15-30 & 3.29 & 1.70 & 0.278 & 0.017 & 0.008 \\
& 30--45 & 3.41 & 1.75 & 0.261 & 0.017 & 0.003 \\
10--20\% 
& 45-60 & 3.55 & 1.82 & 0.264 & 0.018 & 0.003 \\
& 60--75 & 3.69 & 1.87 & 0.241 & 0.015 & 0.009 \\
& 75-90 & 3.78 & 1.91 & 0.240 & 0.015 & 0.012 \\
\hline 
& 0--15 & 2.78 & 1.32 & 0.396 & 0.026 & 0.013 \\
& 15-30 & 2.85 & 1.34 & 0.368 & 0.023 & 0.010 \\
& 30--45 & 2.99 & 1.40 & 0.357 & 0.023 & 0.004 \\
20--30\%
& 45-60 & 3.19 & 1.50 & 0.312 & 0.021 & 0.004 \\
& 60--75 & 3.39 & 1.57 & 0.277 & 0.018 & 0.011 \\
& 75-90 & 3.52 & 1.64 & 0.274 & 0.017 & 0.016 \\
\hline 
& 0--15 & 2.43 & 0.98 & 0.537 & 0.035 & 0.018 \\
& 15-30 & 2.50 & 1.00 & 0.496 & 0.030 & 0.013 \\
& 30--45 & 2.66 & 1.06 & 0.437 & 0.029 & 0.005 \\
30--40\% 
& 45-60 & 2.87 & 1.16 & 0.376 & 0.026 & 0.006 \\
& 60--75 & 3.11 & 1.23 & 0.261 & 0.018 & 0.015 \\
& 75-90 & 3.27 & 1.31 & 0.249 & 0.019 & 0.025 \\
\hline 
& 0--15 & 2.14 & 0.70 & 0.734 & 0.050 & 0.038 \\
& 15-30 & 2.21 & 0.72 & 0.625 & 0.043 & 0.026 \\
& 30--45 & 2.37 & 0.77 & 0.546 & 0.044 & 0.010 \\
40--50\% 
& 45-60 & 2.59 & 0.86 & 0.465 & 0.036 & 0.011 \\
& 60--75 & 2.84 & 0.93 & 0.377 & 0.029 & 0.030 \\
& 75-90 & 3.03 & 1.01 & 0.363 & 0.027 & 0.046 \\
\hline 
& 0--15 & 1.92 & 0.44 & 0.564 & 0.051 & 0.093 \\
& 15-30 & 1.99 & 0.46 & 0.597 & 0.059 & 0.070 \\
& 30--45 & 2.14 & 0.49 & 0.572 & 0.054 & 0.023 \\
50--60\%
& 45-60 & 2.36 & 0.56 & 0.631 & 0.049 & 0.024 \\
& 60--75 & 2.61 & 0.62 & 0.717 & 0.063 & 0.072 \\
& 75-90 & 2.81 & 0.68 & 0.635 & 0.056 & 0.085 \\
\hline
Centrality & $\dphi$ & $\lepsi$ & $\leff$ & $R_{\rm AA}$ & Stat Error(abs) & Sys Error(abs) \\
\hline 
& 0--15 & 3.23 & 1.67 & 0.297 & 0.032 & 0.026 \\
& 15-30 & 3.29 & 1.69 & 0.282 & 0.037 & 0.019 \\
& 30--45 & 3.41 & 1.74 & 0.272 & 0.034 & 0.007 \\
10--20\%
& 45-60 & 3.55 & 1.82 & 0.248 & 0.030 & 0.007 \\
& 60--75 & 3.69 & 1.86 & 0.222 & 0.031 & 0.020 \\
& 75-90 & 3.78 & 1.91 & 0.227 & 0.030 & 0.030 \\
\hline 
& 0--15 & 2.78 & 1.29 & 0.359 & 0.037 & 0.025 \\
& 15-30 & 2.85 & 1.31 & 0.327 & 0.040 & 0.018 \\
& 30--45 & 2.99 & 1.37 & 0.306 & 0.035 & 0.007 \\
20--30\%
& 45-60 & 3.19 & 1.47 & 0.308 & 0.033 & 0.007 \\
& 60--75 & 3.39 & 1.53 & 0.270 & 0.033 & 0.020 \\
& 75-90 & 3.52 & 1.59 & 0.261 & 0.029 & 0.027 \\
\hline 
& 0--15 & 2.43 & 0.99 & 0.458 & 0.055 & 0.042 \\
& 15-30 & 2.50 & 1.02 & 0.482 & 0.052 & 0.033 \\
& 30--45 & 2.66 & 1.08 & 0.429 & 0.063 & 0.012 \\
30--40\% 
& 45-60 & 2.87 & 1.18 & 0.387 & 0.044 & 0.012 \\
& 60--75 & 3.11 & 1.26 & 0.377 & 0.057 & 0.034 \\
& 75-90 & 3.27 & 1.34 & 0.369 & 0.045 & 0.048 \\
\hline 
& 0--15 & 2.14 & 0.69 & 0.639 & 0.086 & 0.091 \\
& 15-30 & 2.21 & 0.71 & 0.527 & 0.079 & 0.058 \\
& 30--45 & 2.37 & 0.75 & 0.492 & 0.062 & 0.022 \\
40--50\% 
& 45-60 & 2.59 & 0.84 & 0.424 & 0.066 & 0.022 \\
& 60--75 & 2.84 & 0.91 & 0.420 & 0.062 & 0.069 \\
& 75-90 & 3.03 & 0.99 & 0.406 & 0.063 & 0.099 \\
\hline 
& 0--15 & 1.92 & 0.45 & 0.922 & 0.161 & 0.149 \\
& 15-30 & 1.99 & 0.46 & 0.895 & 0.150 & 0.118 \\
& 30--45 & 2.14 & 0.50 & 0.651 & 0.106 & 0.039 \\
50--60\%
& 45-60 & 2.36 & 0.56 & 0.598 & 0.078 & 0.048 \\
& 60--75 & 2.61 & 0.62 & 0.424 & 0.089 & 0.128 \\
& 75-90 & 2.81 & 0.69 & 0.351 & 0.042 & 0.182 \\
\end{tabular}\end{ruledtabular}
\end{table*}

\begin{table*}[tbh]
\caption{$R_{\rm AA}$ vs. path length for (upper) $8.0<p_{\rm T}<9.0$ and (lower) $9.0<p_{\rm T}<10.0~{\rm GeV}/c$.
\label{tab:raa_vs_leff_dataK}
} 
\begin{ruledtabular}\begin{tabular}{cclllll}
Centrality & $\dphi$ & $\lepsi$ & $\leff$ & $R_{\rm AA}$ & Stat Error(abs) & Sys Error(abs) \\
\hline 
& 0--15 & 3.23 & 1.67 & 0.296 & 0.030 & 0.017 \\
& 15-30 & 3.29 & 1.68 & 0.280 & 0.029 & 0.013 \\
& 30--45 & 3.41 & 1.73 & 0.255 & 0.026 & 0.005 \\
10--20\% 
& 45-60 & 3.55 & 1.81 & 0.258 & 0.026 & 0.005 \\
& 60--75 & 3.69 & 1.86 & 0.220 & 0.023 & 0.013 \\
& 75-90 & 3.78 & 1.90 & 0.219 & 0.022 & 0.019 \\
\hline 
& 0--15 & 2.78 & 1.33 & 0.415 & 0.043 & 0.024 \\
& 15-30 & 2.85 & 1.36 & 0.356 & 0.038 & 0.016 \\
& 30--45 & 2.99 & 1.42 & 0.359 & 0.038 & 0.006 \\
20--30\%
& 45-60 & 3.19 & 1.52 & 0.329 & 0.035 & 0.006 \\
& 60--75 & 3.39 & 1.59 & 0.288 & 0.032 & 0.017 \\
& 75-90 & 3.52 & 1.66 & 0.320 & 0.033 & 0.027 \\
\hline 
& 0--15 & 2.43 & 0.98 & 0.387 & 0.049 & 0.026 \\
& 15-30 & 2.50 & 1.01 & 0.543 & 0.063 & 0.029 \\
& 30--45 & 2.66 & 1.07 & 0.476 & 0.056 & 0.010 \\
30--40\% 
& 45-60 & 2.87 & 1.17 & 0.396 & 0.048 & 0.010 \\
& 60--75 & 3.11 & 1.25 & 0.318 & 0.041 & 0.026 \\
& 75-90 & 3.27 & 1.32 & 0.308 & 0.039 & 0.037 \\
\hline 
& 0--15 & 2.14 & 0.70 & 0.724 & 0.103 & 0.072 \\
& 15-30 & 2.21 & 0.73 & 0.658 & 0.094 & 0.053 \\
& 30--45 & 2.37 & 0.77 & 0.614 & 0.087 & 0.022 \\
40--50\% 
& 45-60 & 2.59 & 0.87 & 0.535 & 0.076 & 0.025 \\
& 60--75 & 2.84 & 0.94 & 0.382 & 0.058 & 0.065 \\
& 75-90 & 3.03 & 1.03 & 0.310 & 0.048 & 0.086 \\
\hline 
& 0--15 & 1.92 & 0.45 & 0.536 & 0.118 & 0.212 \\
& 15-30 & 1.99 & 0.46 & 0.682 & 0.121 & 0.192 \\
& 30--45 & 2.14 & 0.50 & 0.661 & 0.130 & 0.065 \\
50--60\%
& 45-60 & 2.36 & 0.56 & 0.508 & 0.112 & 0.047 \\
& 60--75 & 2.61 & 0.62 & 0.791 & 0.144 & 0.193 \\
& 75-90 & 2.81 & 0.69 & 0.658 & 0.135 & 0.215 \\
\hline
Centrality & $\dphi$ & $\lepsi$ & $\leff$ & $R_{\rm AA}$ & Stat Error(abs) & Sys Error(abs) \\
\hline
& 0--15 & 3.23 & 1.69 & 0.397 & 0.051 & 0.028 \\
& 15-30 & 3.29 & 1.71 & 0.327 & 0.045 & 0.018 \\
& 30--45 & 3.41 & 1.76 & 0.223 & 0.034 & 0.005 \\
10--20\%
& 45-60 & 3.55 & 1.84 & 0.249 & 0.035 & 0.007 \\
& 60--75 & 3.69 & 1.89 & 0.203 & 0.030 & 0.021 \\
& 75-90 & 3.78 & 1.94 & 0.213 & 0.030 & 0.034 \\
\hline 
& 0--15 & 2.78 & 1.33 & 0.340 & 0.054 & 0.038 \\
& 15-30 & 2.85 & 1.36 & 0.319 & 0.051 & 0.025 \\
& 30--45 & 2.99 & 1.42 & 0.316 & 0.052 & 0.009 \\
20--30\%
& 45-60 & 3.19 & 1.53 & 0.374 & 0.058 & 0.010 \\
& 60--75 & 3.39 & 1.60 & 0.392 & 0.061 & 0.027 \\
& 75-90 & 3.52 & 1.67 & 0.343 & 0.055 & 0.032 \\
\hline 
& 0--15 & 2.43 & 1.02 & 0.637 & 0.108 & 0.059 \\
& 15-30 & 2.50 & 1.05 & 0.588 & 0.103 & 0.044 \\
& 30--45 & 2.66 & 1.11 & 0.519 & 0.090 & 0.018 \\
30--40\% 
& 45-60 & 2.87 & 1.22 & 0.491 & 0.084 & 0.022 \\
& 60--75 & 3.11 & 1.31 & 0.331 & 0.069 & 0.054 \\
& 75-90 & 3.27 & 1.40 & 0.274 & 0.057 & 0.075 \\
\hline 
& 0--15 & 2.14 & 0.70 & 0.428 & 0.115 & 0.158 \\
& 15-30 & 2.21 & 0.72 & 0.612 & 0.139 & 0.155 \\
& 30--45 & 2.37 & 0.77 & 0.316 & 0.102 & 0.026 \\
40--50\% 
& 45-60 & 2.59 & 0.86 & 0.530 & 0.133 & 0.039 \\
& 60--75 & 2.84 & 0.94 & 0.678 & 0.162 & 0.125 \\
& 75-90 & 3.03 & 1.02 & 0.609 & 0.143 & 0.145 \\
\hline 
& 0--15 & 1.92 & 0.45 & 1.866 & 0.486 & 0.498 \\
& 15-30 & 1.99 & 0.47 & 1.150 & 0.310 & 0.273 \\
& 30--45 & 2.14 & 0.51 & 0.702 & 0.282 & 0.091 \\
50--60\%
& 45-60 & 2.36 & 0.58 & 0.305 & 0.192 & 0.077 \\
& 60--75 & 2.61 & 0.64 & 0.172 & 0.073 & 0.369 \\
& 75-90 & 2.81 & 0.71 & -0.005 & -0.002 & -0.812 \\
\end{tabular}\end{ruledtabular}
\end{table*}

\clearpage


\hyphenation{Post-Script Sprin-ger}

\end{document}